%% file: Horndeski_hybrid_stars.tex
\documentclass[12pt]{iopart}

\usepackage{pgfplots}

\usepackage[normalem]{ulem}
\usepackage{graphicx}
\usepackage{dcolumn}
\usepackage{bm}
\usepackage{multirow}
\usepackage{amssymb}
\usepackage{verbatim}
\usepackage{mathrsfs}
\usepackage{amsfonts}
\usepackage{latexsym}
\usepackage{units}
\usepackage{enumerate}

\begin{document}
	%----------------------------------------------------------------------------------------
	%	TITLE AND AUTHOR(S)
	%----------------------------------------------------------------------------------------
	\title[Horndeski fermion-boson stars]{Horndeski fermion-boson stars}
	
	\author{Armando A. Roque$^a$, L. Arturo Ure\~na-López$^a$}
	
	\address{${}^a$Departamento de Física, División de Ciencias e Ingenierías, Campus León, Universidad de Guanajuato, C.P. 37150, León, México}
	
	\vspace{10pt}
	\begin{indented}
		\item[]December 2021
	\end{indented}

	\begin{abstract}
	We establish the existence of static and spherically symmetric fermion-boson stars, in a low energy effective model of (beyond) Horndeski theories. These stars are in equilibrium, and are composed by a mixing of scalar and fermionic matters that only interact gravitationally one with each other. Properties such as mass, radius, and compactness are studied, highlighting the existence of two families of configurations defined by the parameter $c_4$. These families have distinctive properties, although in certain limits both are reduced to their counterparts in General Relativity. Finally, by assuming the same conditions used in General Relativity, we find the maximum compactness of these hybrid stars and determine that it remains below the so-called Buchdahl's limit.
	\end{abstract}

\vspace{2pc}

\vspace{2pc}

\submitto{\CQG}
	
	%----------------------------------------------------------------------------------------
	%	INTRODUCTION
	%----------------------------------------------------------------------------------------

\section{\label{sec:level1} Introduction}
The physics of compact objects such as black holes and neutron stars (NS) has received increased attention since LIGO's first detection of gravitational waves (GW) emitted by a black hole merger~\cite{Abbott:2016blz}. To date, the LIGO-VIRGO collaboration has confirmed around to $50$ merger events~\cite{LIGOScientific:2018mvr, Abbott:2020niy, Abbott:2020gyp, Abbott:2020jks}. Some of these have shown interesting implications, such as GW170817~\cite{TheLIGOScientific:2017qsa}, and GW190521~\cite{Abbott:2020tfl}. The first of this (joined with its electromagnetic counterpart GRB 170817A~\cite{Goldstein:2017mmi}) imposed that the speed of propagation of GWs, is equal to the light speed~\cite{TheLIGOScientific:2017qsa, Goldstein:2017mmi, Monitor:2017mdv}, ruling out (or constrained) a large sector of General Relativity (GR) modifications~\cite{Amendola:2017orw, Ezquiaga:2017ekz, Copeland:2018yuh, Creminelli:2017sry}.\footnote{This condition restricts severely cosmological dark energy models where the scalar field is assumed to be homogeneously distributed in space. This is not the case for this paper, where the scalar field is clumped in localized configurations. However, using some criteria (see e.g.~\cite{Copeland:2018yuh, Bahamonde:2019shr, Bahamonde:2019ipm, Bordin:2020fww}), it is possible to satisfy this restriction at cosmological contexts.} In the case of GW190521 signal, it was interpreted as a quasi-circular merger of black holes, but a recently work~\cite{Bustillo:2020syj} show that it is also consistent with numerically simulated signals from head-on collisions of two horizonless Proca stars. 

In the foreseeable future, with the aid of improved sensitivities of the current and future generations of GW detectors such as: LIGO, Kamioka Gravitational Wave Detector (KAGRA)~\cite{Somiya:2011np, KAGRA:2018plz, KAGRA:2020cvd}, LIGO-Indian (IndIGO)~\cite{Unnikrishnan:2013qwa}, TianQin/Taiji~\cite{2021arXiv210907442G}, and a bigger sample of events, it will be possible to discern between the different possibilities of GW sources, and even to search for the GW signature of a large variety of astrophysical objects, including those that are, or not, predicted within the framework of General Relativity (GR) and the Standard Model of Particle Physics.

For instance, it has long been known that massive scalar fields are able to form self-gravitating configurations without the need of additional matter. These objects, generically known as boson stars (BS), first appeared in the literature in the late sixties~\cite{Ruffini:1969qy}, and have since been widely studied either as astrophysically viable objects, e.g. black hole mimickers, or as tools in mathematical relativity and galactic modelling~\cite{Choptuik:1992jv, Matos:1998vk, UrenaLopez:2010ur, Torres:2000dw, Guzman:2009zz, AmaroSeoane:2010qx}. Additionally, if cosmological scalar fields exist in Nature, one viable possibility is that during their formation neutron stars will interact with a scalar matter cloud (or an extended BS) at least through gravity, resulting in a new type of self-gravitating objects with mixing of scalar and fermionic matter, which we will refer to as hybrid stars. As with any other exotics self-gravitating system, their hybrid features are expected to be reflected in their properties like mass, size, and compactness~\cite{DiGiovanni:2020frc}.

Here, we focus on a subset of the so-called Gleyzes-Langlois-Piazza-Vernizzi (GLPV) models of gravity~\cite{Gleyzes:2014dya, Gleyzes:2014qga}. The GLPV theory is an extension to Horndeski gravity (the most general theory of gravity in four spacetime dimensions with a single scalar field, leading to second-order field equations), has six arbitrary functions of the scalar field and its first derivatives (contracted with the spacetime metric to provide a scalar). It encompasses a series of models that are, in general, non-renormalizable, and that must be seen as a low energy effective field theory~\cite{Georgi:1994qn, Pich:1998xt,Burgess:2003jk, Kaplan:2005es, Manohar:2018aog, Cohen:2019wxr, Burgess:2020tbq, Penco:2020kvy}. Similar to other Horndeski extensions (e.g. DHOST, EST~\cite{Langlois:2015cwa, Crisostomi:2016czh}), it includes higher derivative operators that do not appear in simpler realizations (e.g. the Brans-Dicke model~\cite{Brans:1961sx}), but it is ghost-free and does not propagate additional degrees, apart from the usual spin two field and the scalar~\cite{Gleyzes:2014dya, Gleyzes:2014qga}. 

It is important to point out that Horndeski's theory (and its extensions) is usually introduced as a way of explaining the current accelerating expansion of the Universe. However, in most of these models the internal scalar degree of freedom is massless or quite ultra-light ($\sim 10^{-33}$ eV) in order to be consistent with the current cosmological data on the dark energy component. These ultra-light particles could also lead to hybrid stars, like the ones that we presented in this paper, but they will be so large that they would not correspond to a compact object. This is the reason why, in the present manuscript, we concentrate only on mass values that can be relevant at astrophysical scales (see Fig.~\ref{Fig1.1}). As a matter of fact, these particles could constitute part of, or even all, the dark matter sector (e.g.,~\cite{Cembranos:2008gj}).

An interesting property of the GLPV theory is the possibility to suppress the additional degree of freedom (through any of the known screening mechanisms~\cite{Brax:2013ida, Joyce:2014kja}), which helps to recover GR predictions and to avoid the strong post-Newtonian constraints from the Solar System~\cite{Will_2014}. The Vainshtein screening mechanism~\cite{Vainshtein:1972sx}, driven by derivative self-couplings~\cite{Kimura:2011dc, Narikawa:2013pjr, Koyama:2013paa}\footnote{For the models studied in this manuscript, a screening mechanism is necessary for scalar field masses smaller than $\sim 10^{-3}$eV, for which the Vainshtein mechanism is driven by operators of the form $\partial \phi, \partial^2\phi$. A screening analysis is beyond the purposes of this work, but some more details of the role of these operators can be found in~\cite{Joyce:2014kja, Clifton:2011jh}. Larger masses $\gtrsim 10^{-3}$eV imply that the Compton wavelength is shorter than about a millimeter, a distance that today is inaccessible to gravitational experiments.}, has attracted recent attention on these models due to its interesting phenomenology~\cite{Crisostomi:2017lbg}, such as the relation with consistent non-linear massive gravity theories~\cite{Rubakov:2008nh, deRham:2014zqa, Hinterbichler:2011tt}. It is important to note that in some cases, these models present solid angle deficits that induce a singularity at the center of compact objects~\cite{Kase:2015gxi, DeFelice:2015sya}. However, there is evidence that these singularities can be avoided if the scalar field depends also on time (e.g.~\cite{Kase:2015zva}).

In this paper we explore the existence and properties of hybrid star in a low energy effective gravity theory that includes operators with higher order derivatives presented in~\cite{Barranco:2021auj}, and look for the prevalence the Buchdahl's limit for their compactness. This effective theory correspond to a subset of the GLPV gravity. For previous work with similar motivations in GR and Horndeski gravity (with massless field) see e.g.,~\cite{Henriques:1989ez, Henriques:1990xg, Lopes:1992np, Cisterna:2015yla, Cisterna:2016vdx, Maselli:2016gxk, Babichev:2016jom, Henriques:2003yr, Valdez-Alvarado:2020vqa}. 

The organization of the manuscript is as follows. In Section~\ref{sec2}, we present the model that describes the static and spherically symmetric regime and identify the boundary conditions that allow us to construct the self-gravitating hybrid stars. For their fermionic part we consider two particular equations of state (EOS): a polytropic one, and an incompressible fluid with a constant energy density. In Section~\ref{sect3} we identify the parameter space where these compact objects lie in, as well as describe their main characteristics, establishing the existence of a family of solutions and the limits where typical NS or BS are recovered. In Section~\ref{sec4} we focus on the numerical study of a possible Buchdahl's limit on the compactness of theses stars. Finally, in Section~\ref{sec:conclusions} we give some concluding remarks.\footnote{In this manuscript we use Wald's notation~\cite{Wald:1984rg}: plus signature for the spacetime metric, $(-,+,+,+)$, the definitions $R_{\mu\nu\rho}{}^{\sigma}\equiv \partial_\nu\Gamma^{\sigma}_{\mu\rho}+\Gamma^{\alpha}_{\mu\rho}\Gamma^{\sigma}_{\alpha\nu}-(\nu\leftrightarrow \mu)$ for the Riemann tensor, $R_{\mu\nu}\equiv R_{\mu\alpha\nu}{}^{\alpha}$ for the Ricci tensor, and $R\equiv R_{\mu}{}^{\mu}$ for the Ricci scalar. We work in natural units, $\hbar=c=1$, and the reduced Planck mass is $M_{\rm{Pl}}\equiv 1/\sqrt{8\pi G}=2.431\times 10^{18}\,$GeV. Additionally, we assume a minimal coupling of matter to gravity (i.e. matter fields couple only to the Jordan spacetime metric, $g_{\mu\nu}^{\rm{Jor.}}\equiv g_{\mu\nu}$).}
 
\section{Theoretical Framework}\label{sec2} 
As was pointed out before, the GLPV model contains six arbitrary functions: $G_2, G_3, G_4, G_5, F_4, F_5$, which depend on the scalar field $\phi$ and its first derivatives (written as a standard canonical kinetic term $X\equiv g^{\mu\nu}\phi_{\mu}\phi_{\nu}$). The corresponding action of the model is
\begin{equation}
	S=\int d^{4}x \sqrt{-g}\left(\sum_{i=2}^{5} \mathcal{L}_{i}[g_{\mu\nu},\phi]+\mathcal{L}_m[g_{\mu\nu},\Psi]\right) \, ,
	\label{eq:lag}
\end{equation}
where $\mathcal{L}_m$ is the matter Lagrangian that contains all the standard model fields and their possible extensions, $\mathcal{L}_i$ indicates the gravitational sector and is given as a linear combination of the following Lagrangians,
\numparts
\begin{eqnarray}\label{eq.Lagrangians}
\fl		\mathcal{L}_{2}\equiv& G_{2}(\phi, X),\label{BH1}\\
\fl		\mathcal{L}_{3}\equiv& G_{3}(\phi, X)\Box\phi,\\
\fl		\mathcal{L}_{4}\equiv& G_{4}(\phi, X)R-2G_{4X}(\phi,X)\left[(\Box\phi)^{2}-\phi^{\mu\nu}\phi_{\mu\nu}\right]\nonumber\\
\fl		&+F_{4}(\phi,X){\epsilon^{\mu\nu\rho}}_{\sigma}\epsilon^{\mu'\nu' \rho'\sigma}\phi_{\mu}\phi_{\mu'}\phi_{\nu\nu'}\phi_{\rho\rho'},\label{eq.quartic}\\
\fl		\mathcal{L}_{5}\equiv& G_{5}(\phi, X)G_{\mu\nu}\phi^{\mu\nu}+\frac{1}{3}G_{5X}(\phi,X)\left[(\Box\phi)^{3}-3\Box\phi\phi_{\mu\nu}\phi^{\mu\nu}+2\phi_{\mu\nu}\phi^{\mu\sigma}{\phi^{\nu}}_{\sigma}\right]\nonumber\\
\fl		&+\,F_{5}(\phi,X)\epsilon^{\mu\nu\rho\sigma}\epsilon^{\mu'\nu' \rho'\sigma'}\phi_{\mu}\phi_{\mu'}\phi_{\nu\nu'}\phi_{\rho\rho'}\phi_{\sigma\sigma'}\,.\label{BH2}
	\end{eqnarray}
\endnumparts
In the above equations, $R$ and $G_{\mu\nu}\equiv R_{\mu\nu}-\frac{1}{2}Rg_{\mu\nu}$ are the Ricci scalar and Einstein tensor, respectively. The term $\epsilon^{\mu\nu\rho\sigma}$ correspond to the totally antisymmetric Levi-Civita tensor. To simplify the notation we have used the definition $\phi_{\mu\nu}\equiv\nabla_{\mu}\nabla_{\nu}\phi$, $\Box\phi\equiv\phi_{\mu}{}^{\mu}$ (d'Alembert operator), together with the subindex notation in the functions $G_i$ and $F_i$ (e.g. $G_{i X}$) to denote partial differentiation with respect to their arguments.

Some comments are in turn about the {\it a priori} arbitrary functions $G_i, F_i$. The set of GLPV models with $F_4$ or $F_5$ different from zero, will have higher order differential operators at the level of the equations of motion.\footnote{In some case the $F_4$ and $F_5$ terms in the GLPV theory can be mapped to the pure Horndeski action via disformal transformations (see e.g.~\cite{Crisostomi:2016tcp}), for this case the matter stops being minimally coupled to gravity.  As in this paper we always work in the Jordan frame, the $F_4$ or $F_5$, are different sectors from the pure Horndeski theory.} Nevertheless, the propagating degrees of freedom obey second order equations, avoiding the  so-called Ostrogradski instabilities~\cite{Ostrogradsky:1850fid, Pais:1950za}. Models with $F_4=F_5=0$ represent the Horndeski theory, where the equations of motion remain second order. 

Notice that by choosing by $G_4=M_{\rm{Pl}}^2/2$ with all other functions set to zero (except in the case that we want to include a cosmological constant: $G_2=-M_{\rm{Pl}}^2\Lambda$), GR is recovered. Along the same lines, if $G_2$ is not a constant (or zero), the scalar degree of freedom propagates, but it remains minimally coupled to the metric, and the gravity sector is still described by GR (one could naturally argue whether this represents or not a real modification of gravity). An example of this particular set (and that is addressed in this paper) are the Einstein-Klein-Gordon (EKG) models: $G_2=-\frac{1}{2}X-V(\phi)$. On the contrary, for the cases where the scalar field is non-minimal coupled to the metric (as is our case, see Eq.~\eref{eq.interactions}), an additional scalar mediator is added to the gravity sector, apart from the usual spin two field, and the differences from GR will be manifest~\cite{Barranco:2021auj}.

\subsection{The gravity sector}
The action~\eref{eq:lag} represents a large family of scalar-tensor theories, in particular we are interested in those who introduce infrared modifications of gravity. A simple choice of functions $\mathcal{L}_i$ that represents a low energy effective model was presented in~\cite{Barranco:2021auj} (see also~\cite{Brihaye:2016lin, Chagoya:2018lmv, Chagoya:2020lwy})
\begin{eqnarray}\label{eq.interactions}
\fl	\mathcal{L}_{\rm{grav}} =& \frac{1}{2}M_{\rm{Pl}}^2 R - X -m^2\phi\bar{\phi}\nonumber \\
\fl	& + \frac{M_{\rm{Pl}}}{\Lambda^3}\left[c_4XR - 2c_4[\Box\phi\Box\bar{\phi}-\phi^{\mu\nu}\bar{\phi}_{\mu\nu}]
	+\frac{d_4}{X}{\epsilon^{\mu\nu\rho}}_{\sigma}\epsilon^{\mu'\nu' \rho'\sigma}\phi_{\mu}\bar{\phi}_{\mu'}\phi_{\nu\nu'}\bar{\phi}_{\rho\rho'}
	\right].
\end{eqnarray}
Here, $\bar{\phi}$ denotes the complex conjugate of the scalar field $\phi$, and its mass parameter is $m$. In the first line we have removed the $1/2$ factor in front of the kinetic complex term $X=\phi_{\mu}\bar{\phi}^{\mu}$, in order to get the standard normalization of a complex scalar field.
Note that the whole cubic and quintic sector was eliminated\footnote{The $F_5$ sector includes operators of mass dimension nine (or larger), and it is suppressed at low energies. The contributions from $\mathcal{L}_{3}$ and $\mathcal{L}_{5}$ dissapear by imposing a discrete $\mathbb{Z}_2$ mirror symmetry $\phi\to -\phi$~\cite{Diez-Tejedor:2018fue}.}. 

We will focus our attention on the family of models with the dimensionless parameters $c_4=0, \pm 1/2$, $d_4=0$ (the $F_4$ contributions are off), in the region of parameter space where the effective approximation~\eref{eq.interactions} is valid: $m<\Lambda\ll M_{\rm{Pl}}$~\cite{Barranco:2021auj}. Note that the strength of the higher derivative operators are mediated by inverse powers of $\Lambda$, which would then represent the energy scale at which such operators are relevant (the other scale is $m$). A lower bound on this scale constrains the possible signatures that these terms may leave on observables at low energies.

Before we conclude this section, it is important to make some clarifications regarding the values of the parameters $\Lambda$, $c_4$ and $d_4$. From an analysis of the gravity sector in Eq.~(\ref{eq.interactions}), in the limit $\Lambda\to\infty$ ($\Lambda \gg M_{\rm{Pl}}$ in physical units) it is possible to show that the (beyond) Horndeski contributions are no longer relevant because the operators induced by couplings of the scalar field with gravity are suppressed by powers of $M_{\rm{Pl}}^{1/4}\Lambda^{1/4}$~\cite{Barranco:2021auj}. Additionally, as we take $G_2=-X-m^2\phi\bar{\phi}$ then the effective model becomes equivalent to the EKG model~\cite{Barranco:2021auj}.

As our interest is to find typical signatures of the Horndeski terms, we would need to consider $\Lambda\to 0$ ($\Lambda\ll M_{\rm{Pl}}$ in physical units). However, as indicated in the previous paragraph, we will restrict our work to the region $m<\Lambda\ll M_{\rm{Pl}}$, as the condition $m < \Lambda$ allows the existence of equilibrium configurations in the low energy regimen. One extra condition (the so-called strong field one in~\cite{Barranco:2021auj}), which will be defined in the next section, will help us to avoid the inclusion of the next leading order terms in Eq.~(\ref{eq.interactions}).

Finally, our choice for the values of the coefficients $c_4, d_4$ is motivated by models in the literature with gravitational and cosmological applications that include these terms, e.g. Fab Four~\cite{Charmousis:2011bf, Charmousis:2011ea}. For practical purposes, the $c_4$ coefficient can be always absorbed into the scale $\Lambda$ and fix it without loss of generality to $c_4=\pm 1/2$. In contrast, $d_4$ is arbitrary and we set it to zero, but the analysis presented below is also valid for the case $d_4\neq 0$. For the case $c_4=d_4=0$, it is necessary to consider the next order in the gravity action~(\ref{eq.interactions}), however, as proved in~\cite{Barranco:2021auj}, within our parameter space such a choice of values is equivalent to the EKG model (see Appendix B in~\cite{Barranco:2021auj} for more details).

\subsection{Hybrid stars}
Taking the effective Lagrangian~\eref{eq.interactions} as the gravitational sector in the action~\eref{eq:lag}, and assuming a fermionic field as the only baryonic source, we now proceed to construct compact and localized solutions, by first finding the dynamical equations for the metric components, the scalar field, and the fermionic pressure.

The variation of the action Eq.~(\ref{eq:lag}) with respect to the metric $g^{\mu\nu}$ results in
\numparts
\begin{eqnarray}\label{met_eq}
	G_{\mu\nu}+\frac{c_4}{2M_{\rm{Pl}}\Lambda^{3}}H_{\mu\nu}=\frac{1}{M_{\rm{Pl}}^2}\left(T_{\mu\nu}-T_{\mu\nu}^{(\phi)}\right)\,,
\end{eqnarray}
where the tensor $H_{\mu\nu}$ represents the gravitational modification introduced to GR,
\begin{eqnarray}
\fl	H_{\mu\nu}= &G_{\mu\nu}X+g_{\mu\nu}\left( \bar{\phi}_{\alpha\rho}\phi^{\alpha\rho}-\Box\phi\Box\bar{\phi}+2R_{\alpha\rho}\phi^{\alpha}\bar{\phi}^{\rho}\right) -\phi^{\alpha}\bar{\phi}^{\rho}\left( R_{\mu\alpha\nu\rho}+R_{\mu\rho\nu\alpha}\right)\nonumber\\
\fl	&+\left[ \phi_{\mu}\left( \frac{R}{2}\phi_{\nu}-R_{\nu\alpha}\phi^{\alpha}\right)+\bar{\phi}_{\mu\nu}\Box\phi-R_{\mu\alpha}\bar{\phi}_{\nu}\phi^{\alpha}-\bar{\phi}_{\mu\alpha}{\phi_{\nu}}^{\alpha}+\rm{c.c}\right] \,,
\end{eqnarray}
and $T_{\mu\nu}$, $T_{\mu\nu}^{(\phi)}$ are the stress-energy tensor of the fermionic and bosonic fields, respectively,
\begin{eqnarray}
	T_{\mu\nu}&=&\frac{-2}{\sqrt{-g}}\frac{\delta \mathcal{L}_m}{\delta g^{\mu\nu}}\,,\label{Tenerg}\\
	T_{\mu\nu}^{(\phi)}&=&g_{\mu\nu}\left(X+m^2\phi\bar{\phi} \right)-\phi_{\mu}\bar{\phi}{}_{\nu}-\rm{c.c}\,.
\end{eqnarray}  
Likewise, the variation with respect to $\bar{\phi}$ leads to,
\begin{equation}\label{field_eq}
	\Box\phi-m^{2}\phi+	\frac{2 c_4M_{\rm{Pl}}}{\Lambda^{3}}G^{\mu\nu}\phi_{\mu\nu}=0\,.
\end{equation}
\endnumparts
If we set $c_4=0$ in the foregoing equations the standard GR and Klein-Gordon (KG) equations of motion are readily recovered. A similar result appears in the limit $\Lambda\to\infty$, at which the higher derivative operators vanish.

As we are interested in equilibrium configurations, we assume a static and  spherically symmetric spacetime line element in the form
\begin{equation}
	ds^{2}=-N^{2}(r)dt^{2}+g^{2}(r)dr^{2}+r^{2}d\theta^{2}+r^{2}\sin^2\theta d\varphi^{2} \, , \label{metric}
\end{equation}
and impose a harmonic ansatz for the scalar field,
\begin{equation}
	\phi(t,r)=\sigma(r)e^{i \omega t} \, . \label{ansat}
\end{equation}
The metric functions $N(r)$, $g(r)$ and the radial component of the scalar field $\sigma(r)$, depend only on the radial coordinate $r$, whereas $\omega$, the angular frequency of oscillation of the scalar field, is real and constant.

The harmonic ansatz~\eref{ansat} reduces the equations of motion~\eref{met_eq}-\eref{field_eq} to a simpler time-independent system that is compatible with the static metric~(\ref{metric}). 
After some manipulations, the equations of motion can be written in the form,
\numparts%\label{fieleq}
\begin{eqnarray}
\fl	  \frac{2(1+\alpha)}{r}\frac{g'}{g^3}-\frac{(1-\beta_0)}{M_{\rm{Pl}}^2 g^2}\sigma'^2-\left[m^2+(1-\gamma)\frac{\omega^2}{N^2}\right]\frac{\sigma^2}{M_{\rm{Pl}}^2}+\left(1-\frac{1}{g^2}\right)\frac{1}{r^2}=\frac{T_{00}}{N^2 M_{\rm{Pl}}^2 },\label{fielg}\\[0.2cm]  
\fl	    \frac{2(1+\alpha)}{r}\frac{N'}{N g^2}-\frac{\left(1-\beta_1\right)}{M_{\rm{Pl}}^2 g^2}\sigma'^2+\left[m^2-(1-\gamma)\frac{\omega^2}{N^2}\right]\frac{\sigma^2}{M_{\rm{Pl}}^2}-(1-\frac{1}{g^2})\frac{1}{r^2}=\frac{T_{11}}{g^2 M_{\rm{Pl}}^2 },\label{fielN}\\[0.2cm]
\fl	  (1+\varepsilon)\sigma''+\left[(1-\eta)\left(\frac{N'}{N}-\frac{g'}{g}\right)+\frac{2(1+\zeta)}{r}\right]\sigma'-
	g^2\left(m^2-(1+\theta)\frac{\omega^2}{N^2}\right)\sigma= 0,\label{eq.KG.modified}
\end{eqnarray}
\endnumparts
where $T_{00}$ and $T_{11}$, are the $0-0$ and $1-1$ components of the covariant stress-energy tensor~(\ref{Tenerg}), and $\alpha$, $\beta_0$, $\gamma$, $\beta_1$, $\varepsilon$,  $\eta$, $\zeta$, and $\theta$ are dimensionless functions given by
\numparts
\begin{eqnarray} 
	\alpha=\frac{2 c_4\sigma^2}{M_{\rm{Pl}}\Lambda^3}\left( \frac{\omega^2}{N^2}-\frac{3\sigma'^2}{g^2\sigma^2}\right) \,,&\quad&
	\beta_0=\frac{2 c_4M_{\rm{Pl}}}{\Lambda^3 r^2}\left(1+\frac{1}{g^2}+\frac{4\sigma''r}{g^2\sigma'}\right)\,, \label{eqs.app1a} \\
	\gamma=\frac{2 c_4M_{\rm{Pl}}}{\Lambda^3 r^2}\left(1-\frac{1}{g^2}\right)\,,&\quad&
	\beta_1=\frac{2 c_4M_{\rm{Pl}}}{\Lambda_3^3 r^2}\left(1-\frac{3}{g^2}-\frac{4\omega^2\sigma r}{N^2\sigma'}\right)\,,\label{eqs.app1b} \\	
	\varepsilon=\frac{2 c_4M_{\rm{Pl}}}{\Lambda^3 g^2 r^2}\left(1+\frac{2rN'}{N}-g^2\right)\,, &\quad&
	\eta=\frac{2 c_4M_{\rm{Pl}}}{\Lambda^3 g^2 r^2}\left(g^2-3\right) \,,\\
	\zeta=\frac{2c_4M_{\rm{Pl}}}{\Lambda^3 g^2}\left( \frac{N''}{N}-\frac{3g'}{g}\frac{N'}{N}\right) \,,&\quad&
	\theta=\frac{2 c_4M_{\rm{Pl}}}{\Lambda^3 g^2 r^2}\left( 1-g^2-\frac{2r g'}{g}\right)\,. \label{eqs.app1c}
\end{eqnarray}
\endnumparts
It is necessary to point out that a second order derivatives of the lapse functions $N''$ is implicit in the term $\zeta$. Using the trace of the field equations~\eref{met_eq}, it is possible to remove this dependence from the structure equations, leaving a system that only depends on $N, g, p, \sigma, \sigma'$ and $r$.

From now on it is assumed that the fermionic matter supports no transverse stresses, and has no mass motion, which are consistent with our aim of studying equilibrium configurations. Under these conditions, the fermionic energy-momentum tensor~(\ref{Tenerg}) takes the form of the perfect fluid one~\cite{Oppenheimer:1939ne, Weinberg:1972kfs, Misner:1974qy}
\begin{equation}
	T_{\mu\nu}=(\epsilon+p)u_{\mu}u_{\nu}+p g_{\mu\nu}\,.\label{perfl}
\end{equation}
Here, $p$ is the pressure of the fluid and the total energy density $\epsilon$ corresponds to the sum of the rest mass density of the fluid $\rho$, and its internal energy $\varepsilon$, $\epsilon=\rho+\varepsilon$. The assumptions of time independence and spherical symmetry imply that both $p$ and $\epsilon$ are functions only of the radial coordinate. Finally, the four-velocity is defined as $u^{\nu}=u^{0}(1,0,0,0)$, where the component $u^{0}$ is computed from the normalization $u^{\nu} u_{\nu}=-1$, which leads to $u^{0}=1/N$.

To get the desired hybrid stars profiles, we need to solve the system~\eref{fielg}-\eref{eq.KG.modified}, together with the trace of Eq.~\eref{met_eq}, and the conservation equation $\nabla^{\mu}T_{\mu\nu}=0$, for $N', g', \sigma'',$ and $p'$ functions. The resulting system must be solved numerically, and it is necessary to define a set of boundary conditions in the center and at large distances from the star. At the center, since we want regular spacetime configurations (no divergences of curvature scalars), we have 
\numparts
\begin{eqnarray}
g(r=0)= 1,\hspace{0.95cm} N(r=0)=N_{0},\hspace{0.95cm} p(r=0)=p_0,\label{boundayC1}\\ %\hspace{0.95cm} N'(\bar{r}=0)=0, \label{boundayC1}\\p(r=0)=p_0, \hspace{0.95cm} 
\sigma(r=0)= \sigma_{0},\hspace{1.2cm} \sigma'(r=0)=0,\label{boundayC2}
\end{eqnarray}
\endnumparts
where $\sigma_0$ is the field amplitude at the origin, $p_0$ is the central fermionic pressure of the star, and $N_0$ the lapse function evaluated at the centre of the configuration. These are free and positive constants that one can choose arbitrarily. On other hand, to obtain localized configurations, the boundary condition at infinity must be the same as that for the vacuum state:
\begin{equation}\label{bounday2}
    \lim_{r\to\infty}p(r)= 0,\quad
    \lim_{r\to\infty}\sigma(r)= 0,\quad
    \lim_{r\to\infty}N(r)= N_{\infty},\quad 
    \lim_{r\to\infty}g(r)= 1,
\end{equation}
where $N_{\infty}$ is an arbitrary and positive constant, which is equal to the limit value $1/\lim_{r\to\infty}g(r)$, if we like to recovery the Schwarzschild metric. Notice that it is not necessary to define boundary conditions for $\epsilon$ since these are inferred through the fermionic EOS.

A note on the lapse function is in turn. The value $N_0$ can always be reabsorbed in the definition of the time parameter and fixed to $N_0=1$ with no loss of generality. In such a case, the boundary condition $N_{\infty}=1/\lim_{r\to\infty}g(r)=1$ is not respected by outwards integration starting from $N_0=1$. However, we can make use of the freedom to redefine the time coordinate, and the frequency accordingly, $(N, \omega)\mapsto x (N,\omega)$, in such a way that this condition is satisfied at infinity. 

\begin{figure}
	\centering	
	\scalebox{0.45}{
		\input{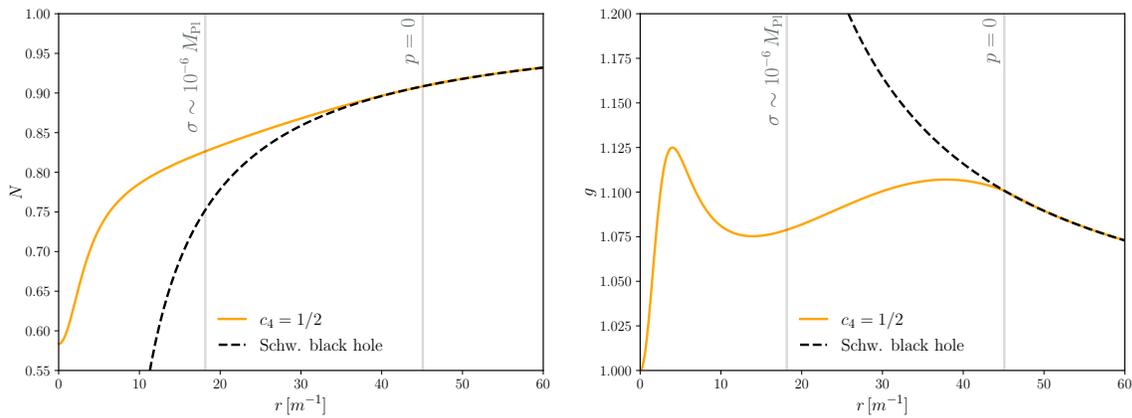}
	}
	\caption{{\bf Metric profiles.} Lapse function $N(r)$ (left panel) and radial component $g(r)$ (right panel), normalized to one at spatial infinity Ec.~(\ref{bounday2}), for a configuration with $\sigma_0=0.25\,M_{\rm{Pl}}$ and, $p=8\times10^{-4}\, m^{2}M_{\rm{Pl}}^2$, in the model: $c_4= 1/2$, with $\Lambda=1.5\,M_{\rm{Pl}}^{1/3}m^{2/3}$. The dashed lines represent the Schwarzschild metric components $N_{\rm{Schw.}}(r)=(1-r_{s}/r)^{1/2}$, and $g_{\rm{Schw.}}(r)=N^{-1}_{\rm{Schw.}}(r)$, where $r_{s}\equiv M/(4\pi M_{\rm{Pl}}^2)$ is the Schwarzschild radius for an object of the same total mass $M=99.24 M_{\mathrm{Pl}}^{2} m^{-1}$. As is possible to see, for $r\to \infty$ the scalar field decays exponentially and it is not possible to differentiate between the two objects. The vertical lines correspond to the radii where $\sigma\sim$O$(10^{-6})$, and $p=0$ (border of the baryonic component of the star).}\label{metric_p}
\end{figure}

To this effect, we first obtain the corresponding hybrid star profile (with $N_{\infty}\neq1$) for the set of initial values $(\sigma_0, p_0, N_0=1)$. To meet the condition $N_\infty=1$, we then redefine the time coordinate as $N^{\rm{new}}(r)=xN(r)$, hence a new frequency $\omega^{\rm{new}}=x\omega$, in such a way that
\begin{equation}
   x N(r_{\rm{max}})=\frac{1}{g(r_{\rm{max}})},\nonumber
\end{equation}
with $r_{\rm{max}}$ the maximum radius of integration in the numerical code. In this manuscript we not write the super-index ``new'' explicitly, and it is understood that only rescaled values are reported. The corresponding profiles of the metric functions associated to one of these configurations once the re-scaling has been carried out are shown in Figure~\ref{metric_p}.

To close the system of equations we need to write down an equation of state for the fluid component, for which we consider two simple cases. First, a polytropic equation of state
\numparts
\begin{eqnarray}
	p=k \rho^{\Gamma}\,,\label{EDE}
\end{eqnarray}
where $k$ is the polytropic constant, $\Gamma$ is the adiabatic index, and $\rho$ is the mass density, all related to the energy density by
\begin{equation}
	\epsilon=\left(\frac{p}{k}\right)^{1/\Gamma}+\frac{p}{\Gamma-1}\,.\label{eqrel}
\end{equation}
\endnumparts
Second, an incompressible fluid with a constant energy density, $\epsilon=\rm{cte}$. This case will allow us to explore numerically the Buchdahl's limit on the compactness of a star. 

\section{Numerical results}\label{sect3}
For the numerical implementation, it is convenient to rewrite the dynamical equations in terms of the new dimensionless variables, 
\begin{eqnarray}
\fl	\bar{r}\equiv m r, \quad \bar{\sigma}\equiv\frac{\sigma}{M_{\rm{Pl}}},\quad \bar{\omega}\equiv\frac{\omega}{m}, \quad \bar{\Lambda}\equiv\frac{\Lambda}{M_{\rm{Pl}}^{1/3}m^{2/3}},\quad\bar{p}\equiv\frac{p}{m^{2}M_{\rm{Pl}}^2}, \quad\bar{\epsilon}\equiv\frac{\epsilon}{m^{2}M_{\rm{Pl}}^2}. \label{eq:dimless}
\end{eqnarray}
This change of variable remove the dependence on the scalar field mass $m$, and the Planck mass $M_{\rm{Pl}}$, from the equations of motion, combining the energy scales ($m, \Lambda$) in $\bar{\Lambda}$.

In terms of these variables we study the behavior of the system~(\ref{fielg})-(\ref{eq.KG.modified}) in two regimes: near the origin, and for large distances. For the former we perform a Taylor expansion around $\bar{r}=0$, and using the boundary conditions~\eref{boundayC1}-\eref{boundayC2} we obtain perturbative solutions which are valid near the origin. We assumed that $\Lambda$ remains large in units of $M_{\rm{Pl}}^{1/3}m^{2/3}$, and after some manipulation we arrive to a set of equations similar to Eqs.~(3.8) in~\cite{Barranco:2021auj}, with the particularity that now we have an extra equation corresponding to the pressure $\bar{p}(\bar{r})$. The found series expansions were used to validated our numerical implementation near the origin.

Likewise, to study the solutions at large distances we use a flat metric. From the equations of motion~(\ref{eq.KG.modified}), and in the limit in which the scalar field remains small and the fermionic density is zero, we find the following asymptotic behaviour of the scalar field profile
\begin{equation}\label{eq.sigma.infinity}
    \bar{\sigma}(\bar r) \sim \frac{1}{\bar{r}}\exp \left[\sqrt{1-\frac{\bar{\omega}^2}{N_{\infty}^2}}\bar{r}\right].
\end{equation}
Similarly to the results in~\cite{Barranco:2021auj}, it is the mass term $m$, together with the condition $\bar{\omega}<1$,   what makes possible the exponential decay of the wave function at spatial infinity. Finally, we study the extreme situation: $\bar{p}_0 \gtrapprox \bar{\sigma}_0$ (hybrid stars with a bosonic core). Assuming $\bar{\omega}<1$ (to guarantee again that the scalar field has an exponential decay), and $|\bar{\sigma}'|\gg|\bar{p}'|$, we arrive to a system similar to the GR Tolman-Oppenheimer-Volkoff one, which implies that in this regime the discrepancies with standard neutron stars are small. These results were numerically validated, see for instance the profile of $\bar{p}(\bar{r})$ shown on the left panel of Fig.~\ref{Fig1} below).

To find the profiles of the hybrid stars, with the boundary behaviors described above, we solved numerically the system of differential equations for $N', g', p', \sigma''$, in terms of the new variables~\eref{eq:dimless} and considering the boundary conditions~\eref{boundayC1}-\eref{boundayC2}, using a shooting method~\cite{Numerical, Dias:2015nua}. At this point it is necessary to point out that given a $(\sigma_0, p_0)$ pair, there can be multiple frequencies that satisfy the conditions in~\eref{bounday2}, and to fix this we only look for scalar field profiles without nodes, that is, the ground state in each case. Also, for simplicity in the notation, we hereafter drop the bar in the variables, and put full units whenever necessary to avoid any confusion.

Figure~\ref{Fig1} shows some illustrative examples of our numerical realizations, in particular two limit cases: i) stars for which the scalar field profile drops more sharply than the pressure one (left panel), that we will call hybrid stars with a bosonic core, and ii) for which the pressure profile drops more sharply that the scalar field one (right panel), that we will call hybrid stars with a fermionic core. The profiles shown correspond to the Horndeski models $c_4=\pm1/2$, and for comparison purposes we include also the standard EKG results ($c_4=0$). It can be seen that a positive (negative) values of the coupling constants $c_4$ will open (close) the respectively $p$ and $\sigma$ profiles, to configurations that are broader (narrower) in comparison with the equivalents EKG hybrid stars. Similar results were reported in~\cite{Barranco:2021auj} (for $p=0$), suggesting that positive (negative) couplings are associated to repulsive (attractive) self-interactions. 

\begin{figure}
	\centering	
	\scalebox{0.45}{
		\input{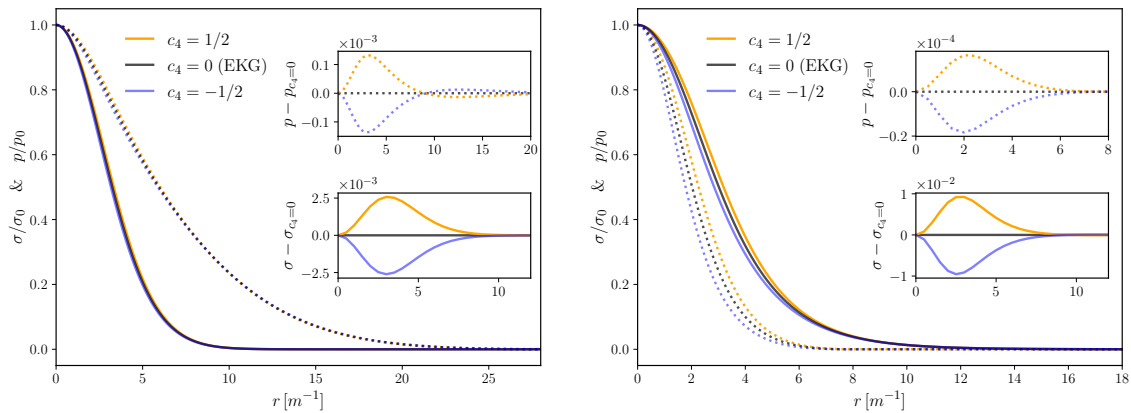}
	}
	\caption{{\bf Pressure and field profiles.} Radial profiles of the fermionic pressure $p$ (dotted curves), see Eq.~(\ref{perfl}), and the scalar field $\phi$ (solid curves), see Eq.~(\ref{ansat}), in Horndeski models with $c_4=\pm 1/2$, $\Lambda=1.5\,M_{\rm{Pl}}^{1/3}m^{2/3}$, and a fermionic fluid described by Eq.~(\ref{EDE}) with $\Gamma=2$, and $k=100\, \left(m^{2}M_{\rm{Pl}}^2\right)^{-1}$. The profiles are normalized with their respective central values: $\sigma_0=0.15\,M_{\rm{Pl}}$ and $p_0=0.01\,m^{2}M_{\rm{Pl}}^2$ (left panel), and $\sigma_0=0.25\, M_{\rm{Pl}}$ and $p_0=2.25\times 10^{-4}\,m^{2}M_{\rm{Pl}}^2$ (right panel). The profiles also exemplify the limit cases i) (left panel) and ii) (right panel) mentioned in the main text. The insets show the differences in each case with respect to the standard EKG results $(c_4=0)$.\label{Fig1}}
\end{figure}

\begin{figure}[tb]
	\centering	
	\scalebox{0.45}{
		\input{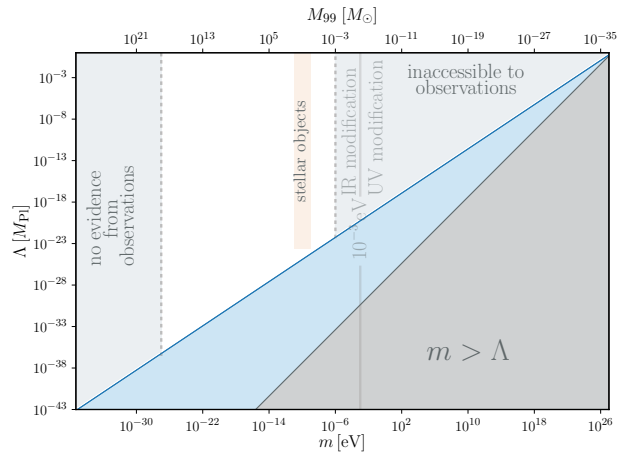}
	}
	\caption{{\bf Parameter space.} The shaded region corresponds to $m>\Lambda$ and denotes the combinations for which hybrid stars are not allowed. The blue straight line $\Lambda=M_{\rm{Pl}}^{1/3}m^{2/3}$ delimits the white ($\Lambda\gtrsim M_{\rm{Pl}}^{1/3}m^{2/3}$) and blue regions ($\Lambda\lesssim M_{\rm{Pl}}^{1/3}m^{2/3}$). Configurations with combinations of parameters inside of the blue sector can be develop distinctive features associated to the higher derivative operators, contrary to the white region. In this manuscript we work in an internal region of the white zone, denoted with an orange rectangle, which corresponds to field masses in the range $10^{-9}-10^{-11}$ eV. For reference, the most compact stable neutron star configuration has $\bar{M}_{99}=206.2$, obtained using Eq.(\ref{eq.mass.radius.HS}). This figure is an adapted version of Figure~2 in~\cite{Barranco:2021auj}.}\label{Fig1.1}
\end{figure}

It is convenient here to take a look at our parameter space. We are interested in exploring astrophysical objects (i.e., with mass $M\sim 1-20\,M_{\odot}$ and size $R\sim 9-10^5$ km), such that the fermionic matter forms objects similar to typical neutron stars, with central density (pressure) of around $10^{17} \textrm{kg/m}^{3}$ ($10^{34}$ Pa)~\cite{Ozel:2016oaf}. The set of central (fermionic) energy densities that we will explore is then $\epsilon_0\in [10^{-4}, 10^{-2}]\,m^{2}M_{\rm{Pl}}^{2}$, which implies using Eq.~(\ref{eqrel}) with $k=100\, m^{-2}M_{\rm{Pl}}^{-2}$ and $\Gamma=2$ that $p_0\in [10^{-6}, 10^{-2}]\,m^{2}M_{\rm{Pl}}^{2}$. Using the conversion factors $\epsilon =1.38 \,\bar{\epsilon}\, m[\textrm{eV}]^{2}\,c^{2}\times 10^{39} \textrm{kg/m}^{3}$ and $p=1.24\,\bar{p}\, m[\textrm{eV}]^{2}\times 10^{56}\textrm{Pa}$ to recover the right units for the physical quantities, it is relatively simple to check that field masses in the range $10^{-9}-10^{-11}$ eV correspond to astrophysical objects. Similarly, the values of the parameters $k, \Gamma$ were chosen to have consistency with the masses and radii accepted for neutron stars~\cite{Lattimer:2000nx}.
Finally, as was previously pointed out, the coupling parameter is constrained to $m<\Lambda\ll M_{\rm{Pl}}$ implying a borderline represented by by $\Lambda[M_{\rm{Pl}}]>m/M_{\rm{Pl}}$. The scenario that we explore in this paper includes an extra constraint $\Lambda\gg M^{1/3}_{\rm{Pl}}m^{2/3}$, that represents a new region (inside of $\Lambda[M_{\rm{Pl}}]>m/M_{\rm{Pl}}$) where higher derivative operators are negligible whatever the amplitude of the central field~\cite{Barranco:2021auj}. All the aforementioned regions are summarized in Figure~\ref{Fig1.1}, where the orange rectangle indicates the one region that is explored in this paper.

Since the hybrid stars are constituted by two components, fermionic and scalar fields, whose densities vanish at a finite and infinite radius, respectively, properties like the mass or size cannot be computed (keeping in mind all contributions) using only the typical argument $p=0$ (e.g. the extreme case ii)). Nevertheless, by construction the total density vanishes asymptotically as the spacetime metric approaches the Schwarzschild solution. Therefore, choosing a sufficiently large radius $r$, it is possible to estimate the mass $M$ of these objects via the Schwarzschild metric:
\begin{equation}\label{eq.mass.function}
    \bar{M} (\bar{r}) = 4\pi \bar{r}\left[ 1-\frac{1}{g^{2}(\bar{r})}\right],\label{mass_eq}
\end{equation}
where $M=\bar{M}M_{\rm{Pl}}^2/m$. 

\begin{figure}
	\centering	
	\scalebox{0.45}{
		\input{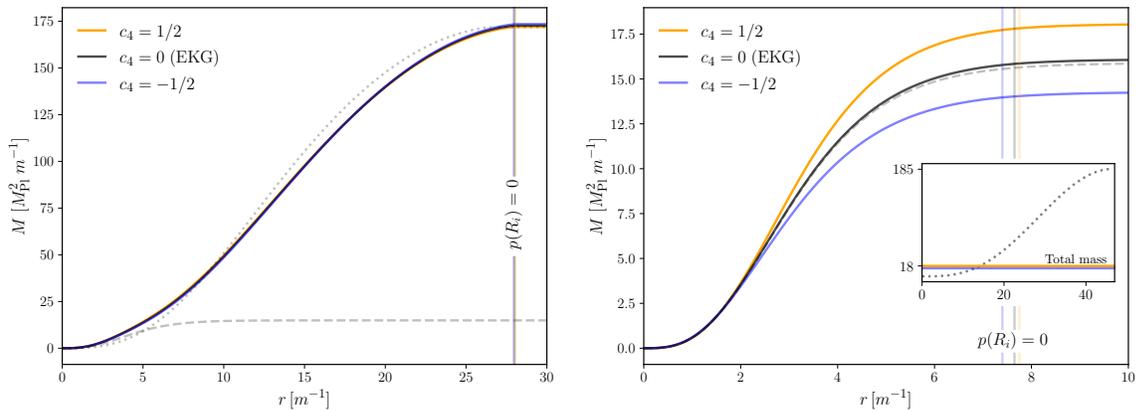}
	}
	\caption{{\bf Mass profiles.} Hybrid star mass profiles $M(r)$, see Eq.~(\ref{eq.mass.function}), as a function of the radial coordinate in Horndeski models with $\Lambda=1.5\,M_{\rm{Pl}}^{1/3}m^{2/3}$ and $c_4=\pm 1/2$, and also for the EKG model ($c_4=0$). The profiles correspond to the same configurations of Figure~\ref{Fig1}, and to the cases i) (left panel) and ii) (right panel). Also shown for comparison is the same configuration in GR with $\sigma=0$ (a neutron star), and $p=0$ (a boson star), represented by the dotted and dashed curves, respectively. The vertical lines indicate the radii where the fermionic pressure is zero for every configuration.}\label{Fig2}
\end{figure}

\setcounter{footnote}{0} % resetting the footnote
Figure~\ref{Fig2} shows the mass profile (computed using Eq.~\eref{mass_eq}) for the extreme cases i), ii). The left panel corresponds to the case i), where the scalar field profile drops more sharply than its pressure profile, and, as we expected, the maximal mass contribution comes from fermionic density. All mass profiles shown are very similar to that of a neutron star in GR with the same central pressure (dotted line). Likewise, in the right panel we show case ii), for which the largest contribution to the mass comes from the scalar field. The fermionic density is suppressed by the scalar degree of freedom, and the mass profile lies below the equivalent neutron star in GR (see the inset). Note that for this case, unlike the previous one, the Hordeski models present differences between their mass profiles, reflecting the effect of having considered a non-minimal coupling between the scalar field and the metric. For $c_4>0,\, p\neq0$ ($c_4<0,\, p\neq0$), $M(r)$ is larger (smaller) than in EKG, $c_4=0,\, p\neq0$.
Additionally, and although the fermionic matter is suppressed, its inclusion is reflected as a slight increase in the mass of the objects, see for example the model $c_4=0,\, p\neq0$, which is slightly higher than the $c_4=0,\, p=0$ (dashed line).

\begin{figure}
	\centering	
	\scalebox{0.5}{
		\input{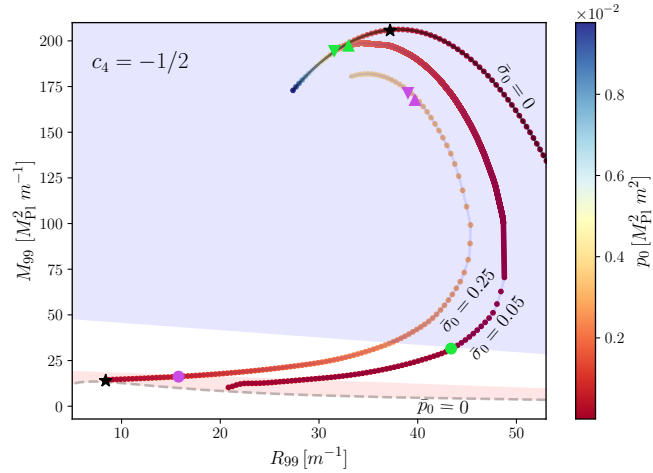}
	}
	\caption{{\bf $M_{99}$ vs $R_{99}$ profiles.} Each curve corresponds to a set of configurations computed for varying the central pressure $p_0$ in the range $[4\times 10^{-6}, 10^{-2}]\, m^{2}M_{\rm{Pl}}^2$ (vertical color bar), for two central amplitudes $\sigma_0=0.25, 0.05 \, M_{\rm{Pl}}$, with $\Lambda= 1.5\, M^{1/3}_{\rm{Pl}}m^{2/3}$ and $c_4=-1/2$. The red (blue) shaded region indicates configurations with a fermionic (bosonic) core for this particular EOS. The limit cases i) and ii) are reached when $p\to 10^{-2}\, m^{2}M_{\rm{Pl}}^2$ and $p\to 4\times 10^{-6}\, m^{2}M_{\rm{Pl}}^2$, respectively. Note that the curves $p=0$ (dashed line) and $\sigma_0=0$ (NS curve) represent the lower and upper configuration limits, respectively. Markers represent borderline configurations analyzed in Figure~\ref{FigA2}. The fermionic matter is described by Eq.~(\ref{EDE}) with $\Gamma=2$, and $k=100\, m^{-2}M_{\rm{Pl}}^{-2}$. See the text for more details.}\label{Fig3}
\end{figure}

Now, we need to defined a criterion for the radius of this hybrid stars. Similarly to a typical BS, the scalar field profile decreases monotonically as $r$ increases, and in some cases more sharply than the fermionic pressure profile. Hence, we define the effective radius of the object, $R_{99}$, as that where 99$\%$ of the total mass $M_{T}$ is contained, that is $M_{99}=0.99 M_{T}$. Using the above definitions, we computed the relation $M_{99}$ {\it vs} $R_{99}$ for a set of hybrid stars with two central scalar amplitudes $\sigma_0=0.25, 0.05\, M_{\rm{Pl}}$, whose fermionic matter is described by Eq.~(\ref{eqrel}), with $k=100\, m^{-2}M_{\rm{Pl}}^{-2}$, $\Gamma=2$, and the central pressures are limited to the range $p_0 \in [4\times 10^{-6}, 10^{-2}]\, m^{2}M_{\rm{Pl}}^2$. The results are shown in Figure~\ref{Fig3}, for one Hordenski model with $\Lambda=1.5\,M_{\rm{Pl}}^{1/3}m^{2/3}$ and $c_4=-1/2$. As can be seen, if we start from a purely scalar configuration, $p = 0$ (dashed line), and we increase the central pressure $p_0$, the new configurations will have larger radii and masses until reaching a maximum point from which both quantities decrease again. It is easy to see when comparing the curves for $\sigma_0=0.25\, M_{\rm{Pl}}$ and $\sigma_0=0.05\, M_{\rm{Pl}}$, that the largest mass point can be reached with a lower central pressure (see color bar) for a smaller scalar field amplitude. 

If we move to the right in the $p=0$ curve, the configurations resulting from increasing the central pressure become more similar to that obtained considering only fermionic matter in the GR (NS curve). Note that the curves $p=0$ and $\sigma_0=0$ are the lower and upper borders, in the sense that all configurations are enclosed between these. We shaded in red (blue) the region where the pressure (scalar field) profile drop more sharply that than the scalar field (pressure) profile. The limit cases i) and ii) are reached when $p\to 10^{-2}\, m^{2}M_{\rm{Pl}}^2$ and $p\to 4\times 10^{-6}\, m^{2}M_{\rm{Pl}}^2$, respectively. Although the results correspond to the a particular model, their described features are valid for the rest of the models ($c_4=0, 1/2$). We also expect the qualitative aspects of our results to hold for other EOS, as a different choice only changes the shape of the curve $M$ vs $R$ (see e.g.,~\cite{DiGiovanni:2021ejn}), and the respective regions indicated in Fig.~\ref{Fig3}, but not the physical behavior of the solutions.

In order to restore the physical quantities for the axes in Figure~\ref{Fig3}, the following relations are needed
\begin{equation}\label{eq.mass.radius.HS}
    M_{99} =  \frac{5.31\bar{M}_{99}}{m[\rm{eV}]}\times 10^{-12}\,M_{\odot} ,\quad
    R_{99} =  \frac{1.97\bar{R}_{99}}{m[\rm{eV}]}\times 10^{-10}\,\rm{km} .
\end{equation}
Table~\ref{tabla1} shows some configurations (and their corresponding parameters) capable to reproduce the mass of some strange star candidates: SAX J1808.4-3658~\cite{2016EPJC...76..693M}, Vela X-1~\cite{2011ApJ...730...25R}, PSRJ 0348+0432~\cite{Maurya:2015wma}, and 4U 1608-52~\cite{Guver:2008gc}. The third column shows the radii predicted for the respective stars, which are of the same order as those reported in~\cite{Guver:2008gc, Maurya:2015wma, 2016EPJC...76..693M, Sharif:2020xwm}. It is necessary to point out that the resultant values can be obtained using others set of parameters ($\sigma_0, p_0, \Lambda, m$), and this degeneration in the free parameters make it necessary to complement our analysis with other physical observables (e.g.~\cite{DiGiovanni:2021ejn}), but this is beyond the scope of the present manuscript.
	
\renewcommand{\tabcolsep}{11pt}
\renewcommand {\arraystretch}{1.5}
\begin{table}[t]
	\centering
	
\scalebox{0.94}{
	\begin{tabular}{@{\hskip .05 in} l@{\hskip .15 in} c@{\hskip .2 in} c@{\hskip .2 in} c@{\hskip .2 in} c@{\hskip .2 in} c@{\hskip .2 in} c@{\hskip .2 in} c@{\hskip .1 in}}
		\hline\hline
		& $M$ & $R_{99}$ & $\sigma_{0}$ & $\rho_0$ & $p_0$ & $\omega$& $C$\\[-0.1cm]
		&$[M_{\odot}]$& $[\textrm{km}]$ & $[M_{\mathrm{Pl}}]$ &$10^{18}[\textrm{kg/m}^{3}]$& $10^{36}$ [Pa] & $[m]$& \\[0.1cm]
		\hline
		SAX J1808.4-3658~\cite{2016EPJC...76..693M} &$1.44$& $8.5$&$0.12$&$8.38$ & $1.73$ &$0.175496$&$ 0.25$\\[0.1cm]
		 Vela X-1~\cite{2011ApJ...730...25R} &$1.788$&$10.1$&$0.10$&$3.45$&$0.29$ &$0.286371$&$0.26$\\[0.1cm]
		 PSRJ 0348+0432~\cite{Maurya:2015wma} &$2.1$& $13.0$&$0.01$&$1.41$& $0.049$& $0.442769$&$0.24 $\\[0.1cm]
		 4U 1608-52~\cite{Guver:2008gc} &$1.74$& $10.9$&$0.35$&$4.88$&$0.589$&$0.286509$&$0.23$\\[0.1cm]
		\hline
	\end{tabular}
	}
	\caption{{\bf Predicted parameters for strange star candidates.} The configurations correspond to a family with $c_4=1/2$, $\Lambda=1.5\,M_{\rm{Pl}}^{1/3}m^{2/3}$, where the scalar field mass is $m=5.135\times 10^{-10}$ eV. The central mass density $\rho_0$ is computed through the EOS~(\ref{EDE}).}\label{tabla1}
\end{table}

\section{Compactness}\label{sec4}
\begin{figure}
	\centering	
	\scalebox{0.45}{
		\input{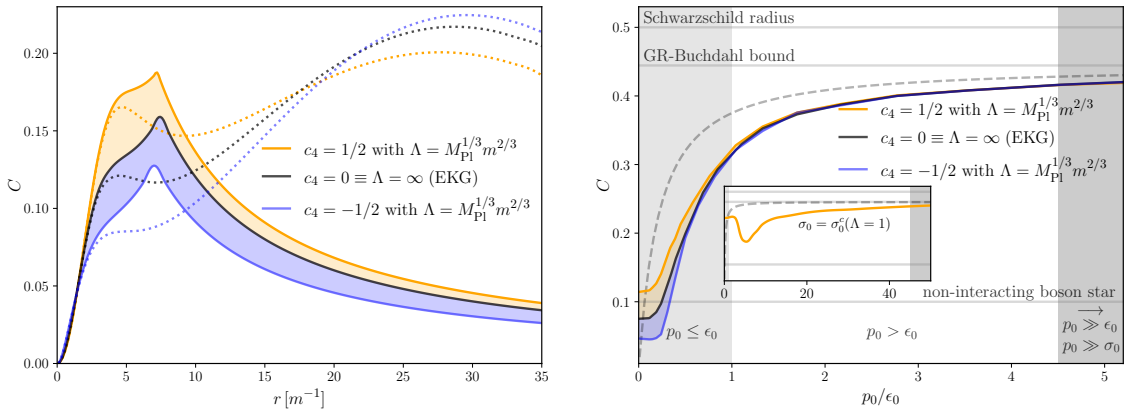}
	}
	\caption{{\bf Hybrid stars compactness.} The compactness~(\ref{eq.compacness}), for the models $c_4=\pm1/2$, with $\sigma_0=0.25 \, M_{\rm{Pl}}$, and a set of values of $\Lambda \in [1, \infty)\, M^{1/3}_{\rm{Pl}}m^{2/3}$. The solid/dotted orange (blue) curve corresponds to the model $c_4=1/2$ ($c_4=-1/2$) with the border value $\Lambda= M^{1/3}_{\rm{Pl}}m^{2/3}$, while the black line correspond to $\Lambda=\infty$. (Left) The compactness profile as function of the radial coordinate for a polytropic EOS (dotted lines) with $k=100\, m^{-2}M_{\rm{Pl}}^{-2}$, $\Gamma=2$, and an incompressible fluid (solid lines), with $\epsilon_0=0.01\, M^{2}_{\rm{Pl}}m^{2}$. (Right) One example of our numerical Buchdahl's limit implementation for the same incompressible fluid. The dashed black line represents the GR case, while the left (right) shape regions denote the hybrid stars with $p_0 \leq \epsilon_0$ ($p_0 \gg \epsilon_0$ and $p_0\gg\sigma_0$). The inset corresponds to the amplitude $\sigma^c_0=0.6  \,  M_{\rm{Pl}}$, it is the last stable configuration for the model with $c_4=1/2$, $\Lambda= M^{1/3}_{\rm{Pl}}m^{2/3}$ and $p=0$. For comparison we also show the following maximum compactness limits: Schwarzschild black hole $C=1/2$, Buchdahl's limit $C_{\rm{max}}=4/9$~\cite{Buchdahl:1959zz}, and that of a BS without self-interactions $C_{\rm{max}}=0.1$.}\label{compac}
\end{figure}

It is usual to define for objects with a sharp border (e.g. fermionic stars, black holes) the compactness as the ratio between their total mass and radius: $C \equiv G M_T/(c^2 R)$. In the case of a hybrid star (in general for stars with a scalar field component), we define the compactness in terms of the $99\%$ quantities in the form
\begin{equation}\label{eq.compacness}
    C\equiv \frac{M_{99}}{8\pi M_{\rm{Pl}}^2 R_{99}} \, ,
\end{equation}
where the factor of $8\pi M_{\rm{Pl}}^2$ is included in such a way that the resulting number is dimensionless. Notice that for the limit case i), which corresponds to the fermionic dominated case, we find $R_{99} \sim R$ where $R$ is the radius at which $p(R)=0$.

In the left panel of Figure~\ref{compac} we show the compactness profile $C(r)$ for a configuration with $\sigma_0=0.25 \, M_{\rm{Pl}}$, $p_0=3.82 \times 10^{-3} m^{2}M_{\rm{Pl}}^2$, $k=100\, m^{-2}M_{\rm{Pl}}^{-2}$ and $\Gamma=2$ for the models $c_4=0,\pm1/2$ (indicated by the different colors). The dotted curves correspond to the polytropic fluid, whereas the solid curves correspond to the pressureless fluid, and we can see that the maximum value the compactness can reach depend on the parameter $c_4$, although the influence of the latter parameter seems to be different for each type of fluid. The hybrid star with a polytropic fluid has a larger compactness, but this appears so for the chosen value of $p_0$, as one can increase its value for the pressureless fluid and reach much larger compactness. We show some illustrative examples of our numerical realization for the models $c_4=\pm1/2$, with $\epsilon_0=0.01\, M^{2}_{\rm{Pl}}m^{2}$, $\sigma_0=0.25 \,  M_{\rm{Pl}}$. The solid orange (blue) curve corresponds to the model $c_4=1/2$ ($c_4=-1/2$) with the border value $\Lambda= M^{1/3}_{\rm{Pl}}m^{2/3}$, while the black curve corresponds to the other border $\Lambda=\infty$ (which is equivalent to EKG model). The shaded orange (blue) region corresponds to the rest of configurations with values of $\Lambda$ in between these borders values.

The changes in the compactness profiles will imply changes in several properties, for example: the possible gravitational radiation emitted by an asymmetric neutron star~\cite{Jaranowski:1998qm} or by compact binary systems~\cite{Hanna:2016uhs, Palenzuela:2017kcg}. The different criteria applied in the selection of the NS (core) EOS (see~\cite{Zdunik:2016vza, Li:2020dst} for a summary) could also be affected, because in the context of the GR the stars cannot reach masses of $2\, M_{\odot}$, could be viable now in the context of hybrid stars. It is necessary to comment that although the results presented so far correspond to a polytropic EOS, the authors hope that this behavior is generic. A study in detail of more realistic EOS would be very interesting, but it is beyond the scope of the present work.

In terms of the compactness~\eref{eq.compacness}, neutron stars in GR may reach values in the range of $C \approx 0.1 - 0.2$~\cite{Xtreme}. For a BS with no self-interactions the compactness can be as large as $C=0.1$~\cite{Liebling:2012fv}, growing up to $C=0.158$ if we include an attractive $\lambda\phi^4$ self-interaction term, and $C \approx 0.33$ is possible considering solitonic potentials for the scalar field~\cite{PhysRevD.35.3658, Cardoso:2016oxy, Palenzuela:2017kcg}. In all these cases the compactness are below the Buchdahl's limit $C=4/9$~\cite{Buchdahl:1959zz}\footnote{ Many are the works that study, and generalize this limit. Some assuming various situations~\cite{Andreasson:2007ck, Karageorgis:2007cy, Andreasson:2012dj, Sharma:2021fda, Sharma:2020ooh, Dadhich:2019jyf}, or extensions to GR~\cite{Dadhich:2010qh, Chakraborty:2020ifg, Garcia-Aspeitia:2014pna, Kumar:2021vqa, Dadhich:2016fku}.}. To close this section we assume a similar criteria (a constant fermionic energy density and isotropic pressure), and we explore whether, in these theories where the scalar field is non-minimally coupled to the metric, there is any change to Buchdahl's limit. 

Due to the intricacy of the system~\eref{fielg}-\eref{eq.KG.modified}, our implementations is numerical. We fix an energy density value $\epsilon_0$, and computed the compactness for a set of hybrid stars profiles with increasing values of $p_0$. This procedure is repeated for different values of $\Lambda \in [1, \infty)\, M^{1/3}_{\rm{Pl}}m^{2/3}$ and $\sigma_0 \in [10^{-2}, \sigma^c_0(\Lambda)] M_{\rm{Pl}}$, where $\sigma^c_0(\Lambda)$ corresponds to the last stable configuration with $p=0$ for a given value of $\Lambda$. In the right panel of Figure~\ref{compac} we show our numerical study of the Buchdahl's limit. The dashed black line represents the GR case, while the left (right) shaded regions denote the hybrid stars with $p_0 \leq \epsilon_0$ ($p_0 \gg \epsilon_0$ and $p_0 \gg \sigma_0$). Note that for large values of $p_0$, all the compactness profiles converge to a limit value smaller than Buchdahl's limit. We have verified that for the extreme case: $c_4=1/2$, $\Lambda= M^{1/3}_{\rm{Pl}}m^{2/3}$, $\sigma=\sigma^c_0=0.6  \,  M_{\rm{Pl}}$ (see the inset in Fig.~\ref{compac}), this behavior is still true\footnote{One might think that for a large $p_0$ value, this conclusion is not valid. However, analyzing Figure~\ref{Fig3} we conclude that for such cases the hybrid stars can be seen as a typical NS, therefore the conclusion is still valid.}. That is, the Buchdahl's limit remains as the upper bound in the compactness for the hybrid stars. Note that considering another value of $\epsilon_0$ does not affect our conclusion, it will only imply that the shaded regions move to the left/right depending on the chosen value.

\section{Concluding remarks}\label{sec:conclusions}
In this paper we have shown that is possible to obtain self-gravitating hybrid objects in a low energy effective model~\cite{Barranco:2021auj}. This model represents a sub-set of the (beyond) Horndeski family that introduces infrared modifications of gravity, and in which scalar field is non-minimally coupled to the metric.

The constructed hybrid stars are composed by a mixing of scalar and fermionic matter whose only interaction is gravitational. Unlike previous works (see e.g.~\cite{Maselli:2016gxk, Babichev:2016jom}), we considered a massive scalar field with a time dependence and a mass range between $10^{-9}-10^{-11}$ eV that correspond to astrophysical objects (see Figure~\ref{Fig1.1}). These objects present some differences and similarities with respect to their EKG counterpart. On the one hand, configurations with a negative (positive) dimensionless coupling $c_4$ have scalar/masses profiles that are smaller (larger) than their counterparts in GR. On the other hand, similarly to the EKG counterpart, for a fixed scalar field amplitude $\sigma_0$, and a large (small) value of the central fermionic pressure $p_0$, the equivalents NS (BS) profiles are recovered. Two limit cases were identified: i) stars with a bosonic core, and ii) the opposite case of stars with a fermionic core. Despite that we used a polytropic EOS to describe the fermionic matter, the authors consider that our concluding remarks would be valid for any other EOS, as the aforementioned behaviors are due the gravitational model and not of the EOS used. However, a future study with more realistic EOS could help to further validate our results.

Finally, we studied the compactness of these stars, and our results show that, unlike typical NS and BS, the hybrids stars present a local and global maximum in the compactness profile (as a function of the radial distance). This change could imply possible signatures in some astrophysical observables. In the limit cases i), ii), their respective compactness profiles are equivalent to GR results. Additionally, an incompressible fluid with a constant energy density was considered, and we implemented a numerical Buchdahl's limit. We showed that for our parameter space the compactness of these stars will always be less than $4/9$, which leads us to conclude that the standard Buchdahl's limit is still valid for this family of Hordenski models. It is important to note that this conclusion about the Buchdahl's limit can be altered if the solutions are coupled strongly and $\Lambda\leq  M^{1/3}_{\rm{Pl}}m^{2/3}$ (blue region in Fig.~\ref{Fig1.1}). However, these cases are model dependent and beyond the scope of the present manuscript, and we shall report about them elsewhere.

% Acknowledgments
\ack{We thank Alberto Diez-Tejedor for several enlightening discussions about some parts of this paper. This work was partially supported by Programa para el Desarrollo Profesional Docente; Direcci\'on de Apoyo a la Investigaci\'on y al Posgrado, Universidad de Guanajuato; CONACyT M\'exico under Grants No. A1-S-17899, No. 286897, No. 297771, No. 304001; CONACyT ``Ciencia de Frontera”, projects No. 376127 “Sombras, lentes y ondas gravitatorias generadas por objetos compactos astrofísicos'', and the Instituto Avanzado de Cosmolog\'ia Collaboration. We acknowledge the use of the COUGHs server at the Universidad de Guanajuato.}

\appendix
\section{Stability analysis}\label{apendix1}

The analysis of the stability of the Horndeski fermion-boson stars is more complicated than in the standard GR cases, whether boson stars~\cite{Gleiser:1988rq, Gleiser:1988ih}, fermion stars~\cite{Friedman:1988er, Cook:1993qr} or fermion-boson stars~\cite{Henriques:1989ez, Henriques:1990xg, Jetzer:1990xa}.
There are stability theorems that indicate the existence of a critical mass such that $dM/d\rho_0=0$, with $\rho_0$ the central value of the scalar or fermion density, and these critical points indicate the transition between the stable and unstable configurations. For fermion-boson stars, stability can still be analysed using the binding energy and the number of bosonic and fermionic particles as a function of the two free parameters $(\sigma_0, p_0)$~\cite{Henriques:1989ez, Henriques:1990xg, Jetzer:1990xa}. 

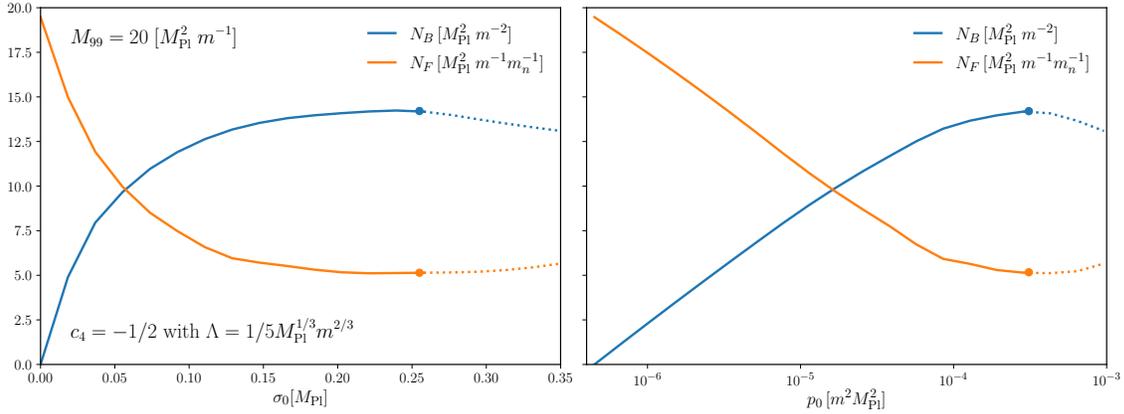
\begin{figure}
	\centering	
	\scalebox{0.45}{
		\input{Figuras/estab1.pgf}
	}
	\caption{{\bf $N_{B}$ and $N_{F}$ vs $(\sigma_0,p_0)$.} The number of bosons $N_B$ (Eq.~(\ref{EqNB})) and fermions $N_F$ (Eq.~(\ref{EqNF})) as function of the central values $\sigma_0$ (left panel) and $p_0$ (right panel), for a family of configurations with a fixed mass $M_{99}=20\; M_{\mathrm{Pl}}^{2}\; m^{-1}$, corresponding to the model $c_4=-1/2$ with $\Lambda= (1/5) M^{1/3}_{\mathrm{Pl}}m^{2/3}$. Note that the maximum of $N_B$ coincides with the occurrence of the minimum of $N_F$. The equilibrium solutions to the right of the critical points (dotted lines) correspond to unstable configurations. }\label{FigA1}
\end{figure}

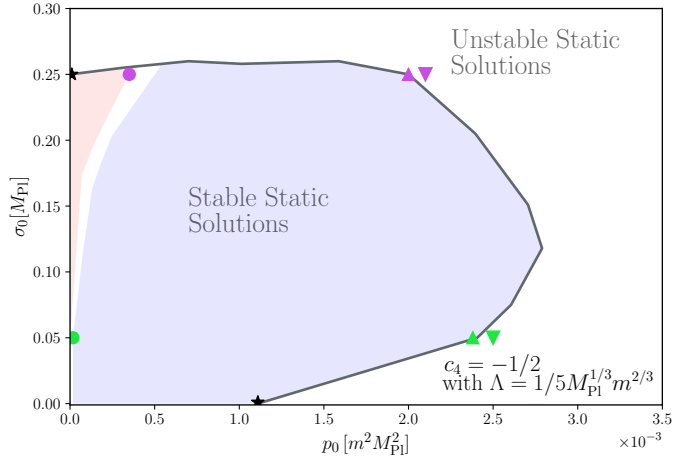
\begin{figure}
	\centering	
	\scalebox{0.5}{
		\input{Figuras/estab2.pgf}
	}
	\caption{{\bf Stability/instability regions.} Stability results for the model $c_4=-1/2$ with $\Lambda= (1/5) M^{1/3}_{\mathrm{Pl}}m^{2/3}$, $\Gamma=2$, and $k=100\, m^{-2}M_{\rm{Pl}}^{-2}$. The black solid line delimits the stable/unstable region on the plane $(\sigma_0,p_0)$. The markers and shaded regions correspond to those also indicated in Fig.~\ref{eq.mass.radius.HS}.}\label{FigA2}
\end{figure}

Here, we use the criterion developed in~\cite{Valdez-Alvarado:2012rct} (an alternative criterion to that of the original papers~\cite{Henriques:1989ez, Henriques:1990xg}) to find the critical points (values of the pair $(\sigma_0, p_0)$), and the stability regions of the Horndenski stars.\footnote{The stability theorem for boson stars indicates the existence of a critical point where $dN_{B}/d\rho_0=dM/d\rho_0=0$, in which the scalar field is minimally coupled to gravity~\cite{Gleiser:1988ih}. It has recently been validated that a similar criterion applies for Horndeski stars described by Eq.~(\ref{eq.interactions}), see the discussion in Section~4.1.1 of Ref.~\cite{Barranco:2021auj} for more details.} We follow the recipe of~\cite{Valdez-Alvarado:2020vqa}, which is summarized below.
\begin{enumerate}
    \item First, the number of bosons $N_B$, and fermions $N_F$, are computed using equations defined below for a family of Horndeski fermion-boson stars that have the same $M_0$ mass.
    \item Later, the critical pair $(\sigma_0, p_0)$ is identified as the point that satisfies the conditions:
    \begin{eqnarray}\label{estCond}
    \frac{\partial N_{B}}{\partial\sigma_0}\Bigg|_{M=M_0}=\frac{\partial N_{B}}{\partial p_0}\Bigg|_{M=M_0}=0\,,\\
    \frac{\partial N_{F}}{\partial\sigma_0}\Bigg|_{M=M_0}=\frac{\partial N_{F}}{\partial p_0}\Bigg|_{M=M_0}=0\,.\label{estCond2}
    \end{eqnarray}
    Configurations located to the left (right) of the point where the maximum and minimum of $N_B$, $N_F$ coalesce (defined by Eqs.~(\ref{estCond}, \ref{estCond2})), are considered stable (unstable) configurations. An illustrative case for $M_0=M_{99}=20 M_{\rm{Pl}}^2 m^{-1}$ is shown in Figure~\ref{FigA1}, where the stable (unstable) configurations are represented by a solid (dotted) line.
    \item Finally, step (ii) is repeated for a set of $M_0$ masses. Identifying the respective critical points, we can construct the boundary curve that splits the parameter space in two well-defined regions on the plane $(\sigma_0,p_0)$: inside the border the configurations are stable, whereas those outside are unstable, see Figure~\ref{FigA2}. 
\end{enumerate}

The boson number $N_B$ is defined by the Noether conserved charge associated to the global $U(1)$ transformation $\phi\to\phi\, e^{i\theta}$ of the Lagrangian~(\ref{eq.interactions}). In general, it is straightforward to show that the conserved current is given by
\begin{equation}\label{Noether_Eq}
 j_{\mu}= i\left[\frac{\partial\mathcal{L}}{\partial(\nabla^{\mu}\nabla^{\nu}\phi)}\nabla^{\nu}\phi-\nabla^{\nu}\left(\frac{\partial\mathcal{L}}{\partial(\nabla^{\mu}\nabla^{\nu}\phi)}\right)\phi  
 +\frac{\partial\mathcal{L}}{\partial(\nabla^{\mu}\phi)}\phi\right]+\rm{c.c.}\,.
\end{equation}
Associated to this current there is a conserved charge
\begin{equation}
    N_B=-\int_{\Sigma_t}j_{\mu}n^\mu d\gamma = -4\pi\int_0^\infty \frac{g}{N} r^2 j_0 dr,\label{EqNB}
\end{equation}
where $n^{\mu}=(1/N,0,0,0)$ is the future-directed time-like unit normal vector to the Cauchy hypersurface $\Sigma_t$, $d\gamma=\sqrt{\textrm{det}(\gamma_{ij})} d^3x$ is the volume element on the hypersurface, and the last integral in Eq.~(\ref{EqNB}) assumes a static spherically symmetric spacetime line-element of the form~(\ref{metric}).

Using the Lagrangian~(\ref{eq.interactions}) (with $d_4=0$) in the conserved current~(\ref{Noether_Eq}) we find
\begin{equation}
    j_{\mu}=i J_{\mu}\nonumber+\frac{2 i c_4 M_{\rm{Pl}}}{\Lambda}\Big(G_{\mu\nu}J^{\nu}-\big[\nabla_{\mu}\phi\Box\bar{\phi}-\rm{c.c}\big]-\big[\nabla_\nu\nabla_\mu\phi\nabla^{\nu}\bar{\phi}-\rm{c.c}\big]\Big)\,,\nonumber
\end{equation}
where $J_{\psi}\equiv\bar{\phi}\nabla_\psi\phi-\phi\nabla_\psi\bar{\phi}$. Note that the second term appears from the second order derivatives in the Lagrangian~(\ref{eq.interactions}). For a static, spherically-symmetric spacetime~(\ref{metric}), and a harmonic ansatz for the scalar field, see Eq.~(\ref{ansat}), we get in particular that
\begin{equation}
\fl	j_{0}=-2 \omega \sigma^2\Bigg[1+\frac{2 c_4 M_{\rm{Pl}}}{\Lambda^3 r^2 g^2}\Bigg(1-g^{2}-2r \mathcal{G}\left(1-\frac{r\chi}{2}\right)+2r\chi\left\lbrace 1+\frac{r\Theta}{2}+\frac{r\chi}{2}-\frac{r \mathcal{N}}{2}\right\rbrace \Bigg)\Bigg]\nonumber\,,\label{j0}
\end{equation}
where $\mathcal{G}=\frac{d\ln g}{dr}$, $\mathcal{N}=\frac{d\ln N}{dr}$, $\Theta=\frac{d \ln \sigma'}{dr}$, and $\chi=\frac{d\ln \sigma}{dr}$. Finally, the boson number $N_B$ is computed from Eq.~(\ref{EqNB}) using $j_0$ from Eq.~(\ref{j0}). 

To compute the fermion number $N_F$, we start from the rest mass density expression: $\rho\equiv m_i N/V$, where $N$ is the number of particles of mass $m_i$ in a volume $V$. Using the fact that the fermionic number is conserved, we arrive to~\cite{Misner:1974qy, Valdez-Alvarado:2012rct}
\begin{equation}
    N_F= \frac{4\pi}{m_n}\int_0^\infty g \rho r^2  dr,\label{EqNF}
\end{equation}
where the rest mass density profile is given by the equation of state (e.g., Eq.~(\ref{EDE})). 

Figure~\ref{FigA2} shows the stability results for the model $c_4=-1/2$ with $\Lambda= (1/5) M^{1/3}_{\mathrm{Pl}}m^{2/3}$, $\Gamma=2$, and $k=100\, m^{-2}M_{\rm{Pl}}^{-2}$. The black solid line represents the border that delimit the stable/unstable region on the plane $(\sigma_0,p_0)$, and the markers correspond to the same configurations indicated in Figure~\ref{eq.mass.radius.HS}. Notice that the dark star markers are purely bosonic or fermionic configurations, and represent the most massive configurations that are stable of such configurations. Similarly to Figure~\ref{Fig3}, the red (blue) shaded region denote configurations with a fermionic (bosonic) core for the above EOS. 

As previously commented, for the determination of the stability border, the bosonic/fermionic number is computed for different total masses $M_0$, and then one looks for the configuration that meets the conditions Eqs.~(\ref{estCond}) and~(\ref{estCond2}). Typical behaviors of the profiles of $N_B, N_F$ are shown in Figure~\ref{FigA1} for the case $M_{99}=20 M_{\rm{Pl}}^2 m^{-1}$. The stable (unstable) configurations are represented by a solid (dotted) line. Notice that a configuration with $M_{99}=20 M_{\rm{Pl}}^2 m^{-1}$ is not possible for purely bosonic stars (right panel). Our results in general are equivalent to those reported in Figure~4 of Ref.~\cite{Valdez-Alvarado:2012rct} within the GR context. Nonetheless, it is important to point out that stable hybrid stars are possible for central values $(\sigma_0,p_0)$ above the critical values corresponding to their respective purely bosonic or fermionic configurations (represented by the dark star markers in Fig.~\ref{FigA2}).

\section*{References}
\bibliographystyle{unsrt}

\bibliography{ref.bib}

\end{document}

%% file: Figuras/estab1.pgf
%% Creator: Matplotlib, PGF backend
%%
%% To include the figure in your LaTeX document, write
%%   \input{<filename>.pgf}
%%
%% Make sure the required packages are loaded in your preamble
%%   \usepackage{pgf}
%%
%% and, on pdftex
%%   \usepackage[utf8]{inputenc}\DeclareUnicodeCharacter{2212}{-}
%%
%% or, on luatex and xetex
%%   \usepackage{unicode-math}
%%
%% Figures using additional raster images can only be included by \input if
%% they are in the same directory as the main LaTeX file. For loading figures
%% from other directories you can use the `import` package
%%   \usepackage{import}
%%
%% and then include the figures with
%%   \import{<path to file>}{<filename>.pgf}
%%
%% Matplotlib used the following preamble
%%   \usepackage{amssymb}
%%
\begingroup%
\makeatletter%
\begin{pgfpicture}%
\pgfpathrectangle{\pgfpointorigin}{\pgfqpoint{13.154490in}{4.935584in}}%
\pgfusepath{use as bounding box, clip}%
\begin{pgfscope}%
\pgfsetbuttcap%
\pgfsetmiterjoin%
\definecolor{currentfill}{rgb}{1.000000,1.000000,1.000000}%
\pgfsetfillcolor{currentfill}%
\pgfsetlinewidth{0.000000pt}%
\definecolor{currentstroke}{rgb}{1.000000,1.000000,1.000000}%
\pgfsetstrokecolor{currentstroke}%
\pgfsetdash{}{0pt}%
\pgfpathmoveto{\pgfqpoint{0.000000in}{0.000000in}}%
\pgfpathlineto{\pgfqpoint{13.154490in}{0.000000in}}%
\pgfpathlineto{\pgfqpoint{13.154490in}{4.935584in}}%
\pgfpathlineto{\pgfqpoint{0.000000in}{4.935584in}}%
\pgfpathclose%
\pgfusepath{fill}%
\end{pgfscope}%
\begin{pgfscope}%
\pgfsetbuttcap%
\pgfsetmiterjoin%
\definecolor{currentfill}{rgb}{1.000000,1.000000,1.000000}%
\pgfsetfillcolor{currentfill}%
\pgfsetlinewidth{0.000000pt}%
\definecolor{currentstroke}{rgb}{0.000000,0.000000,0.000000}%
\pgfsetstrokecolor{currentstroke}%
\pgfsetstrokeopacity{0.000000}%
\pgfsetdash{}{0pt}%
\pgfpathmoveto{\pgfqpoint{0.487343in}{0.625214in}}%
\pgfpathlineto{\pgfqpoint{6.536123in}{0.625214in}}%
\pgfpathlineto{\pgfqpoint{6.536123in}{4.777714in}}%
\pgfpathlineto{\pgfqpoint{0.487343in}{4.777714in}}%
\pgfpathclose%
\pgfusepath{fill}%
\end{pgfscope}%
\begin{pgfscope}%
\pgfsetbuttcap%
\pgfsetroundjoin%
\definecolor{currentfill}{rgb}{0.000000,0.000000,0.000000}%
\pgfsetfillcolor{currentfill}%
\pgfsetlinewidth{0.803000pt}%
\definecolor{currentstroke}{rgb}{0.000000,0.000000,0.000000}%
\pgfsetstrokecolor{currentstroke}%
\pgfsetdash{}{0pt}%
\pgfsys@defobject{currentmarker}{\pgfqpoint{0.000000in}{-0.048611in}}{\pgfqpoint{0.000000in}{0.000000in}}{%
\pgfpathmoveto{\pgfqpoint{0.000000in}{0.000000in}}%
\pgfpathlineto{\pgfqpoint{0.000000in}{-0.048611in}}%
\pgfusepath{stroke,fill}%
}%
\begin{pgfscope}%
\pgfsys@transformshift{0.487343in}{0.625214in}%
\pgfsys@useobject{currentmarker}{}%
\end{pgfscope}%
\end{pgfscope}%
\begin{pgfscope}%
\definecolor{textcolor}{rgb}{0.000000,0.000000,0.000000}%
\pgfsetstrokecolor{textcolor}%
\pgfsetfillcolor{textcolor}%
\pgftext[x=0.487343in,y=0.527992in,,top]{\color{textcolor}\rmfamily\fontsize{13.000000}{15.600000}\selectfont \(\displaystyle {0.00}\)}%
\end{pgfscope}%
\begin{pgfscope}%
\pgfsetbuttcap%
\pgfsetroundjoin%
\definecolor{currentfill}{rgb}{0.000000,0.000000,0.000000}%
\pgfsetfillcolor{currentfill}%
\pgfsetlinewidth{0.803000pt}%
\definecolor{currentstroke}{rgb}{0.000000,0.000000,0.000000}%
\pgfsetstrokecolor{currentstroke}%
\pgfsetdash{}{0pt}%
\pgfsys@defobject{currentmarker}{\pgfqpoint{0.000000in}{-0.048611in}}{\pgfqpoint{0.000000in}{0.000000in}}{%
\pgfpathmoveto{\pgfqpoint{0.000000in}{0.000000in}}%
\pgfpathlineto{\pgfqpoint{0.000000in}{-0.048611in}}%
\pgfusepath{stroke,fill}%
}%
\begin{pgfscope}%
\pgfsys@transformshift{1.351454in}{0.625214in}%
\pgfsys@useobject{currentmarker}{}%
\end{pgfscope}%
\end{pgfscope}%
\begin{pgfscope}%
\definecolor{textcolor}{rgb}{0.000000,0.000000,0.000000}%
\pgfsetstrokecolor{textcolor}%
\pgfsetfillcolor{textcolor}%
\pgftext[x=1.351454in,y=0.527992in,,top]{\color{textcolor}\rmfamily\fontsize{13.000000}{15.600000}\selectfont \(\displaystyle {0.05}\)}%
\end{pgfscope}%
\begin{pgfscope}%
\pgfsetbuttcap%
\pgfsetroundjoin%
\definecolor{currentfill}{rgb}{0.000000,0.000000,0.000000}%
\pgfsetfillcolor{currentfill}%
\pgfsetlinewidth{0.803000pt}%
\definecolor{currentstroke}{rgb}{0.000000,0.000000,0.000000}%
\pgfsetstrokecolor{currentstroke}%
\pgfsetdash{}{0pt}%
\pgfsys@defobject{currentmarker}{\pgfqpoint{0.000000in}{-0.048611in}}{\pgfqpoint{0.000000in}{0.000000in}}{%
\pgfpathmoveto{\pgfqpoint{0.000000in}{0.000000in}}%
\pgfpathlineto{\pgfqpoint{0.000000in}{-0.048611in}}%
\pgfusepath{stroke,fill}%
}%
\begin{pgfscope}%
\pgfsys@transformshift{2.215566in}{0.625214in}%
\pgfsys@useobject{currentmarker}{}%
\end{pgfscope}%
\end{pgfscope}%
\begin{pgfscope}%
\definecolor{textcolor}{rgb}{0.000000,0.000000,0.000000}%
\pgfsetstrokecolor{textcolor}%
\pgfsetfillcolor{textcolor}%
\pgftext[x=2.215566in,y=0.527992in,,top]{\color{textcolor}\rmfamily\fontsize{13.000000}{15.600000}\selectfont \(\displaystyle {0.10}\)}%
\end{pgfscope}%
\begin{pgfscope}%
\pgfsetbuttcap%
\pgfsetroundjoin%
\definecolor{currentfill}{rgb}{0.000000,0.000000,0.000000}%
\pgfsetfillcolor{currentfill}%
\pgfsetlinewidth{0.803000pt}%
\definecolor{currentstroke}{rgb}{0.000000,0.000000,0.000000}%
\pgfsetstrokecolor{currentstroke}%
\pgfsetdash{}{0pt}%
\pgfsys@defobject{currentmarker}{\pgfqpoint{0.000000in}{-0.048611in}}{\pgfqpoint{0.000000in}{0.000000in}}{%
\pgfpathmoveto{\pgfqpoint{0.000000in}{0.000000in}}%
\pgfpathlineto{\pgfqpoint{0.000000in}{-0.048611in}}%
\pgfusepath{stroke,fill}%
}%
\begin{pgfscope}%
\pgfsys@transformshift{3.079677in}{0.625214in}%
\pgfsys@useobject{currentmarker}{}%
\end{pgfscope}%
\end{pgfscope}%
\begin{pgfscope}%
\definecolor{textcolor}{rgb}{0.000000,0.000000,0.000000}%
\pgfsetstrokecolor{textcolor}%
\pgfsetfillcolor{textcolor}%
\pgftext[x=3.079677in,y=0.527992in,,top]{\color{textcolor}\rmfamily\fontsize{13.000000}{15.600000}\selectfont \(\displaystyle {0.15}\)}%
\end{pgfscope}%
\begin{pgfscope}%
\pgfsetbuttcap%
\pgfsetroundjoin%
\definecolor{currentfill}{rgb}{0.000000,0.000000,0.000000}%
\pgfsetfillcolor{currentfill}%
\pgfsetlinewidth{0.803000pt}%
\definecolor{currentstroke}{rgb}{0.000000,0.000000,0.000000}%
\pgfsetstrokecolor{currentstroke}%
\pgfsetdash{}{0pt}%
\pgfsys@defobject{currentmarker}{\pgfqpoint{0.000000in}{-0.048611in}}{\pgfqpoint{0.000000in}{0.000000in}}{%
\pgfpathmoveto{\pgfqpoint{0.000000in}{0.000000in}}%
\pgfpathlineto{\pgfqpoint{0.000000in}{-0.048611in}}%
\pgfusepath{stroke,fill}%
}%
\begin{pgfscope}%
\pgfsys@transformshift{3.943789in}{0.625214in}%
\pgfsys@useobject{currentmarker}{}%
\end{pgfscope}%
\end{pgfscope}%
\begin{pgfscope}%
\definecolor{textcolor}{rgb}{0.000000,0.000000,0.000000}%
\pgfsetstrokecolor{textcolor}%
\pgfsetfillcolor{textcolor}%
\pgftext[x=3.943789in,y=0.527992in,,top]{\color{textcolor}\rmfamily\fontsize{13.000000}{15.600000}\selectfont \(\displaystyle {0.20}\)}%
\end{pgfscope}%
\begin{pgfscope}%
\pgfsetbuttcap%
\pgfsetroundjoin%
\definecolor{currentfill}{rgb}{0.000000,0.000000,0.000000}%
\pgfsetfillcolor{currentfill}%
\pgfsetlinewidth{0.803000pt}%
\definecolor{currentstroke}{rgb}{0.000000,0.000000,0.000000}%
\pgfsetstrokecolor{currentstroke}%
\pgfsetdash{}{0pt}%
\pgfsys@defobject{currentmarker}{\pgfqpoint{0.000000in}{-0.048611in}}{\pgfqpoint{0.000000in}{0.000000in}}{%
\pgfpathmoveto{\pgfqpoint{0.000000in}{0.000000in}}%
\pgfpathlineto{\pgfqpoint{0.000000in}{-0.048611in}}%
\pgfusepath{stroke,fill}%
}%
\begin{pgfscope}%
\pgfsys@transformshift{4.807900in}{0.625214in}%
\pgfsys@useobject{currentmarker}{}%
\end{pgfscope}%
\end{pgfscope}%
\begin{pgfscope}%
\definecolor{textcolor}{rgb}{0.000000,0.000000,0.000000}%
\pgfsetstrokecolor{textcolor}%
\pgfsetfillcolor{textcolor}%
\pgftext[x=4.807900in,y=0.527992in,,top]{\color{textcolor}\rmfamily\fontsize{13.000000}{15.600000}\selectfont \(\displaystyle {0.25}\)}%
\end{pgfscope}%
\begin{pgfscope}%
\pgfsetbuttcap%
\pgfsetroundjoin%
\definecolor{currentfill}{rgb}{0.000000,0.000000,0.000000}%
\pgfsetfillcolor{currentfill}%
\pgfsetlinewidth{0.803000pt}%
\definecolor{currentstroke}{rgb}{0.000000,0.000000,0.000000}%
\pgfsetstrokecolor{currentstroke}%
\pgfsetdash{}{0pt}%
\pgfsys@defobject{currentmarker}{\pgfqpoint{0.000000in}{-0.048611in}}{\pgfqpoint{0.000000in}{0.000000in}}{%
\pgfpathmoveto{\pgfqpoint{0.000000in}{0.000000in}}%
\pgfpathlineto{\pgfqpoint{0.000000in}{-0.048611in}}%
\pgfusepath{stroke,fill}%
}%
\begin{pgfscope}%
\pgfsys@transformshift{5.672012in}{0.625214in}%
\pgfsys@useobject{currentmarker}{}%
\end{pgfscope}%
\end{pgfscope}%
\begin{pgfscope}%
\definecolor{textcolor}{rgb}{0.000000,0.000000,0.000000}%
\pgfsetstrokecolor{textcolor}%
\pgfsetfillcolor{textcolor}%
\pgftext[x=5.672012in,y=0.527992in,,top]{\color{textcolor}\rmfamily\fontsize{13.000000}{15.600000}\selectfont \(\displaystyle {0.30}\)}%
\end{pgfscope}%
\begin{pgfscope}%
\pgfsetbuttcap%
\pgfsetroundjoin%
\definecolor{currentfill}{rgb}{0.000000,0.000000,0.000000}%
\pgfsetfillcolor{currentfill}%
\pgfsetlinewidth{0.803000pt}%
\definecolor{currentstroke}{rgb}{0.000000,0.000000,0.000000}%
\pgfsetstrokecolor{currentstroke}%
\pgfsetdash{}{0pt}%
\pgfsys@defobject{currentmarker}{\pgfqpoint{0.000000in}{-0.048611in}}{\pgfqpoint{0.000000in}{0.000000in}}{%
\pgfpathmoveto{\pgfqpoint{0.000000in}{0.000000in}}%
\pgfpathlineto{\pgfqpoint{0.000000in}{-0.048611in}}%
\pgfusepath{stroke,fill}%
}%
\begin{pgfscope}%
\pgfsys@transformshift{6.536123in}{0.625214in}%
\pgfsys@useobject{currentmarker}{}%
\end{pgfscope}%
\end{pgfscope}%
\begin{pgfscope}%
\definecolor{textcolor}{rgb}{0.000000,0.000000,0.000000}%
\pgfsetstrokecolor{textcolor}%
\pgfsetfillcolor{textcolor}%
\pgftext[x=6.536123in,y=0.527992in,,top]{\color{textcolor}\rmfamily\fontsize{13.000000}{15.600000}\selectfont \(\displaystyle {0.35}\)}%
\end{pgfscope}%
\begin{pgfscope}%
\definecolor{textcolor}{rgb}{0.000000,0.000000,0.000000}%
\pgfsetstrokecolor{textcolor}%
\pgfsetfillcolor{textcolor}%
\pgftext[x=3.511733in,y=0.338177in,,top]{\color{textcolor}\rmfamily\fontsize{15.000000}{18.000000}\selectfont \(\displaystyle \sigma_0 [M_{\mathrm{Pl}}]\)}%
\end{pgfscope}%
\begin{pgfscope}%
\pgfsetbuttcap%
\pgfsetroundjoin%
\definecolor{currentfill}{rgb}{0.000000,0.000000,0.000000}%
\pgfsetfillcolor{currentfill}%
\pgfsetlinewidth{0.803000pt}%
\definecolor{currentstroke}{rgb}{0.000000,0.000000,0.000000}%
\pgfsetstrokecolor{currentstroke}%
\pgfsetdash{}{0pt}%
\pgfsys@defobject{currentmarker}{\pgfqpoint{-0.048611in}{0.000000in}}{\pgfqpoint{-0.000000in}{0.000000in}}{%
\pgfpathmoveto{\pgfqpoint{-0.000000in}{0.000000in}}%
\pgfpathlineto{\pgfqpoint{-0.048611in}{0.000000in}}%
\pgfusepath{stroke,fill}%
}%
\begin{pgfscope}%
\pgfsys@transformshift{0.487343in}{0.625214in}%
\pgfsys@useobject{currentmarker}{}%
\end{pgfscope}%
\end{pgfscope}%
\begin{pgfscope}%
\definecolor{textcolor}{rgb}{0.000000,0.000000,0.000000}%
\pgfsetstrokecolor{textcolor}%
\pgfsetfillcolor{textcolor}%
\pgftext[x=0.181596in, y=0.567344in, left, base]{\color{textcolor}\rmfamily\fontsize{13.000000}{15.600000}\selectfont \(\displaystyle {0.0}\)}%
\end{pgfscope}%
\begin{pgfscope}%
\pgfsetbuttcap%
\pgfsetroundjoin%
\definecolor{currentfill}{rgb}{0.000000,0.000000,0.000000}%
\pgfsetfillcolor{currentfill}%
\pgfsetlinewidth{0.803000pt}%
\definecolor{currentstroke}{rgb}{0.000000,0.000000,0.000000}%
\pgfsetstrokecolor{currentstroke}%
\pgfsetdash{}{0pt}%
\pgfsys@defobject{currentmarker}{\pgfqpoint{-0.048611in}{0.000000in}}{\pgfqpoint{-0.000000in}{0.000000in}}{%
\pgfpathmoveto{\pgfqpoint{-0.000000in}{0.000000in}}%
\pgfpathlineto{\pgfqpoint{-0.048611in}{0.000000in}}%
\pgfusepath{stroke,fill}%
}%
\begin{pgfscope}%
\pgfsys@transformshift{0.487343in}{1.144276in}%
\pgfsys@useobject{currentmarker}{}%
\end{pgfscope}%
\end{pgfscope}%
\begin{pgfscope}%
\definecolor{textcolor}{rgb}{0.000000,0.000000,0.000000}%
\pgfsetstrokecolor{textcolor}%
\pgfsetfillcolor{textcolor}%
\pgftext[x=0.181596in, y=1.086406in, left, base]{\color{textcolor}\rmfamily\fontsize{13.000000}{15.600000}\selectfont \(\displaystyle {2.5}\)}%
\end{pgfscope}%
\begin{pgfscope}%
\pgfsetbuttcap%
\pgfsetroundjoin%
\definecolor{currentfill}{rgb}{0.000000,0.000000,0.000000}%
\pgfsetfillcolor{currentfill}%
\pgfsetlinewidth{0.803000pt}%
\definecolor{currentstroke}{rgb}{0.000000,0.000000,0.000000}%
\pgfsetstrokecolor{currentstroke}%
\pgfsetdash{}{0pt}%
\pgfsys@defobject{currentmarker}{\pgfqpoint{-0.048611in}{0.000000in}}{\pgfqpoint{-0.000000in}{0.000000in}}{%
\pgfpathmoveto{\pgfqpoint{-0.000000in}{0.000000in}}%
\pgfpathlineto{\pgfqpoint{-0.048611in}{0.000000in}}%
\pgfusepath{stroke,fill}%
}%
\begin{pgfscope}%
\pgfsys@transformshift{0.487343in}{1.663339in}%
\pgfsys@useobject{currentmarker}{}%
\end{pgfscope}%
\end{pgfscope}%
\begin{pgfscope}%
\definecolor{textcolor}{rgb}{0.000000,0.000000,0.000000}%
\pgfsetstrokecolor{textcolor}%
\pgfsetfillcolor{textcolor}%
\pgftext[x=0.181596in, y=1.605469in, left, base]{\color{textcolor}\rmfamily\fontsize{13.000000}{15.600000}\selectfont \(\displaystyle {5.0}\)}%
\end{pgfscope}%
\begin{pgfscope}%
\pgfsetbuttcap%
\pgfsetroundjoin%
\definecolor{currentfill}{rgb}{0.000000,0.000000,0.000000}%
\pgfsetfillcolor{currentfill}%
\pgfsetlinewidth{0.803000pt}%
\definecolor{currentstroke}{rgb}{0.000000,0.000000,0.000000}%
\pgfsetstrokecolor{currentstroke}%
\pgfsetdash{}{0pt}%
\pgfsys@defobject{currentmarker}{\pgfqpoint{-0.048611in}{0.000000in}}{\pgfqpoint{-0.000000in}{0.000000in}}{%
\pgfpathmoveto{\pgfqpoint{-0.000000in}{0.000000in}}%
\pgfpathlineto{\pgfqpoint{-0.048611in}{0.000000in}}%
\pgfusepath{stroke,fill}%
}%
\begin{pgfscope}%
\pgfsys@transformshift{0.487343in}{2.182401in}%
\pgfsys@useobject{currentmarker}{}%
\end{pgfscope}%
\end{pgfscope}%
\begin{pgfscope}%
\definecolor{textcolor}{rgb}{0.000000,0.000000,0.000000}%
\pgfsetstrokecolor{textcolor}%
\pgfsetfillcolor{textcolor}%
\pgftext[x=0.181596in, y=2.124531in, left, base]{\color{textcolor}\rmfamily\fontsize{13.000000}{15.600000}\selectfont \(\displaystyle {7.5}\)}%
\end{pgfscope}%
\begin{pgfscope}%
\pgfsetbuttcap%
\pgfsetroundjoin%
\definecolor{currentfill}{rgb}{0.000000,0.000000,0.000000}%
\pgfsetfillcolor{currentfill}%
\pgfsetlinewidth{0.803000pt}%
\definecolor{currentstroke}{rgb}{0.000000,0.000000,0.000000}%
\pgfsetstrokecolor{currentstroke}%
\pgfsetdash{}{0pt}%
\pgfsys@defobject{currentmarker}{\pgfqpoint{-0.048611in}{0.000000in}}{\pgfqpoint{-0.000000in}{0.000000in}}{%
\pgfpathmoveto{\pgfqpoint{-0.000000in}{0.000000in}}%
\pgfpathlineto{\pgfqpoint{-0.048611in}{0.000000in}}%
\pgfusepath{stroke,fill}%
}%
\begin{pgfscope}%
\pgfsys@transformshift{0.487343in}{2.701464in}%
\pgfsys@useobject{currentmarker}{}%
\end{pgfscope}%
\end{pgfscope}%
\begin{pgfscope}%
\definecolor{textcolor}{rgb}{0.000000,0.000000,0.000000}%
\pgfsetstrokecolor{textcolor}%
\pgfsetfillcolor{textcolor}%
\pgftext[x=0.100000in, y=2.643594in, left, base]{\color{textcolor}\rmfamily\fontsize{13.000000}{15.600000}\selectfont \(\displaystyle {10.0}\)}%
\end{pgfscope}%
\begin{pgfscope}%
\pgfsetbuttcap%
\pgfsetroundjoin%
\definecolor{currentfill}{rgb}{0.000000,0.000000,0.000000}%
\pgfsetfillcolor{currentfill}%
\pgfsetlinewidth{0.803000pt}%
\definecolor{currentstroke}{rgb}{0.000000,0.000000,0.000000}%
\pgfsetstrokecolor{currentstroke}%
\pgfsetdash{}{0pt}%
\pgfsys@defobject{currentmarker}{\pgfqpoint{-0.048611in}{0.000000in}}{\pgfqpoint{-0.000000in}{0.000000in}}{%
\pgfpathmoveto{\pgfqpoint{-0.000000in}{0.000000in}}%
\pgfpathlineto{\pgfqpoint{-0.048611in}{0.000000in}}%
\pgfusepath{stroke,fill}%
}%
\begin{pgfscope}%
\pgfsys@transformshift{0.487343in}{3.220526in}%
\pgfsys@useobject{currentmarker}{}%
\end{pgfscope}%
\end{pgfscope}%
\begin{pgfscope}%
\definecolor{textcolor}{rgb}{0.000000,0.000000,0.000000}%
\pgfsetstrokecolor{textcolor}%
\pgfsetfillcolor{textcolor}%
\pgftext[x=0.100000in, y=3.162656in, left, base]{\color{textcolor}\rmfamily\fontsize{13.000000}{15.600000}\selectfont \(\displaystyle {12.5}\)}%
\end{pgfscope}%
\begin{pgfscope}%
\pgfsetbuttcap%
\pgfsetroundjoin%
\definecolor{currentfill}{rgb}{0.000000,0.000000,0.000000}%
\pgfsetfillcolor{currentfill}%
\pgfsetlinewidth{0.803000pt}%
\definecolor{currentstroke}{rgb}{0.000000,0.000000,0.000000}%
\pgfsetstrokecolor{currentstroke}%
\pgfsetdash{}{0pt}%
\pgfsys@defobject{currentmarker}{\pgfqpoint{-0.048611in}{0.000000in}}{\pgfqpoint{-0.000000in}{0.000000in}}{%
\pgfpathmoveto{\pgfqpoint{-0.000000in}{0.000000in}}%
\pgfpathlineto{\pgfqpoint{-0.048611in}{0.000000in}}%
\pgfusepath{stroke,fill}%
}%
\begin{pgfscope}%
\pgfsys@transformshift{0.487343in}{3.739589in}%
\pgfsys@useobject{currentmarker}{}%
\end{pgfscope}%
\end{pgfscope}%
\begin{pgfscope}%
\definecolor{textcolor}{rgb}{0.000000,0.000000,0.000000}%
\pgfsetstrokecolor{textcolor}%
\pgfsetfillcolor{textcolor}%
\pgftext[x=0.100000in, y=3.681719in, left, base]{\color{textcolor}\rmfamily\fontsize{13.000000}{15.600000}\selectfont \(\displaystyle {15.0}\)}%
\end{pgfscope}%
\begin{pgfscope}%
\pgfsetbuttcap%
\pgfsetroundjoin%
\definecolor{currentfill}{rgb}{0.000000,0.000000,0.000000}%
\pgfsetfillcolor{currentfill}%
\pgfsetlinewidth{0.803000pt}%
\definecolor{currentstroke}{rgb}{0.000000,0.000000,0.000000}%
\pgfsetstrokecolor{currentstroke}%
\pgfsetdash{}{0pt}%
\pgfsys@defobject{currentmarker}{\pgfqpoint{-0.048611in}{0.000000in}}{\pgfqpoint{-0.000000in}{0.000000in}}{%
\pgfpathmoveto{\pgfqpoint{-0.000000in}{0.000000in}}%
\pgfpathlineto{\pgfqpoint{-0.048611in}{0.000000in}}%
\pgfusepath{stroke,fill}%
}%
\begin{pgfscope}%
\pgfsys@transformshift{0.487343in}{4.258651in}%
\pgfsys@useobject{currentmarker}{}%
\end{pgfscope}%
\end{pgfscope}%
\begin{pgfscope}%
\definecolor{textcolor}{rgb}{0.000000,0.000000,0.000000}%
\pgfsetstrokecolor{textcolor}%
\pgfsetfillcolor{textcolor}%
\pgftext[x=0.100000in, y=4.200781in, left, base]{\color{textcolor}\rmfamily\fontsize{13.000000}{15.600000}\selectfont \(\displaystyle {17.5}\)}%
\end{pgfscope}%
\begin{pgfscope}%
\pgfsetbuttcap%
\pgfsetroundjoin%
\definecolor{currentfill}{rgb}{0.000000,0.000000,0.000000}%
\pgfsetfillcolor{currentfill}%
\pgfsetlinewidth{0.803000pt}%
\definecolor{currentstroke}{rgb}{0.000000,0.000000,0.000000}%
\pgfsetstrokecolor{currentstroke}%
\pgfsetdash{}{0pt}%
\pgfsys@defobject{currentmarker}{\pgfqpoint{-0.048611in}{0.000000in}}{\pgfqpoint{-0.000000in}{0.000000in}}{%
\pgfpathmoveto{\pgfqpoint{-0.000000in}{0.000000in}}%
\pgfpathlineto{\pgfqpoint{-0.048611in}{0.000000in}}%
\pgfusepath{stroke,fill}%
}%
\begin{pgfscope}%
\pgfsys@transformshift{0.487343in}{4.777714in}%
\pgfsys@useobject{currentmarker}{}%
\end{pgfscope}%
\end{pgfscope}%
\begin{pgfscope}%
\definecolor{textcolor}{rgb}{0.000000,0.000000,0.000000}%
\pgfsetstrokecolor{textcolor}%
\pgfsetfillcolor{textcolor}%
\pgftext[x=0.100000in, y=4.719844in, left, base]{\color{textcolor}\rmfamily\fontsize{13.000000}{15.600000}\selectfont \(\displaystyle {20.0}\)}%
\end{pgfscope}%
\begin{pgfscope}%
\pgfpathrectangle{\pgfqpoint{0.487343in}{0.625214in}}{\pgfqpoint{6.048780in}{4.152500in}}%
\pgfusepath{clip}%
\pgfsetrectcap%
\pgfsetroundjoin%
\pgfsetlinewidth{2.007500pt}%
\definecolor{currentstroke}{rgb}{0.121569,0.466667,0.705882}%
\pgfsetstrokecolor{currentstroke}%
\pgfsetdash{}{0pt}%
\pgfpathmoveto{\pgfqpoint{0.487343in}{0.625214in}}%
\pgfpathlineto{\pgfqpoint{0.805700in}{1.641163in}}%
\pgfpathlineto{\pgfqpoint{1.124056in}{2.276432in}}%
\pgfpathlineto{\pgfqpoint{1.442413in}{2.633132in}}%
\pgfpathlineto{\pgfqpoint{1.760770in}{2.902251in}}%
\pgfpathlineto{\pgfqpoint{2.079127in}{3.097100in}}%
\pgfpathlineto{\pgfqpoint{2.397484in}{3.247719in}}%
\pgfpathlineto{\pgfqpoint{2.715841in}{3.359699in}}%
\pgfpathlineto{\pgfqpoint{3.034198in}{3.437093in}}%
\pgfpathlineto{\pgfqpoint{3.352554in}{3.491257in}}%
\pgfpathlineto{\pgfqpoint{3.670911in}{3.525366in}}%
\pgfpathlineto{\pgfqpoint{3.989268in}{3.550129in}}%
\pgfpathlineto{\pgfqpoint{4.307625in}{3.570256in}}%
\pgfpathlineto{\pgfqpoint{4.625982in}{3.581343in}}%
\pgfpathlineto{\pgfqpoint{4.944339in}{3.568218in}}%
\pgfusepath{stroke}%
\end{pgfscope}%
\begin{pgfscope}%
\pgfpathrectangle{\pgfqpoint{0.487343in}{0.625214in}}{\pgfqpoint{6.048780in}{4.152500in}}%
\pgfusepath{clip}%
\pgfsetbuttcap%
\pgfsetroundjoin%
\pgfsetlinewidth{2.007500pt}%
\definecolor{currentstroke}{rgb}{0.121569,0.466667,0.705882}%
\pgfsetstrokecolor{currentstroke}%
\pgfsetdash{{2.000000pt}{3.300000pt}}{0.000000pt}%
\pgfpathmoveto{\pgfqpoint{4.944339in}{3.568218in}}%
\pgfpathlineto{\pgfqpoint{5.262696in}{3.529603in}}%
\pgfpathlineto{\pgfqpoint{5.581053in}{3.478121in}}%
\pgfpathlineto{\pgfqpoint{5.899409in}{3.429922in}}%
\pgfpathlineto{\pgfqpoint{6.217766in}{3.385066in}}%
\pgfpathlineto{\pgfqpoint{6.536123in}{3.343553in}}%
\pgfusepath{stroke}%
\end{pgfscope}%
\begin{pgfscope}%
\pgfpathrectangle{\pgfqpoint{0.487343in}{0.625214in}}{\pgfqpoint{6.048780in}{4.152500in}}%
\pgfusepath{clip}%
\pgfsetbuttcap%
\pgfsetroundjoin%
\definecolor{currentfill}{rgb}{0.121569,0.466667,0.705882}%
\pgfsetfillcolor{currentfill}%
\pgfsetlinewidth{1.003750pt}%
\definecolor{currentstroke}{rgb}{0.121569,0.466667,0.705882}%
\pgfsetstrokecolor{currentstroke}%
\pgfsetdash{}{0pt}%
\pgfsys@defobject{currentmarker}{\pgfqpoint{-0.041667in}{-0.041667in}}{\pgfqpoint{0.041667in}{0.041667in}}{%
\pgfpathmoveto{\pgfqpoint{0.000000in}{-0.041667in}}%
\pgfpathcurveto{\pgfqpoint{0.011050in}{-0.041667in}}{\pgfqpoint{0.021649in}{-0.037276in}}{\pgfqpoint{0.029463in}{-0.029463in}}%
\pgfpathcurveto{\pgfqpoint{0.037276in}{-0.021649in}}{\pgfqpoint{0.041667in}{-0.011050in}}{\pgfqpoint{0.041667in}{0.000000in}}%
\pgfpathcurveto{\pgfqpoint{0.041667in}{0.011050in}}{\pgfqpoint{0.037276in}{0.021649in}}{\pgfqpoint{0.029463in}{0.029463in}}%
\pgfpathcurveto{\pgfqpoint{0.021649in}{0.037276in}}{\pgfqpoint{0.011050in}{0.041667in}}{\pgfqpoint{0.000000in}{0.041667in}}%
\pgfpathcurveto{\pgfqpoint{-0.011050in}{0.041667in}}{\pgfqpoint{-0.021649in}{0.037276in}}{\pgfqpoint{-0.029463in}{0.029463in}}%
\pgfpathcurveto{\pgfqpoint{-0.037276in}{0.021649in}}{\pgfqpoint{-0.041667in}{0.011050in}}{\pgfqpoint{-0.041667in}{0.000000in}}%
\pgfpathcurveto{\pgfqpoint{-0.041667in}{-0.011050in}}{\pgfqpoint{-0.037276in}{-0.021649in}}{\pgfqpoint{-0.029463in}{-0.029463in}}%
\pgfpathcurveto{\pgfqpoint{-0.021649in}{-0.037276in}}{\pgfqpoint{-0.011050in}{-0.041667in}}{\pgfqpoint{0.000000in}{-0.041667in}}%
\pgfpathclose%
\pgfusepath{stroke,fill}%
}%
\begin{pgfscope}%
\pgfsys@transformshift{4.894311in}{3.575413in}%
\pgfsys@useobject{currentmarker}{}%
\end{pgfscope}%
\end{pgfscope}%
\begin{pgfscope}%
\pgfpathrectangle{\pgfqpoint{0.487343in}{0.625214in}}{\pgfqpoint{6.048780in}{4.152500in}}%
\pgfusepath{clip}%
\pgfsetrectcap%
\pgfsetroundjoin%
\pgfsetlinewidth{2.007500pt}%
\definecolor{currentstroke}{rgb}{1.000000,0.498039,0.054902}%
\pgfsetstrokecolor{currentstroke}%
\pgfsetdash{}{0pt}%
\pgfpathmoveto{\pgfqpoint{0.487343in}{4.668057in}}%
\pgfpathlineto{\pgfqpoint{0.805700in}{3.737454in}}%
\pgfpathlineto{\pgfqpoint{1.124056in}{3.100402in}}%
\pgfpathlineto{\pgfqpoint{1.442413in}{2.692109in}}%
\pgfpathlineto{\pgfqpoint{1.760770in}{2.391389in}}%
\pgfpathlineto{\pgfqpoint{2.079127in}{2.177750in}}%
\pgfpathlineto{\pgfqpoint{2.397484in}{1.991696in}}%
\pgfpathlineto{\pgfqpoint{2.715841in}{1.861824in}}%
\pgfpathlineto{\pgfqpoint{3.034198in}{1.810783in}}%
\pgfpathlineto{\pgfqpoint{3.352554in}{1.771706in}}%
\pgfpathlineto{\pgfqpoint{3.670911in}{1.728936in}}%
\pgfpathlineto{\pgfqpoint{3.989268in}{1.698190in}}%
\pgfpathlineto{\pgfqpoint{4.307625in}{1.686382in}}%
\pgfpathlineto{\pgfqpoint{4.625982in}{1.689407in}}%
\pgfpathlineto{\pgfqpoint{4.944339in}{1.693131in}}%
\pgfusepath{stroke}%
\end{pgfscope}%
\begin{pgfscope}%
\pgfpathrectangle{\pgfqpoint{0.487343in}{0.625214in}}{\pgfqpoint{6.048780in}{4.152500in}}%
\pgfusepath{clip}%
\pgfsetbuttcap%
\pgfsetroundjoin%
\pgfsetlinewidth{2.007500pt}%
\definecolor{currentstroke}{rgb}{1.000000,0.498039,0.054902}%
\pgfsetstrokecolor{currentstroke}%
\pgfsetdash{{2.000000pt}{3.300000pt}}{0.000000pt}%
\pgfpathmoveto{\pgfqpoint{4.944339in}{1.693131in}}%
\pgfpathlineto{\pgfqpoint{5.262696in}{1.696364in}}%
\pgfpathlineto{\pgfqpoint{5.581053in}{1.704488in}}%
\pgfpathlineto{\pgfqpoint{5.899409in}{1.724390in}}%
\pgfpathlineto{\pgfqpoint{6.217766in}{1.756095in}}%
\pgfpathlineto{\pgfqpoint{6.536123in}{1.799603in}}%
\pgfusepath{stroke}%
\end{pgfscope}%
\begin{pgfscope}%
\pgfpathrectangle{\pgfqpoint{0.487343in}{0.625214in}}{\pgfqpoint{6.048780in}{4.152500in}}%
\pgfusepath{clip}%
\pgfsetbuttcap%
\pgfsetroundjoin%
\definecolor{currentfill}{rgb}{1.000000,0.498039,0.054902}%
\pgfsetfillcolor{currentfill}%
\pgfsetlinewidth{1.003750pt}%
\definecolor{currentstroke}{rgb}{1.000000,0.498039,0.054902}%
\pgfsetstrokecolor{currentstroke}%
\pgfsetdash{}{0pt}%
\pgfsys@defobject{currentmarker}{\pgfqpoint{-0.041667in}{-0.041667in}}{\pgfqpoint{0.041667in}{0.041667in}}{%
\pgfpathmoveto{\pgfqpoint{0.000000in}{-0.041667in}}%
\pgfpathcurveto{\pgfqpoint{0.011050in}{-0.041667in}}{\pgfqpoint{0.021649in}{-0.037276in}}{\pgfqpoint{0.029463in}{-0.029463in}}%
\pgfpathcurveto{\pgfqpoint{0.037276in}{-0.021649in}}{\pgfqpoint{0.041667in}{-0.011050in}}{\pgfqpoint{0.041667in}{0.000000in}}%
\pgfpathcurveto{\pgfqpoint{0.041667in}{0.011050in}}{\pgfqpoint{0.037276in}{0.021649in}}{\pgfqpoint{0.029463in}{0.029463in}}%
\pgfpathcurveto{\pgfqpoint{0.021649in}{0.037276in}}{\pgfqpoint{0.011050in}{0.041667in}}{\pgfqpoint{0.000000in}{0.041667in}}%
\pgfpathcurveto{\pgfqpoint{-0.011050in}{0.041667in}}{\pgfqpoint{-0.021649in}{0.037276in}}{\pgfqpoint{-0.029463in}{0.029463in}}%
\pgfpathcurveto{\pgfqpoint{-0.037276in}{0.021649in}}{\pgfqpoint{-0.041667in}{0.011050in}}{\pgfqpoint{-0.041667in}{0.000000in}}%
\pgfpathcurveto{\pgfqpoint{-0.041667in}{-0.011050in}}{\pgfqpoint{-0.037276in}{-0.021649in}}{\pgfqpoint{-0.029463in}{-0.029463in}}%
\pgfpathcurveto{\pgfqpoint{-0.021649in}{-0.037276in}}{\pgfqpoint{-0.011050in}{-0.041667in}}{\pgfqpoint{0.000000in}{-0.041667in}}%
\pgfpathclose%
\pgfusepath{stroke,fill}%
}%
\begin{pgfscope}%
\pgfsys@transformshift{4.894311in}{1.692579in}%
\pgfsys@useobject{currentmarker}{}%
\end{pgfscope}%
\end{pgfscope}%
\begin{pgfscope}%
\pgfsetrectcap%
\pgfsetmiterjoin%
\pgfsetlinewidth{1.003750pt}%
\definecolor{currentstroke}{rgb}{0.000000,0.000000,0.000000}%
\pgfsetstrokecolor{currentstroke}%
\pgfsetdash{}{0pt}%
\pgfpathmoveto{\pgfqpoint{0.487343in}{0.625214in}}%
\pgfpathlineto{\pgfqpoint{0.487343in}{4.777714in}}%
\pgfusepath{stroke}%
\end{pgfscope}%
\begin{pgfscope}%
\pgfsetrectcap%
\pgfsetmiterjoin%
\pgfsetlinewidth{1.003750pt}%
\definecolor{currentstroke}{rgb}{0.000000,0.000000,0.000000}%
\pgfsetstrokecolor{currentstroke}%
\pgfsetdash{}{0pt}%
\pgfpathmoveto{\pgfqpoint{6.536123in}{0.625214in}}%
\pgfpathlineto{\pgfqpoint{6.536123in}{4.777714in}}%
\pgfusepath{stroke}%
\end{pgfscope}%
\begin{pgfscope}%
\pgfsetrectcap%
\pgfsetmiterjoin%
\pgfsetlinewidth{1.003750pt}%
\definecolor{currentstroke}{rgb}{0.000000,0.000000,0.000000}%
\pgfsetstrokecolor{currentstroke}%
\pgfsetdash{}{0pt}%
\pgfpathmoveto{\pgfqpoint{0.487343in}{0.625214in}}%
\pgfpathlineto{\pgfqpoint{6.536123in}{0.625214in}}%
\pgfusepath{stroke}%
\end{pgfscope}%
\begin{pgfscope}%
\pgfsetrectcap%
\pgfsetmiterjoin%
\pgfsetlinewidth{1.003750pt}%
\definecolor{currentstroke}{rgb}{0.000000,0.000000,0.000000}%
\pgfsetstrokecolor{currentstroke}%
\pgfsetdash{}{0pt}%
\pgfpathmoveto{\pgfqpoint{0.487343in}{4.777714in}}%
\pgfpathlineto{\pgfqpoint{6.536123in}{4.777714in}}%
\pgfusepath{stroke}%
\end{pgfscope}%
\begin{pgfscope}%
\definecolor{textcolor}{rgb}{0.000000,0.000000,0.000000}%
\pgfsetstrokecolor{textcolor}%
\pgfsetfillcolor{textcolor}%
\pgftext[x=0.832987in,y=0.936651in,left,base]{\color{textcolor}\rmfamily\fontsize{16.000000}{19.200000}\selectfont \(\displaystyle c_4=-1/2\) with \(\displaystyle \Lambda= 1/5 M^{1/3}_{\mathrm{Pl}}m^{2/3}\)}%
\end{pgfscope}%
\begin{pgfscope}%
\definecolor{textcolor}{rgb}{0.000000,0.000000,0.000000}%
\pgfsetstrokecolor{textcolor}%
\pgfsetfillcolor{textcolor}%
\pgftext[x=0.832987in,y=4.362464in,left,base]{\color{textcolor}\rmfamily\fontsize{16.000000}{19.200000}\selectfont \(\displaystyle M_{99}= 20\; [M_{\mathrm{Pl}}^{2}\; m^{-1}]\)}%
\end{pgfscope}%
\begin{pgfscope}%
\pgfsetrectcap%
\pgfsetroundjoin%
\pgfsetlinewidth{2.007500pt}%
\definecolor{currentstroke}{rgb}{0.121569,0.466667,0.705882}%
\pgfsetstrokecolor{currentstroke}%
\pgfsetdash{}{0pt}%
\pgfpathmoveto{\pgfqpoint{4.301812in}{4.482543in}}%
\pgfpathlineto{\pgfqpoint{4.614312in}{4.482543in}}%
\pgfusepath{stroke}%
\end{pgfscope}%
\begin{pgfscope}%
\definecolor{textcolor}{rgb}{0.000000,0.000000,0.000000}%
\pgfsetstrokecolor{textcolor}%
\pgfsetfillcolor{textcolor}%
\pgftext[x=4.780978in,y=4.409627in,left,base]{\color{textcolor}\rmfamily\fontsize{15.000000}{18.000000}\selectfont \(\displaystyle N_B \,[M_{\mathrm{Pl}}^{2}\; m^{-2}]\)}%
\end{pgfscope}%
\begin{pgfscope}%
\pgfsetrectcap%
\pgfsetroundjoin%
\pgfsetlinewidth{2.007500pt}%
\definecolor{currentstroke}{rgb}{1.000000,0.498039,0.054902}%
\pgfsetstrokecolor{currentstroke}%
\pgfsetdash{}{0pt}%
\pgfpathmoveto{\pgfqpoint{4.301812in}{4.141422in}}%
\pgfpathlineto{\pgfqpoint{4.614312in}{4.141422in}}%
\pgfusepath{stroke}%
\end{pgfscope}%
\begin{pgfscope}%
\definecolor{textcolor}{rgb}{0.000000,0.000000,0.000000}%
\pgfsetstrokecolor{textcolor}%
\pgfsetfillcolor{textcolor}%
\pgftext[x=4.780978in,y=4.068505in,left,base]{\color{textcolor}\rmfamily\fontsize{15.000000}{18.000000}\selectfont \(\displaystyle N_F \,[M_{\mathrm{Pl}}^{2}\; m^{-1} m_{n}^{-1}]\)}%
\end{pgfscope}%
\begin{pgfscope}%
\pgfsetbuttcap%
\pgfsetmiterjoin%
\definecolor{currentfill}{rgb}{1.000000,1.000000,1.000000}%
\pgfsetfillcolor{currentfill}%
\pgfsetlinewidth{0.000000pt}%
\definecolor{currentstroke}{rgb}{0.000000,0.000000,0.000000}%
\pgfsetstrokecolor{currentstroke}%
\pgfsetstrokeopacity{0.000000}%
\pgfsetdash{}{0pt}%
\pgfpathmoveto{\pgfqpoint{6.838562in}{0.625214in}}%
\pgfpathlineto{\pgfqpoint{12.887343in}{0.625214in}}%
\pgfpathlineto{\pgfqpoint{12.887343in}{4.777714in}}%
\pgfpathlineto{\pgfqpoint{6.838562in}{4.777714in}}%
\pgfpathclose%
\pgfusepath{fill}%
\end{pgfscope}%
\begin{pgfscope}%
\pgfsetbuttcap%
\pgfsetroundjoin%
\definecolor{currentfill}{rgb}{0.000000,0.000000,0.000000}%
\pgfsetfillcolor{currentfill}%
\pgfsetlinewidth{0.803000pt}%
\definecolor{currentstroke}{rgb}{0.000000,0.000000,0.000000}%
\pgfsetstrokecolor{currentstroke}%
\pgfsetdash{}{0pt}%
\pgfsys@defobject{currentmarker}{\pgfqpoint{0.000000in}{-0.048611in}}{\pgfqpoint{0.000000in}{0.000000in}}{%
\pgfpathmoveto{\pgfqpoint{0.000000in}{0.000000in}}%
\pgfpathlineto{\pgfqpoint{0.000000in}{-0.048611in}}%
\pgfusepath{stroke,fill}%
}%
\begin{pgfscope}%
\pgfsys@transformshift{7.546948in}{0.625214in}%
\pgfsys@useobject{currentmarker}{}%
\end{pgfscope}%
\end{pgfscope}%
\begin{pgfscope}%
\definecolor{textcolor}{rgb}{0.000000,0.000000,0.000000}%
\pgfsetstrokecolor{textcolor}%
\pgfsetfillcolor{textcolor}%
\pgftext[x=7.546948in,y=0.527992in,,top]{\color{textcolor}\rmfamily\fontsize{13.000000}{15.600000}\selectfont \(\displaystyle {10^{-6}}\)}%
\end{pgfscope}%
\begin{pgfscope}%
\pgfsetbuttcap%
\pgfsetroundjoin%
\definecolor{currentfill}{rgb}{0.000000,0.000000,0.000000}%
\pgfsetfillcolor{currentfill}%
\pgfsetlinewidth{0.803000pt}%
\definecolor{currentstroke}{rgb}{0.000000,0.000000,0.000000}%
\pgfsetstrokecolor{currentstroke}%
\pgfsetdash{}{0pt}%
\pgfsys@defobject{currentmarker}{\pgfqpoint{0.000000in}{-0.048611in}}{\pgfqpoint{0.000000in}{0.000000in}}{%
\pgfpathmoveto{\pgfqpoint{0.000000in}{0.000000in}}%
\pgfpathlineto{\pgfqpoint{0.000000in}{-0.048611in}}%
\pgfusepath{stroke,fill}%
}%
\begin{pgfscope}%
\pgfsys@transformshift{9.327079in}{0.625214in}%
\pgfsys@useobject{currentmarker}{}%
\end{pgfscope}%
\end{pgfscope}%
\begin{pgfscope}%
\definecolor{textcolor}{rgb}{0.000000,0.000000,0.000000}%
\pgfsetstrokecolor{textcolor}%
\pgfsetfillcolor{textcolor}%
\pgftext[x=9.327079in,y=0.527992in,,top]{\color{textcolor}\rmfamily\fontsize{13.000000}{15.600000}\selectfont \(\displaystyle {10^{-5}}\)}%
\end{pgfscope}%
\begin{pgfscope}%
\pgfsetbuttcap%
\pgfsetroundjoin%
\definecolor{currentfill}{rgb}{0.000000,0.000000,0.000000}%
\pgfsetfillcolor{currentfill}%
\pgfsetlinewidth{0.803000pt}%
\definecolor{currentstroke}{rgb}{0.000000,0.000000,0.000000}%
\pgfsetstrokecolor{currentstroke}%
\pgfsetdash{}{0pt}%
\pgfsys@defobject{currentmarker}{\pgfqpoint{0.000000in}{-0.048611in}}{\pgfqpoint{0.000000in}{0.000000in}}{%
\pgfpathmoveto{\pgfqpoint{0.000000in}{0.000000in}}%
\pgfpathlineto{\pgfqpoint{0.000000in}{-0.048611in}}%
\pgfusepath{stroke,fill}%
}%
\begin{pgfscope}%
\pgfsys@transformshift{11.107211in}{0.625214in}%
\pgfsys@useobject{currentmarker}{}%
\end{pgfscope}%
\end{pgfscope}%
\begin{pgfscope}%
\definecolor{textcolor}{rgb}{0.000000,0.000000,0.000000}%
\pgfsetstrokecolor{textcolor}%
\pgfsetfillcolor{textcolor}%
\pgftext[x=11.107211in,y=0.527992in,,top]{\color{textcolor}\rmfamily\fontsize{13.000000}{15.600000}\selectfont \(\displaystyle {10^{-4}}\)}%
\end{pgfscope}%
\begin{pgfscope}%
\pgfsetbuttcap%
\pgfsetroundjoin%
\definecolor{currentfill}{rgb}{0.000000,0.000000,0.000000}%
\pgfsetfillcolor{currentfill}%
\pgfsetlinewidth{0.803000pt}%
\definecolor{currentstroke}{rgb}{0.000000,0.000000,0.000000}%
\pgfsetstrokecolor{currentstroke}%
\pgfsetdash{}{0pt}%
\pgfsys@defobject{currentmarker}{\pgfqpoint{0.000000in}{-0.048611in}}{\pgfqpoint{0.000000in}{0.000000in}}{%
\pgfpathmoveto{\pgfqpoint{0.000000in}{0.000000in}}%
\pgfpathlineto{\pgfqpoint{0.000000in}{-0.048611in}}%
\pgfusepath{stroke,fill}%
}%
\begin{pgfscope}%
\pgfsys@transformshift{12.887343in}{0.625214in}%
\pgfsys@useobject{currentmarker}{}%
\end{pgfscope}%
\end{pgfscope}%
\begin{pgfscope}%
\definecolor{textcolor}{rgb}{0.000000,0.000000,0.000000}%
\pgfsetstrokecolor{textcolor}%
\pgfsetfillcolor{textcolor}%
\pgftext[x=12.887343in,y=0.527992in,,top]{\color{textcolor}\rmfamily\fontsize{13.000000}{15.600000}\selectfont \(\displaystyle {10^{-3}}\)}%
\end{pgfscope}%
\begin{pgfscope}%
\definecolor{textcolor}{rgb}{0.000000,0.000000,0.000000}%
\pgfsetstrokecolor{textcolor}%
\pgfsetfillcolor{textcolor}%
\pgftext[x=9.862952in,y=0.336954in,,top]{\color{textcolor}\rmfamily\fontsize{15.000000}{18.000000}\selectfont \(\displaystyle p_0\,[m^{2}M_{\mathrm{Pl}}^2]\)}%
\end{pgfscope}%
\begin{pgfscope}%
\pgfsetbuttcap%
\pgfsetroundjoin%
\definecolor{currentfill}{rgb}{0.000000,0.000000,0.000000}%
\pgfsetfillcolor{currentfill}%
\pgfsetlinewidth{0.803000pt}%
\definecolor{currentstroke}{rgb}{0.000000,0.000000,0.000000}%
\pgfsetstrokecolor{currentstroke}%
\pgfsetdash{}{0pt}%
\pgfsys@defobject{currentmarker}{\pgfqpoint{-0.048611in}{0.000000in}}{\pgfqpoint{-0.000000in}{0.000000in}}{%
\pgfpathmoveto{\pgfqpoint{-0.000000in}{0.000000in}}%
\pgfpathlineto{\pgfqpoint{-0.048611in}{0.000000in}}%
\pgfusepath{stroke,fill}%
}%
\begin{pgfscope}%
\pgfsys@transformshift{6.838562in}{0.625214in}%
\pgfsys@useobject{currentmarker}{}%
\end{pgfscope}%
\end{pgfscope}%
\begin{pgfscope}%
\pgfsetbuttcap%
\pgfsetroundjoin%
\definecolor{currentfill}{rgb}{0.000000,0.000000,0.000000}%
\pgfsetfillcolor{currentfill}%
\pgfsetlinewidth{0.803000pt}%
\definecolor{currentstroke}{rgb}{0.000000,0.000000,0.000000}%
\pgfsetstrokecolor{currentstroke}%
\pgfsetdash{}{0pt}%
\pgfsys@defobject{currentmarker}{\pgfqpoint{-0.048611in}{0.000000in}}{\pgfqpoint{-0.000000in}{0.000000in}}{%
\pgfpathmoveto{\pgfqpoint{-0.000000in}{0.000000in}}%
\pgfpathlineto{\pgfqpoint{-0.048611in}{0.000000in}}%
\pgfusepath{stroke,fill}%
}%
\begin{pgfscope}%
\pgfsys@transformshift{6.838562in}{1.144276in}%
\pgfsys@useobject{currentmarker}{}%
\end{pgfscope}%
\end{pgfscope}%
\begin{pgfscope}%
\pgfsetbuttcap%
\pgfsetroundjoin%
\definecolor{currentfill}{rgb}{0.000000,0.000000,0.000000}%
\pgfsetfillcolor{currentfill}%
\pgfsetlinewidth{0.803000pt}%
\definecolor{currentstroke}{rgb}{0.000000,0.000000,0.000000}%
\pgfsetstrokecolor{currentstroke}%
\pgfsetdash{}{0pt}%
\pgfsys@defobject{currentmarker}{\pgfqpoint{-0.048611in}{0.000000in}}{\pgfqpoint{-0.000000in}{0.000000in}}{%
\pgfpathmoveto{\pgfqpoint{-0.000000in}{0.000000in}}%
\pgfpathlineto{\pgfqpoint{-0.048611in}{0.000000in}}%
\pgfusepath{stroke,fill}%
}%
\begin{pgfscope}%
\pgfsys@transformshift{6.838562in}{1.663339in}%
\pgfsys@useobject{currentmarker}{}%
\end{pgfscope}%
\end{pgfscope}%
\begin{pgfscope}%
\pgfsetbuttcap%
\pgfsetroundjoin%
\definecolor{currentfill}{rgb}{0.000000,0.000000,0.000000}%
\pgfsetfillcolor{currentfill}%
\pgfsetlinewidth{0.803000pt}%
\definecolor{currentstroke}{rgb}{0.000000,0.000000,0.000000}%
\pgfsetstrokecolor{currentstroke}%
\pgfsetdash{}{0pt}%
\pgfsys@defobject{currentmarker}{\pgfqpoint{-0.048611in}{0.000000in}}{\pgfqpoint{-0.000000in}{0.000000in}}{%
\pgfpathmoveto{\pgfqpoint{-0.000000in}{0.000000in}}%
\pgfpathlineto{\pgfqpoint{-0.048611in}{0.000000in}}%
\pgfusepath{stroke,fill}%
}%
\begin{pgfscope}%
\pgfsys@transformshift{6.838562in}{2.182401in}%
\pgfsys@useobject{currentmarker}{}%
\end{pgfscope}%
\end{pgfscope}%
\begin{pgfscope}%
\pgfsetbuttcap%
\pgfsetroundjoin%
\definecolor{currentfill}{rgb}{0.000000,0.000000,0.000000}%
\pgfsetfillcolor{currentfill}%
\pgfsetlinewidth{0.803000pt}%
\definecolor{currentstroke}{rgb}{0.000000,0.000000,0.000000}%
\pgfsetstrokecolor{currentstroke}%
\pgfsetdash{}{0pt}%
\pgfsys@defobject{currentmarker}{\pgfqpoint{-0.048611in}{0.000000in}}{\pgfqpoint{-0.000000in}{0.000000in}}{%
\pgfpathmoveto{\pgfqpoint{-0.000000in}{0.000000in}}%
\pgfpathlineto{\pgfqpoint{-0.048611in}{0.000000in}}%
\pgfusepath{stroke,fill}%
}%
\begin{pgfscope}%
\pgfsys@transformshift{6.838562in}{2.701464in}%
\pgfsys@useobject{currentmarker}{}%
\end{pgfscope}%
\end{pgfscope}%
\begin{pgfscope}%
\pgfsetbuttcap%
\pgfsetroundjoin%
\definecolor{currentfill}{rgb}{0.000000,0.000000,0.000000}%
\pgfsetfillcolor{currentfill}%
\pgfsetlinewidth{0.803000pt}%
\definecolor{currentstroke}{rgb}{0.000000,0.000000,0.000000}%
\pgfsetstrokecolor{currentstroke}%
\pgfsetdash{}{0pt}%
\pgfsys@defobject{currentmarker}{\pgfqpoint{-0.048611in}{0.000000in}}{\pgfqpoint{-0.000000in}{0.000000in}}{%
\pgfpathmoveto{\pgfqpoint{-0.000000in}{0.000000in}}%
\pgfpathlineto{\pgfqpoint{-0.048611in}{0.000000in}}%
\pgfusepath{stroke,fill}%
}%
\begin{pgfscope}%
\pgfsys@transformshift{6.838562in}{3.220526in}%
\pgfsys@useobject{currentmarker}{}%
\end{pgfscope}%
\end{pgfscope}%
\begin{pgfscope}%
\pgfsetbuttcap%
\pgfsetroundjoin%
\definecolor{currentfill}{rgb}{0.000000,0.000000,0.000000}%
\pgfsetfillcolor{currentfill}%
\pgfsetlinewidth{0.803000pt}%
\definecolor{currentstroke}{rgb}{0.000000,0.000000,0.000000}%
\pgfsetstrokecolor{currentstroke}%
\pgfsetdash{}{0pt}%
\pgfsys@defobject{currentmarker}{\pgfqpoint{-0.048611in}{0.000000in}}{\pgfqpoint{-0.000000in}{0.000000in}}{%
\pgfpathmoveto{\pgfqpoint{-0.000000in}{0.000000in}}%
\pgfpathlineto{\pgfqpoint{-0.048611in}{0.000000in}}%
\pgfusepath{stroke,fill}%
}%
\begin{pgfscope}%
\pgfsys@transformshift{6.838562in}{3.739589in}%
\pgfsys@useobject{currentmarker}{}%
\end{pgfscope}%
\end{pgfscope}%
\begin{pgfscope}%
\pgfsetbuttcap%
\pgfsetroundjoin%
\definecolor{currentfill}{rgb}{0.000000,0.000000,0.000000}%
\pgfsetfillcolor{currentfill}%
\pgfsetlinewidth{0.803000pt}%
\definecolor{currentstroke}{rgb}{0.000000,0.000000,0.000000}%
\pgfsetstrokecolor{currentstroke}%
\pgfsetdash{}{0pt}%
\pgfsys@defobject{currentmarker}{\pgfqpoint{-0.048611in}{0.000000in}}{\pgfqpoint{-0.000000in}{0.000000in}}{%
\pgfpathmoveto{\pgfqpoint{-0.000000in}{0.000000in}}%
\pgfpathlineto{\pgfqpoint{-0.048611in}{0.000000in}}%
\pgfusepath{stroke,fill}%
}%
\begin{pgfscope}%
\pgfsys@transformshift{6.838562in}{4.258651in}%
\pgfsys@useobject{currentmarker}{}%
\end{pgfscope}%
\end{pgfscope}%
\begin{pgfscope}%
\pgfsetbuttcap%
\pgfsetroundjoin%
\definecolor{currentfill}{rgb}{0.000000,0.000000,0.000000}%
\pgfsetfillcolor{currentfill}%
\pgfsetlinewidth{0.803000pt}%
\definecolor{currentstroke}{rgb}{0.000000,0.000000,0.000000}%
\pgfsetstrokecolor{currentstroke}%
\pgfsetdash{}{0pt}%
\pgfsys@defobject{currentmarker}{\pgfqpoint{-0.048611in}{0.000000in}}{\pgfqpoint{-0.000000in}{0.000000in}}{%
\pgfpathmoveto{\pgfqpoint{-0.000000in}{0.000000in}}%
\pgfpathlineto{\pgfqpoint{-0.048611in}{0.000000in}}%
\pgfusepath{stroke,fill}%
}%
\begin{pgfscope}%
\pgfsys@transformshift{6.838562in}{4.777714in}%
\pgfsys@useobject{currentmarker}{}%
\end{pgfscope}%
\end{pgfscope}%
\begin{pgfscope}%
\pgfpathrectangle{\pgfqpoint{6.838562in}{0.625214in}}{\pgfqpoint{6.048780in}{4.152500in}}%
\pgfusepath{clip}%
\pgfsetrectcap%
\pgfsetroundjoin%
\pgfsetlinewidth{2.007500pt}%
\definecolor{currentstroke}{rgb}{0.121569,0.466667,0.705882}%
\pgfsetstrokecolor{currentstroke}%
\pgfsetdash{}{0pt}%
\pgfpathmoveto{\pgfqpoint{6.929620in}{0.625214in}}%
\pgfpathlineto{\pgfqpoint{7.241524in}{0.867208in}}%
\pgfpathlineto{\pgfqpoint{7.553427in}{1.106531in}}%
\pgfpathlineto{\pgfqpoint{7.865330in}{1.343183in}}%
\pgfpathlineto{\pgfqpoint{8.177234in}{1.577163in}}%
\pgfpathlineto{\pgfqpoint{8.489137in}{1.808472in}}%
\pgfpathlineto{\pgfqpoint{8.801040in}{2.037109in}}%
\pgfpathlineto{\pgfqpoint{9.112943in}{2.262969in}}%
\pgfpathlineto{\pgfqpoint{9.424847in}{2.479351in}}%
\pgfpathlineto{\pgfqpoint{9.736750in}{2.682273in}}%
\pgfpathlineto{\pgfqpoint{10.048653in}{2.871734in}}%
\pgfpathlineto{\pgfqpoint{10.360557in}{3.049412in}}%
\pgfpathlineto{\pgfqpoint{10.672460in}{3.222448in}}%
\pgfpathlineto{\pgfqpoint{10.984363in}{3.369239in}}%
\pgfpathlineto{\pgfqpoint{11.296267in}{3.462257in}}%
\pgfpathlineto{\pgfqpoint{11.608170in}{3.524751in}}%
\pgfpathlineto{\pgfqpoint{11.920073in}{3.568710in}}%
\pgfusepath{stroke}%
\end{pgfscope}%
\begin{pgfscope}%
\pgfpathrectangle{\pgfqpoint{6.838562in}{0.625214in}}{\pgfqpoint{6.048780in}{4.152500in}}%
\pgfusepath{clip}%
\pgfsetbuttcap%
\pgfsetroundjoin%
\pgfsetlinewidth{2.007500pt}%
\definecolor{currentstroke}{rgb}{0.121569,0.466667,0.705882}%
\pgfsetstrokecolor{currentstroke}%
\pgfsetdash{{2.000000pt}{3.300000pt}}{0.000000pt}%
\pgfpathmoveto{\pgfqpoint{11.920073in}{3.568710in}}%
\pgfpathlineto{\pgfqpoint{12.231976in}{3.548335in}}%
\pgfpathlineto{\pgfqpoint{12.543880in}{3.461512in}}%
\pgfpathlineto{\pgfqpoint{12.855783in}{3.343553in}}%
\pgfusepath{stroke}%
\end{pgfscope}%
\begin{pgfscope}%
\pgfpathrectangle{\pgfqpoint{6.838562in}{0.625214in}}{\pgfqpoint{6.048780in}{4.152500in}}%
\pgfusepath{clip}%
\pgfsetbuttcap%
\pgfsetroundjoin%
\definecolor{currentfill}{rgb}{0.121569,0.466667,0.705882}%
\pgfsetfillcolor{currentfill}%
\pgfsetlinewidth{1.003750pt}%
\definecolor{currentstroke}{rgb}{0.121569,0.466667,0.705882}%
\pgfsetstrokecolor{currentstroke}%
\pgfsetdash{}{0pt}%
\pgfsys@defobject{currentmarker}{\pgfqpoint{-0.041667in}{-0.041667in}}{\pgfqpoint{0.041667in}{0.041667in}}{%
\pgfpathmoveto{\pgfqpoint{0.000000in}{-0.041667in}}%
\pgfpathcurveto{\pgfqpoint{0.011050in}{-0.041667in}}{\pgfqpoint{0.021649in}{-0.037276in}}{\pgfqpoint{0.029463in}{-0.029463in}}%
\pgfpathcurveto{\pgfqpoint{0.037276in}{-0.021649in}}{\pgfqpoint{0.041667in}{-0.011050in}}{\pgfqpoint{0.041667in}{0.000000in}}%
\pgfpathcurveto{\pgfqpoint{0.041667in}{0.011050in}}{\pgfqpoint{0.037276in}{0.021649in}}{\pgfqpoint{0.029463in}{0.029463in}}%
\pgfpathcurveto{\pgfqpoint{0.021649in}{0.037276in}}{\pgfqpoint{0.011050in}{0.041667in}}{\pgfqpoint{0.000000in}{0.041667in}}%
\pgfpathcurveto{\pgfqpoint{-0.011050in}{0.041667in}}{\pgfqpoint{-0.021649in}{0.037276in}}{\pgfqpoint{-0.029463in}{0.029463in}}%
\pgfpathcurveto{\pgfqpoint{-0.037276in}{0.021649in}}{\pgfqpoint{-0.041667in}{0.011050in}}{\pgfqpoint{-0.041667in}{0.000000in}}%
\pgfpathcurveto{\pgfqpoint{-0.041667in}{-0.011050in}}{\pgfqpoint{-0.037276in}{-0.021649in}}{\pgfqpoint{-0.029463in}{-0.029463in}}%
\pgfpathcurveto{\pgfqpoint{-0.021649in}{-0.037276in}}{\pgfqpoint{-0.011050in}{-0.041667in}}{\pgfqpoint{0.000000in}{-0.041667in}}%
\pgfpathclose%
\pgfusepath{stroke,fill}%
}%
\begin{pgfscope}%
\pgfsys@transformshift{11.981725in}{3.574052in}%
\pgfsys@useobject{currentmarker}{}%
\end{pgfscope}%
\end{pgfscope}%
\begin{pgfscope}%
\pgfpathrectangle{\pgfqpoint{6.838562in}{0.625214in}}{\pgfqpoint{6.048780in}{4.152500in}}%
\pgfusepath{clip}%
\pgfsetrectcap%
\pgfsetroundjoin%
\pgfsetlinewidth{2.007500pt}%
\definecolor{currentstroke}{rgb}{1.000000,0.498039,0.054902}%
\pgfsetstrokecolor{currentstroke}%
\pgfsetdash{}{0pt}%
\pgfpathmoveto{\pgfqpoint{6.929620in}{4.668057in}}%
\pgfpathlineto{\pgfqpoint{7.241524in}{4.461281in}}%
\pgfpathlineto{\pgfqpoint{7.553427in}{4.248308in}}%
\pgfpathlineto{\pgfqpoint{7.865330in}{4.029140in}}%
\pgfpathlineto{\pgfqpoint{8.177234in}{3.803777in}}%
\pgfpathlineto{\pgfqpoint{8.489137in}{3.572217in}}%
\pgfpathlineto{\pgfqpoint{8.801040in}{3.334462in}}%
\pgfpathlineto{\pgfqpoint{9.112943in}{3.090731in}}%
\pgfpathlineto{\pgfqpoint{9.424847in}{2.854837in}}%
\pgfpathlineto{\pgfqpoint{9.736750in}{2.634995in}}%
\pgfpathlineto{\pgfqpoint{10.048653in}{2.431204in}}%
\pgfpathlineto{\pgfqpoint{10.360557in}{2.236841in}}%
\pgfpathlineto{\pgfqpoint{10.672460in}{2.023715in}}%
\pgfpathlineto{\pgfqpoint{10.984363in}{1.854954in}}%
\pgfpathlineto{\pgfqpoint{11.296267in}{1.796483in}}%
\pgfpathlineto{\pgfqpoint{11.608170in}{1.724737in}}%
\pgfpathlineto{\pgfqpoint{11.920073in}{1.693571in}}%
\pgfusepath{stroke}%
\end{pgfscope}%
\begin{pgfscope}%
\pgfpathrectangle{\pgfqpoint{6.838562in}{0.625214in}}{\pgfqpoint{6.048780in}{4.152500in}}%
\pgfusepath{clip}%
\pgfsetbuttcap%
\pgfsetroundjoin%
\pgfsetlinewidth{2.007500pt}%
\definecolor{currentstroke}{rgb}{1.000000,0.498039,0.054902}%
\pgfsetstrokecolor{currentstroke}%
\pgfsetdash{{2.000000pt}{3.300000pt}}{0.000000pt}%
\pgfpathmoveto{\pgfqpoint{11.920073in}{1.693571in}}%
\pgfpathlineto{\pgfqpoint{12.231976in}{1.688163in}}%
\pgfpathlineto{\pgfqpoint{12.543880in}{1.710309in}}%
\pgfpathlineto{\pgfqpoint{12.855783in}{1.799603in}}%
\pgfusepath{stroke}%
\end{pgfscope}%
\begin{pgfscope}%
\pgfpathrectangle{\pgfqpoint{6.838562in}{0.625214in}}{\pgfqpoint{6.048780in}{4.152500in}}%
\pgfusepath{clip}%
\pgfsetbuttcap%
\pgfsetroundjoin%
\definecolor{currentfill}{rgb}{1.000000,0.498039,0.054902}%
\pgfsetfillcolor{currentfill}%
\pgfsetlinewidth{1.003750pt}%
\definecolor{currentstroke}{rgb}{1.000000,0.498039,0.054902}%
\pgfsetstrokecolor{currentstroke}%
\pgfsetdash{}{0pt}%
\pgfsys@defobject{currentmarker}{\pgfqpoint{-0.041667in}{-0.041667in}}{\pgfqpoint{0.041667in}{0.041667in}}{%
\pgfpathmoveto{\pgfqpoint{0.000000in}{-0.041667in}}%
\pgfpathcurveto{\pgfqpoint{0.011050in}{-0.041667in}}{\pgfqpoint{0.021649in}{-0.037276in}}{\pgfqpoint{0.029463in}{-0.029463in}}%
\pgfpathcurveto{\pgfqpoint{0.037276in}{-0.021649in}}{\pgfqpoint{0.041667in}{-0.011050in}}{\pgfqpoint{0.041667in}{0.000000in}}%
\pgfpathcurveto{\pgfqpoint{0.041667in}{0.011050in}}{\pgfqpoint{0.037276in}{0.021649in}}{\pgfqpoint{0.029463in}{0.029463in}}%
\pgfpathcurveto{\pgfqpoint{0.021649in}{0.037276in}}{\pgfqpoint{0.011050in}{0.041667in}}{\pgfqpoint{0.000000in}{0.041667in}}%
\pgfpathcurveto{\pgfqpoint{-0.011050in}{0.041667in}}{\pgfqpoint{-0.021649in}{0.037276in}}{\pgfqpoint{-0.029463in}{0.029463in}}%
\pgfpathcurveto{\pgfqpoint{-0.037276in}{0.021649in}}{\pgfqpoint{-0.041667in}{0.011050in}}{\pgfqpoint{-0.041667in}{0.000000in}}%
\pgfpathcurveto{\pgfqpoint{-0.041667in}{-0.011050in}}{\pgfqpoint{-0.037276in}{-0.021649in}}{\pgfqpoint{-0.029463in}{-0.029463in}}%
\pgfpathcurveto{\pgfqpoint{-0.021649in}{-0.037276in}}{\pgfqpoint{-0.011050in}{-0.041667in}}{\pgfqpoint{0.000000in}{-0.041667in}}%
\pgfpathclose%
\pgfusepath{stroke,fill}%
}%
\begin{pgfscope}%
\pgfsys@transformshift{11.981725in}{1.698441in}%
\pgfsys@useobject{currentmarker}{}%
\end{pgfscope}%
\end{pgfscope}%
\begin{pgfscope}%
\pgfsetrectcap%
\pgfsetmiterjoin%
\pgfsetlinewidth{1.003750pt}%
\definecolor{currentstroke}{rgb}{0.000000,0.000000,0.000000}%
\pgfsetstrokecolor{currentstroke}%
\pgfsetdash{}{0pt}%
\pgfpathmoveto{\pgfqpoint{6.838562in}{0.625214in}}%
\pgfpathlineto{\pgfqpoint{6.838562in}{4.777714in}}%
\pgfusepath{stroke}%
\end{pgfscope}%
\begin{pgfscope}%
\pgfsetrectcap%
\pgfsetmiterjoin%
\pgfsetlinewidth{1.003750pt}%
\definecolor{currentstroke}{rgb}{0.000000,0.000000,0.000000}%
\pgfsetstrokecolor{currentstroke}%
\pgfsetdash{}{0pt}%
\pgfpathmoveto{\pgfqpoint{12.887343in}{0.625214in}}%
\pgfpathlineto{\pgfqpoint{12.887343in}{4.777714in}}%
\pgfusepath{stroke}%
\end{pgfscope}%
\begin{pgfscope}%
\pgfsetrectcap%
\pgfsetmiterjoin%
\pgfsetlinewidth{1.003750pt}%
\definecolor{currentstroke}{rgb}{0.000000,0.000000,0.000000}%
\pgfsetstrokecolor{currentstroke}%
\pgfsetdash{}{0pt}%
\pgfpathmoveto{\pgfqpoint{6.838562in}{0.625214in}}%
\pgfpathlineto{\pgfqpoint{12.887343in}{0.625214in}}%
\pgfusepath{stroke}%
\end{pgfscope}%
\begin{pgfscope}%
\pgfsetrectcap%
\pgfsetmiterjoin%
\pgfsetlinewidth{1.003750pt}%
\definecolor{currentstroke}{rgb}{0.000000,0.000000,0.000000}%
\pgfsetstrokecolor{currentstroke}%
\pgfsetdash{}{0pt}%
\pgfpathmoveto{\pgfqpoint{6.838562in}{4.777714in}}%
\pgfpathlineto{\pgfqpoint{12.887343in}{4.777714in}}%
\pgfusepath{stroke}%
\end{pgfscope}%
\begin{pgfscope}%
\pgfsetrectcap%
\pgfsetroundjoin%
\pgfsetlinewidth{2.007500pt}%
\definecolor{currentstroke}{rgb}{0.121569,0.466667,0.705882}%
\pgfsetstrokecolor{currentstroke}%
\pgfsetdash{}{0pt}%
\pgfpathmoveto{\pgfqpoint{10.653031in}{4.482543in}}%
\pgfpathlineto{\pgfqpoint{10.965531in}{4.482543in}}%
\pgfusepath{stroke}%
\end{pgfscope}%
\begin{pgfscope}%
\definecolor{textcolor}{rgb}{0.000000,0.000000,0.000000}%
\pgfsetstrokecolor{textcolor}%
\pgfsetfillcolor{textcolor}%
\pgftext[x=11.132198in,y=4.409627in,left,base]{\color{textcolor}\rmfamily\fontsize{15.000000}{18.000000}\selectfont \(\displaystyle N_B \,[M_{\mathrm{Pl}}^{2}\; m^{-2}]\)}%
\end{pgfscope}%
\begin{pgfscope}%
\pgfsetrectcap%
\pgfsetroundjoin%
\pgfsetlinewidth{2.007500pt}%
\definecolor{currentstroke}{rgb}{1.000000,0.498039,0.054902}%
\pgfsetstrokecolor{currentstroke}%
\pgfsetdash{}{0pt}%
\pgfpathmoveto{\pgfqpoint{10.653031in}{4.141422in}}%
\pgfpathlineto{\pgfqpoint{10.965531in}{4.141422in}}%
\pgfusepath{stroke}%
\end{pgfscope}%
\begin{pgfscope}%
\definecolor{textcolor}{rgb}{0.000000,0.000000,0.000000}%
\pgfsetstrokecolor{textcolor}%
\pgfsetfillcolor{textcolor}%
\pgftext[x=11.132198in,y=4.068505in,left,base]{\color{textcolor}\rmfamily\fontsize{15.000000}{18.000000}\selectfont \(\displaystyle N_F \,[M_{\mathrm{Pl}}^{2}\; m^{-1} m_{n}^{-1}]\)}%
\end{pgfscope}%
\end{pgfpicture}%
\makeatother%
\endgroup%

%% file: Figuras/estab2.pgf
%% Creator: Matplotlib, PGF backend
%%
%% To include the figure in your LaTeX document, write
%%   \input{<filename>.pgf}
%%
%% Make sure the required packages are loaded in your preamble
%%   \usepackage{pgf}
%%
%% and, on pdftex
%%   \usepackage[utf8]{inputenc}\DeclareUnicodeCharacter{2212}{-}
%%
%% or, on luatex and xetex
%%   \usepackage{unicode-math}
%%
%% Figures using additional raster images can only be included by \input if
%% they are in the same directory as the main LaTeX file. For loading figures
%% from other directories you can use the `import` package
%%   \usepackage{import}
%%
%% and then include the figures with
%%   \import{<path to file>}{<filename>.pgf}
%%
%% Matplotlib used the following preamble
%%   \usepackage{amssymb}
%%
\begingroup%
\makeatletter%
\begin{pgfpicture}%
\pgfpathrectangle{\pgfpointorigin}{\pgfqpoint{7.133271in}{4.934361in}}%
\pgfusepath{use as bounding box, clip}%
\begin{pgfscope}%
\pgfsetbuttcap%
\pgfsetmiterjoin%
\definecolor{currentfill}{rgb}{1.000000,1.000000,1.000000}%
\pgfsetfillcolor{currentfill}%
\pgfsetlinewidth{0.000000pt}%
\definecolor{currentstroke}{rgb}{1.000000,1.000000,1.000000}%
\pgfsetstrokecolor{currentstroke}%
\pgfsetdash{}{0pt}%
\pgfpathmoveto{\pgfqpoint{0.000000in}{0.000000in}}%
\pgfpathlineto{\pgfqpoint{7.133271in}{0.000000in}}%
\pgfpathlineto{\pgfqpoint{7.133271in}{4.934361in}}%
\pgfpathlineto{\pgfqpoint{0.000000in}{4.934361in}}%
\pgfpathclose%
\pgfusepath{fill}%
\end{pgfscope}%
\begin{pgfscope}%
\pgfsetbuttcap%
\pgfsetmiterjoin%
\definecolor{currentfill}{rgb}{1.000000,1.000000,1.000000}%
\pgfsetfillcolor{currentfill}%
\pgfsetlinewidth{0.000000pt}%
\definecolor{currentstroke}{rgb}{0.000000,0.000000,0.000000}%
\pgfsetstrokecolor{currentstroke}%
\pgfsetstrokeopacity{0.000000}%
\pgfsetdash{}{0pt}%
\pgfpathmoveto{\pgfqpoint{0.729009in}{0.623991in}}%
\pgfpathlineto{\pgfqpoint{6.929009in}{0.623991in}}%
\pgfpathlineto{\pgfqpoint{6.929009in}{4.776491in}}%
\pgfpathlineto{\pgfqpoint{0.729009in}{4.776491in}}%
\pgfpathclose%
\pgfusepath{fill}%
\end{pgfscope}%
\begin{pgfscope}%
\pgfpathrectangle{\pgfqpoint{0.729009in}{0.623991in}}{\pgfqpoint{6.200000in}{4.152500in}}%
\pgfusepath{clip}%
\pgfsetbuttcap%
\pgfsetroundjoin%
\definecolor{currentfill}{rgb}{1.000000,0.000000,0.000000}%
\pgfsetfillcolor{currentfill}%
\pgfsetfillopacity{0.100000}%
\pgfsetlinewidth{0.000000pt}%
\definecolor{currentstroke}{rgb}{1.000000,0.000000,0.000000}%
\pgfsetstrokecolor{currentstroke}%
\pgfsetstrokeopacity{0.100000}%
\pgfsetdash{}{0pt}%
\pgfsys@defobject{currentmarker}{\pgfqpoint{0.735209in}{1.394004in}}{\pgfqpoint{1.436340in}{4.171489in}}{%
\pgfpathmoveto{\pgfqpoint{0.735209in}{4.087486in}}%
\pgfpathlineto{\pgfqpoint{0.735209in}{1.394004in}}%
\pgfpathlineto{\pgfqpoint{0.852064in}{3.033939in}}%
\pgfpathlineto{\pgfqpoint{0.968920in}{3.331318in}}%
\pgfpathlineto{\pgfqpoint{1.085775in}{3.574245in}}%
\pgfpathlineto{\pgfqpoint{1.202630in}{3.801740in}}%
\pgfpathlineto{\pgfqpoint{1.319485in}{4.029236in}}%
\pgfpathlineto{\pgfqpoint{1.436340in}{4.155685in}}%
\pgfpathlineto{\pgfqpoint{1.436340in}{4.171489in}}%
\pgfpathlineto{\pgfqpoint{1.436340in}{4.171489in}}%
\pgfpathlineto{\pgfqpoint{1.319485in}{4.159824in}}%
\pgfpathlineto{\pgfqpoint{1.202630in}{4.146212in}}%
\pgfpathlineto{\pgfqpoint{1.085775in}{4.131531in}}%
\pgfpathlineto{\pgfqpoint{0.968920in}{4.116849in}}%
\pgfpathlineto{\pgfqpoint{0.852064in}{4.102167in}}%
\pgfpathlineto{\pgfqpoint{0.735209in}{4.087486in}}%
\pgfpathclose%
\pgfusepath{fill}%
}%
\begin{pgfscope}%
\pgfsys@transformshift{0.000000in}{0.000000in}%
\pgfsys@useobject{currentmarker}{}%
\end{pgfscope}%
\end{pgfscope}%
\begin{pgfscope}%
\pgfpathrectangle{\pgfqpoint{0.729009in}{0.623991in}}{\pgfqpoint{6.200000in}{4.152500in}}%
\pgfusepath{clip}%
\pgfsetbuttcap%
\pgfsetroundjoin%
\definecolor{currentfill}{rgb}{0.000000,0.000000,1.000000}%
\pgfsetfillcolor{currentfill}%
\pgfsetfillopacity{0.100000}%
\pgfsetlinewidth{0.000000pt}%
\definecolor{currentstroke}{rgb}{0.000000,0.000000,1.000000}%
\pgfsetstrokecolor{currentstroke}%
\pgfsetstrokeopacity{0.100000}%
\pgfsetdash{}{0pt}%
\pgfsys@defobject{currentmarker}{\pgfqpoint{0.759123in}{0.637787in}}{\pgfqpoint{1.667866in}{4.155685in}}{%
\pgfpathmoveto{\pgfqpoint{0.759123in}{0.637787in}}%
\pgfpathlineto{\pgfqpoint{0.759123in}{1.327571in}}%
\pgfpathlineto{\pgfqpoint{0.860095in}{2.196089in}}%
\pgfpathlineto{\pgfqpoint{0.961066in}{2.894904in}}%
\pgfpathlineto{\pgfqpoint{1.062038in}{3.183981in}}%
\pgfpathlineto{\pgfqpoint{1.163009in}{3.435303in}}%
\pgfpathlineto{\pgfqpoint{1.263980in}{3.580912in}}%
\pgfpathlineto{\pgfqpoint{1.364952in}{3.726522in}}%
\pgfpathlineto{\pgfqpoint{1.465923in}{3.872131in}}%
\pgfpathlineto{\pgfqpoint{1.566895in}{4.017740in}}%
\pgfpathlineto{\pgfqpoint{1.667866in}{4.155685in}}%
\pgfpathlineto{\pgfqpoint{1.667866in}{0.637787in}}%
\pgfpathlineto{\pgfqpoint{1.667866in}{0.637787in}}%
\pgfpathlineto{\pgfqpoint{1.566895in}{0.637787in}}%
\pgfpathlineto{\pgfqpoint{1.465923in}{0.637787in}}%
\pgfpathlineto{\pgfqpoint{1.364952in}{0.637787in}}%
\pgfpathlineto{\pgfqpoint{1.263980in}{0.637787in}}%
\pgfpathlineto{\pgfqpoint{1.163009in}{0.637787in}}%
\pgfpathlineto{\pgfqpoint{1.062038in}{0.637787in}}%
\pgfpathlineto{\pgfqpoint{0.961066in}{0.637787in}}%
\pgfpathlineto{\pgfqpoint{0.860095in}{0.637787in}}%
\pgfpathlineto{\pgfqpoint{0.759123in}{0.637787in}}%
\pgfpathclose%
\pgfusepath{fill}%
}%
\begin{pgfscope}%
\pgfsys@transformshift{0.000000in}{0.000000in}%
\pgfsys@useobject{currentmarker}{}%
\end{pgfscope}%
\end{pgfscope}%
\begin{pgfscope}%
\pgfpathrectangle{\pgfqpoint{0.729009in}{0.623991in}}{\pgfqpoint{6.200000in}{4.152500in}}%
\pgfusepath{clip}%
\pgfsetbuttcap%
\pgfsetroundjoin%
\definecolor{currentfill}{rgb}{0.000000,0.000000,1.000000}%
\pgfsetfillcolor{currentfill}%
\pgfsetfillopacity{0.100000}%
\pgfsetlinewidth{0.000000pt}%
\definecolor{currentstroke}{rgb}{0.000000,0.000000,1.000000}%
\pgfsetstrokecolor{currentstroke}%
\pgfsetstrokeopacity{0.100000}%
\pgfsetdash{}{0pt}%
\pgfsys@defobject{currentmarker}{\pgfqpoint{1.667866in}{0.637787in}}{\pgfqpoint{5.670409in}{4.224664in}}{%
\pgfpathmoveto{\pgfqpoint{1.667866in}{0.637787in}}%
\pgfpathlineto{\pgfqpoint{1.667866in}{4.197072in}}%
\pgfpathlineto{\pgfqpoint{1.969009in}{4.224664in}}%
\pgfpathlineto{\pgfqpoint{2.523466in}{4.197072in}}%
\pgfpathlineto{\pgfqpoint{3.540266in}{4.224664in}}%
\pgfpathlineto{\pgfqpoint{4.271866in}{4.086707in}}%
\pgfpathlineto{\pgfqpoint{4.971580in}{3.465901in}}%
\pgfpathlineto{\pgfqpoint{5.522831in}{2.720934in}}%
\pgfpathlineto{\pgfqpoint{5.670409in}{2.264225in}}%
\pgfpathlineto{\pgfqpoint{5.345688in}{1.672463in}}%
\pgfpathlineto{\pgfqpoint{4.991403in}{1.327571in}}%
\pgfpathlineto{\pgfqpoint{2.695024in}{0.637787in}}%
\pgfpathlineto{\pgfqpoint{2.695024in}{0.637787in}}%
\pgfpathlineto{\pgfqpoint{2.695024in}{0.637787in}}%
\pgfpathlineto{\pgfqpoint{4.991403in}{0.637787in}}%
\pgfpathlineto{\pgfqpoint{5.345688in}{0.637787in}}%
\pgfpathlineto{\pgfqpoint{5.670409in}{0.637787in}}%
\pgfpathlineto{\pgfqpoint{5.522831in}{0.637787in}}%
\pgfpathlineto{\pgfqpoint{4.971580in}{0.637787in}}%
\pgfpathlineto{\pgfqpoint{4.271866in}{0.637787in}}%
\pgfpathlineto{\pgfqpoint{3.540266in}{0.637787in}}%
\pgfpathlineto{\pgfqpoint{2.523466in}{0.637787in}}%
\pgfpathlineto{\pgfqpoint{1.969009in}{0.637787in}}%
\pgfpathlineto{\pgfqpoint{1.667866in}{0.637787in}}%
\pgfpathclose%
\pgfusepath{fill}%
}%
\begin{pgfscope}%
\pgfsys@transformshift{0.000000in}{0.000000in}%
\pgfsys@useobject{currentmarker}{}%
\end{pgfscope}%
\end{pgfscope}%
\begin{pgfscope}%
\pgfsetbuttcap%
\pgfsetroundjoin%
\definecolor{currentfill}{rgb}{0.000000,0.000000,0.000000}%
\pgfsetfillcolor{currentfill}%
\pgfsetlinewidth{0.803000pt}%
\definecolor{currentstroke}{rgb}{0.000000,0.000000,0.000000}%
\pgfsetstrokecolor{currentstroke}%
\pgfsetdash{}{0pt}%
\pgfsys@defobject{currentmarker}{\pgfqpoint{0.000000in}{-0.048611in}}{\pgfqpoint{0.000000in}{0.000000in}}{%
\pgfpathmoveto{\pgfqpoint{0.000000in}{0.000000in}}%
\pgfpathlineto{\pgfqpoint{0.000000in}{-0.048611in}}%
\pgfusepath{stroke,fill}%
}%
\begin{pgfscope}%
\pgfsys@transformshift{0.729009in}{0.623991in}%
\pgfsys@useobject{currentmarker}{}%
\end{pgfscope}%
\end{pgfscope}%
\begin{pgfscope}%
\definecolor{textcolor}{rgb}{0.000000,0.000000,0.000000}%
\pgfsetstrokecolor{textcolor}%
\pgfsetfillcolor{textcolor}%
\pgftext[x=0.729009in,y=0.526769in,,top]{\color{textcolor}\rmfamily\fontsize{13.000000}{15.600000}\selectfont 0.0}%
\end{pgfscope}%
\begin{pgfscope}%
\pgfsetbuttcap%
\pgfsetroundjoin%
\definecolor{currentfill}{rgb}{0.000000,0.000000,0.000000}%
\pgfsetfillcolor{currentfill}%
\pgfsetlinewidth{0.803000pt}%
\definecolor{currentstroke}{rgb}{0.000000,0.000000,0.000000}%
\pgfsetstrokecolor{currentstroke}%
\pgfsetdash{}{0pt}%
\pgfsys@defobject{currentmarker}{\pgfqpoint{0.000000in}{-0.048611in}}{\pgfqpoint{0.000000in}{0.000000in}}{%
\pgfpathmoveto{\pgfqpoint{0.000000in}{0.000000in}}%
\pgfpathlineto{\pgfqpoint{0.000000in}{-0.048611in}}%
\pgfusepath{stroke,fill}%
}%
\begin{pgfscope}%
\pgfsys@transformshift{1.614723in}{0.623991in}%
\pgfsys@useobject{currentmarker}{}%
\end{pgfscope}%
\end{pgfscope}%
\begin{pgfscope}%
\definecolor{textcolor}{rgb}{0.000000,0.000000,0.000000}%
\pgfsetstrokecolor{textcolor}%
\pgfsetfillcolor{textcolor}%
\pgftext[x=1.614723in,y=0.526769in,,top]{\color{textcolor}\rmfamily\fontsize{13.000000}{15.600000}\selectfont 0.5}%
\end{pgfscope}%
\begin{pgfscope}%
\pgfsetbuttcap%
\pgfsetroundjoin%
\definecolor{currentfill}{rgb}{0.000000,0.000000,0.000000}%
\pgfsetfillcolor{currentfill}%
\pgfsetlinewidth{0.803000pt}%
\definecolor{currentstroke}{rgb}{0.000000,0.000000,0.000000}%
\pgfsetstrokecolor{currentstroke}%
\pgfsetdash{}{0pt}%
\pgfsys@defobject{currentmarker}{\pgfqpoint{0.000000in}{-0.048611in}}{\pgfqpoint{0.000000in}{0.000000in}}{%
\pgfpathmoveto{\pgfqpoint{0.000000in}{0.000000in}}%
\pgfpathlineto{\pgfqpoint{0.000000in}{-0.048611in}}%
\pgfusepath{stroke,fill}%
}%
\begin{pgfscope}%
\pgfsys@transformshift{2.500438in}{0.623991in}%
\pgfsys@useobject{currentmarker}{}%
\end{pgfscope}%
\end{pgfscope}%
\begin{pgfscope}%
\definecolor{textcolor}{rgb}{0.000000,0.000000,0.000000}%
\pgfsetstrokecolor{textcolor}%
\pgfsetfillcolor{textcolor}%
\pgftext[x=2.500438in,y=0.526769in,,top]{\color{textcolor}\rmfamily\fontsize{13.000000}{15.600000}\selectfont 1.0}%
\end{pgfscope}%
\begin{pgfscope}%
\pgfsetbuttcap%
\pgfsetroundjoin%
\definecolor{currentfill}{rgb}{0.000000,0.000000,0.000000}%
\pgfsetfillcolor{currentfill}%
\pgfsetlinewidth{0.803000pt}%
\definecolor{currentstroke}{rgb}{0.000000,0.000000,0.000000}%
\pgfsetstrokecolor{currentstroke}%
\pgfsetdash{}{0pt}%
\pgfsys@defobject{currentmarker}{\pgfqpoint{0.000000in}{-0.048611in}}{\pgfqpoint{0.000000in}{0.000000in}}{%
\pgfpathmoveto{\pgfqpoint{0.000000in}{0.000000in}}%
\pgfpathlineto{\pgfqpoint{0.000000in}{-0.048611in}}%
\pgfusepath{stroke,fill}%
}%
\begin{pgfscope}%
\pgfsys@transformshift{3.386152in}{0.623991in}%
\pgfsys@useobject{currentmarker}{}%
\end{pgfscope}%
\end{pgfscope}%
\begin{pgfscope}%
\definecolor{textcolor}{rgb}{0.000000,0.000000,0.000000}%
\pgfsetstrokecolor{textcolor}%
\pgfsetfillcolor{textcolor}%
\pgftext[x=3.386152in,y=0.526769in,,top]{\color{textcolor}\rmfamily\fontsize{13.000000}{15.600000}\selectfont 1.5}%
\end{pgfscope}%
\begin{pgfscope}%
\pgfsetbuttcap%
\pgfsetroundjoin%
\definecolor{currentfill}{rgb}{0.000000,0.000000,0.000000}%
\pgfsetfillcolor{currentfill}%
\pgfsetlinewidth{0.803000pt}%
\definecolor{currentstroke}{rgb}{0.000000,0.000000,0.000000}%
\pgfsetstrokecolor{currentstroke}%
\pgfsetdash{}{0pt}%
\pgfsys@defobject{currentmarker}{\pgfqpoint{0.000000in}{-0.048611in}}{\pgfqpoint{0.000000in}{0.000000in}}{%
\pgfpathmoveto{\pgfqpoint{0.000000in}{0.000000in}}%
\pgfpathlineto{\pgfqpoint{0.000000in}{-0.048611in}}%
\pgfusepath{stroke,fill}%
}%
\begin{pgfscope}%
\pgfsys@transformshift{4.271866in}{0.623991in}%
\pgfsys@useobject{currentmarker}{}%
\end{pgfscope}%
\end{pgfscope}%
\begin{pgfscope}%
\definecolor{textcolor}{rgb}{0.000000,0.000000,0.000000}%
\pgfsetstrokecolor{textcolor}%
\pgfsetfillcolor{textcolor}%
\pgftext[x=4.271866in,y=0.526769in,,top]{\color{textcolor}\rmfamily\fontsize{13.000000}{15.600000}\selectfont 2.0}%
\end{pgfscope}%
\begin{pgfscope}%
\pgfsetbuttcap%
\pgfsetroundjoin%
\definecolor{currentfill}{rgb}{0.000000,0.000000,0.000000}%
\pgfsetfillcolor{currentfill}%
\pgfsetlinewidth{0.803000pt}%
\definecolor{currentstroke}{rgb}{0.000000,0.000000,0.000000}%
\pgfsetstrokecolor{currentstroke}%
\pgfsetdash{}{0pt}%
\pgfsys@defobject{currentmarker}{\pgfqpoint{0.000000in}{-0.048611in}}{\pgfqpoint{0.000000in}{0.000000in}}{%
\pgfpathmoveto{\pgfqpoint{0.000000in}{0.000000in}}%
\pgfpathlineto{\pgfqpoint{0.000000in}{-0.048611in}}%
\pgfusepath{stroke,fill}%
}%
\begin{pgfscope}%
\pgfsys@transformshift{5.157580in}{0.623991in}%
\pgfsys@useobject{currentmarker}{}%
\end{pgfscope}%
\end{pgfscope}%
\begin{pgfscope}%
\definecolor{textcolor}{rgb}{0.000000,0.000000,0.000000}%
\pgfsetstrokecolor{textcolor}%
\pgfsetfillcolor{textcolor}%
\pgftext[x=5.157580in,y=0.526769in,,top]{\color{textcolor}\rmfamily\fontsize{13.000000}{15.600000}\selectfont 2.5}%
\end{pgfscope}%
\begin{pgfscope}%
\pgfsetbuttcap%
\pgfsetroundjoin%
\definecolor{currentfill}{rgb}{0.000000,0.000000,0.000000}%
\pgfsetfillcolor{currentfill}%
\pgfsetlinewidth{0.803000pt}%
\definecolor{currentstroke}{rgb}{0.000000,0.000000,0.000000}%
\pgfsetstrokecolor{currentstroke}%
\pgfsetdash{}{0pt}%
\pgfsys@defobject{currentmarker}{\pgfqpoint{0.000000in}{-0.048611in}}{\pgfqpoint{0.000000in}{0.000000in}}{%
\pgfpathmoveto{\pgfqpoint{0.000000in}{0.000000in}}%
\pgfpathlineto{\pgfqpoint{0.000000in}{-0.048611in}}%
\pgfusepath{stroke,fill}%
}%
\begin{pgfscope}%
\pgfsys@transformshift{6.043295in}{0.623991in}%
\pgfsys@useobject{currentmarker}{}%
\end{pgfscope}%
\end{pgfscope}%
\begin{pgfscope}%
\definecolor{textcolor}{rgb}{0.000000,0.000000,0.000000}%
\pgfsetstrokecolor{textcolor}%
\pgfsetfillcolor{textcolor}%
\pgftext[x=6.043295in,y=0.526769in,,top]{\color{textcolor}\rmfamily\fontsize{13.000000}{15.600000}\selectfont 3.0}%
\end{pgfscope}%
\begin{pgfscope}%
\pgfsetbuttcap%
\pgfsetroundjoin%
\definecolor{currentfill}{rgb}{0.000000,0.000000,0.000000}%
\pgfsetfillcolor{currentfill}%
\pgfsetlinewidth{0.803000pt}%
\definecolor{currentstroke}{rgb}{0.000000,0.000000,0.000000}%
\pgfsetstrokecolor{currentstroke}%
\pgfsetdash{}{0pt}%
\pgfsys@defobject{currentmarker}{\pgfqpoint{0.000000in}{-0.048611in}}{\pgfqpoint{0.000000in}{0.000000in}}{%
\pgfpathmoveto{\pgfqpoint{0.000000in}{0.000000in}}%
\pgfpathlineto{\pgfqpoint{0.000000in}{-0.048611in}}%
\pgfusepath{stroke,fill}%
}%
\begin{pgfscope}%
\pgfsys@transformshift{6.929009in}{0.623991in}%
\pgfsys@useobject{currentmarker}{}%
\end{pgfscope}%
\end{pgfscope}%
\begin{pgfscope}%
\definecolor{textcolor}{rgb}{0.000000,0.000000,0.000000}%
\pgfsetstrokecolor{textcolor}%
\pgfsetfillcolor{textcolor}%
\pgftext[x=6.929009in,y=0.526769in,,top]{\color{textcolor}\rmfamily\fontsize{13.000000}{15.600000}\selectfont 3.5}%
\end{pgfscope}%
\begin{pgfscope}%
\definecolor{textcolor}{rgb}{0.000000,0.000000,0.000000}%
\pgfsetstrokecolor{textcolor}%
\pgfsetfillcolor{textcolor}%
\pgftext[x=3.829009in,y=0.336954in,,top]{\color{textcolor}\rmfamily\fontsize{15.000000}{18.000000}\selectfont \(\displaystyle p_0\,[m^{2}M_{\mathrm{Pl}}^2]\)}%
\end{pgfscope}%
\begin{pgfscope}%
\definecolor{textcolor}{rgb}{0.000000,0.000000,0.000000}%
\pgfsetstrokecolor{textcolor}%
\pgfsetfillcolor{textcolor}%
\pgftext[x=6.929009in,y=0.336954in,right,top]{\color{textcolor}\rmfamily\fontsize{13.000000}{15.600000}\selectfont \(\displaystyle \times{10^{-3}}{}\)}%
\end{pgfscope}%
\begin{pgfscope}%
\pgfsetbuttcap%
\pgfsetroundjoin%
\definecolor{currentfill}{rgb}{0.000000,0.000000,0.000000}%
\pgfsetfillcolor{currentfill}%
\pgfsetlinewidth{0.803000pt}%
\definecolor{currentstroke}{rgb}{0.000000,0.000000,0.000000}%
\pgfsetstrokecolor{currentstroke}%
\pgfsetdash{}{0pt}%
\pgfsys@defobject{currentmarker}{\pgfqpoint{-0.048611in}{0.000000in}}{\pgfqpoint{-0.000000in}{0.000000in}}{%
\pgfpathmoveto{\pgfqpoint{-0.000000in}{0.000000in}}%
\pgfpathlineto{\pgfqpoint{-0.048611in}{0.000000in}}%
\pgfusepath{stroke,fill}%
}%
\begin{pgfscope}%
\pgfsys@transformshift{0.729009in}{0.637787in}%
\pgfsys@useobject{currentmarker}{}%
\end{pgfscope}%
\end{pgfscope}%
\begin{pgfscope}%
\definecolor{textcolor}{rgb}{0.000000,0.000000,0.000000}%
\pgfsetstrokecolor{textcolor}%
\pgfsetfillcolor{textcolor}%
\pgftext[x=0.341666in, y=0.579916in, left, base]{\color{textcolor}\rmfamily\fontsize{13.000000}{15.600000}\selectfont \(\displaystyle {0.00}\)}%
\end{pgfscope}%
\begin{pgfscope}%
\pgfsetbuttcap%
\pgfsetroundjoin%
\definecolor{currentfill}{rgb}{0.000000,0.000000,0.000000}%
\pgfsetfillcolor{currentfill}%
\pgfsetlinewidth{0.803000pt}%
\definecolor{currentstroke}{rgb}{0.000000,0.000000,0.000000}%
\pgfsetstrokecolor{currentstroke}%
\pgfsetdash{}{0pt}%
\pgfsys@defobject{currentmarker}{\pgfqpoint{-0.048611in}{0.000000in}}{\pgfqpoint{-0.000000in}{0.000000in}}{%
\pgfpathmoveto{\pgfqpoint{-0.000000in}{0.000000in}}%
\pgfpathlineto{\pgfqpoint{-0.048611in}{0.000000in}}%
\pgfusepath{stroke,fill}%
}%
\begin{pgfscope}%
\pgfsys@transformshift{0.729009in}{1.327571in}%
\pgfsys@useobject{currentmarker}{}%
\end{pgfscope}%
\end{pgfscope}%
\begin{pgfscope}%
\definecolor{textcolor}{rgb}{0.000000,0.000000,0.000000}%
\pgfsetstrokecolor{textcolor}%
\pgfsetfillcolor{textcolor}%
\pgftext[x=0.341666in, y=1.269700in, left, base]{\color{textcolor}\rmfamily\fontsize{13.000000}{15.600000}\selectfont \(\displaystyle {0.05}\)}%
\end{pgfscope}%
\begin{pgfscope}%
\pgfsetbuttcap%
\pgfsetroundjoin%
\definecolor{currentfill}{rgb}{0.000000,0.000000,0.000000}%
\pgfsetfillcolor{currentfill}%
\pgfsetlinewidth{0.803000pt}%
\definecolor{currentstroke}{rgb}{0.000000,0.000000,0.000000}%
\pgfsetstrokecolor{currentstroke}%
\pgfsetdash{}{0pt}%
\pgfsys@defobject{currentmarker}{\pgfqpoint{-0.048611in}{0.000000in}}{\pgfqpoint{-0.000000in}{0.000000in}}{%
\pgfpathmoveto{\pgfqpoint{-0.000000in}{0.000000in}}%
\pgfpathlineto{\pgfqpoint{-0.048611in}{0.000000in}}%
\pgfusepath{stroke,fill}%
}%
\begin{pgfscope}%
\pgfsys@transformshift{0.729009in}{2.017355in}%
\pgfsys@useobject{currentmarker}{}%
\end{pgfscope}%
\end{pgfscope}%
\begin{pgfscope}%
\definecolor{textcolor}{rgb}{0.000000,0.000000,0.000000}%
\pgfsetstrokecolor{textcolor}%
\pgfsetfillcolor{textcolor}%
\pgftext[x=0.341666in, y=1.959484in, left, base]{\color{textcolor}\rmfamily\fontsize{13.000000}{15.600000}\selectfont \(\displaystyle {0.10}\)}%
\end{pgfscope}%
\begin{pgfscope}%
\pgfsetbuttcap%
\pgfsetroundjoin%
\definecolor{currentfill}{rgb}{0.000000,0.000000,0.000000}%
\pgfsetfillcolor{currentfill}%
\pgfsetlinewidth{0.803000pt}%
\definecolor{currentstroke}{rgb}{0.000000,0.000000,0.000000}%
\pgfsetstrokecolor{currentstroke}%
\pgfsetdash{}{0pt}%
\pgfsys@defobject{currentmarker}{\pgfqpoint{-0.048611in}{0.000000in}}{\pgfqpoint{-0.000000in}{0.000000in}}{%
\pgfpathmoveto{\pgfqpoint{-0.000000in}{0.000000in}}%
\pgfpathlineto{\pgfqpoint{-0.048611in}{0.000000in}}%
\pgfusepath{stroke,fill}%
}%
\begin{pgfscope}%
\pgfsys@transformshift{0.729009in}{2.707139in}%
\pgfsys@useobject{currentmarker}{}%
\end{pgfscope}%
\end{pgfscope}%
\begin{pgfscope}%
\definecolor{textcolor}{rgb}{0.000000,0.000000,0.000000}%
\pgfsetstrokecolor{textcolor}%
\pgfsetfillcolor{textcolor}%
\pgftext[x=0.341666in, y=2.649268in, left, base]{\color{textcolor}\rmfamily\fontsize{13.000000}{15.600000}\selectfont \(\displaystyle {0.15}\)}%
\end{pgfscope}%
\begin{pgfscope}%
\pgfsetbuttcap%
\pgfsetroundjoin%
\definecolor{currentfill}{rgb}{0.000000,0.000000,0.000000}%
\pgfsetfillcolor{currentfill}%
\pgfsetlinewidth{0.803000pt}%
\definecolor{currentstroke}{rgb}{0.000000,0.000000,0.000000}%
\pgfsetstrokecolor{currentstroke}%
\pgfsetdash{}{0pt}%
\pgfsys@defobject{currentmarker}{\pgfqpoint{-0.048611in}{0.000000in}}{\pgfqpoint{-0.000000in}{0.000000in}}{%
\pgfpathmoveto{\pgfqpoint{-0.000000in}{0.000000in}}%
\pgfpathlineto{\pgfqpoint{-0.048611in}{0.000000in}}%
\pgfusepath{stroke,fill}%
}%
\begin{pgfscope}%
\pgfsys@transformshift{0.729009in}{3.396923in}%
\pgfsys@useobject{currentmarker}{}%
\end{pgfscope}%
\end{pgfscope}%
\begin{pgfscope}%
\definecolor{textcolor}{rgb}{0.000000,0.000000,0.000000}%
\pgfsetstrokecolor{textcolor}%
\pgfsetfillcolor{textcolor}%
\pgftext[x=0.341666in, y=3.339053in, left, base]{\color{textcolor}\rmfamily\fontsize{13.000000}{15.600000}\selectfont \(\displaystyle {0.20}\)}%
\end{pgfscope}%
\begin{pgfscope}%
\pgfsetbuttcap%
\pgfsetroundjoin%
\definecolor{currentfill}{rgb}{0.000000,0.000000,0.000000}%
\pgfsetfillcolor{currentfill}%
\pgfsetlinewidth{0.803000pt}%
\definecolor{currentstroke}{rgb}{0.000000,0.000000,0.000000}%
\pgfsetstrokecolor{currentstroke}%
\pgfsetdash{}{0pt}%
\pgfsys@defobject{currentmarker}{\pgfqpoint{-0.048611in}{0.000000in}}{\pgfqpoint{-0.000000in}{0.000000in}}{%
\pgfpathmoveto{\pgfqpoint{-0.000000in}{0.000000in}}%
\pgfpathlineto{\pgfqpoint{-0.048611in}{0.000000in}}%
\pgfusepath{stroke,fill}%
}%
\begin{pgfscope}%
\pgfsys@transformshift{0.729009in}{4.086707in}%
\pgfsys@useobject{currentmarker}{}%
\end{pgfscope}%
\end{pgfscope}%
\begin{pgfscope}%
\definecolor{textcolor}{rgb}{0.000000,0.000000,0.000000}%
\pgfsetstrokecolor{textcolor}%
\pgfsetfillcolor{textcolor}%
\pgftext[x=0.341666in, y=4.028837in, left, base]{\color{textcolor}\rmfamily\fontsize{13.000000}{15.600000}\selectfont \(\displaystyle {0.25}\)}%
\end{pgfscope}%
\begin{pgfscope}%
\pgfsetbuttcap%
\pgfsetroundjoin%
\definecolor{currentfill}{rgb}{0.000000,0.000000,0.000000}%
\pgfsetfillcolor{currentfill}%
\pgfsetlinewidth{0.803000pt}%
\definecolor{currentstroke}{rgb}{0.000000,0.000000,0.000000}%
\pgfsetstrokecolor{currentstroke}%
\pgfsetdash{}{0pt}%
\pgfsys@defobject{currentmarker}{\pgfqpoint{-0.048611in}{0.000000in}}{\pgfqpoint{-0.000000in}{0.000000in}}{%
\pgfpathmoveto{\pgfqpoint{-0.000000in}{0.000000in}}%
\pgfpathlineto{\pgfqpoint{-0.048611in}{0.000000in}}%
\pgfusepath{stroke,fill}%
}%
\begin{pgfscope}%
\pgfsys@transformshift{0.729009in}{4.776491in}%
\pgfsys@useobject{currentmarker}{}%
\end{pgfscope}%
\end{pgfscope}%
\begin{pgfscope}%
\definecolor{textcolor}{rgb}{0.000000,0.000000,0.000000}%
\pgfsetstrokecolor{textcolor}%
\pgfsetfillcolor{textcolor}%
\pgftext[x=0.341666in, y=4.718621in, left, base]{\color{textcolor}\rmfamily\fontsize{13.000000}{15.600000}\selectfont \(\displaystyle {0.30}\)}%
\end{pgfscope}%
\begin{pgfscope}%
\definecolor{textcolor}{rgb}{0.000000,0.000000,0.000000}%
\pgfsetstrokecolor{textcolor}%
\pgfsetfillcolor{textcolor}%
\pgftext[x=0.300000in,y=2.700241in,,bottom,rotate=90.000000]{\color{textcolor}\rmfamily\fontsize{15.000000}{18.000000}\selectfont \(\displaystyle \sigma_0 [M_{\mathrm{Pl}}]\)}%
\end{pgfscope}%
\begin{pgfscope}%
\pgfpathrectangle{\pgfqpoint{0.729009in}{0.623991in}}{\pgfqpoint{6.200000in}{4.152500in}}%
\pgfusepath{clip}%
\pgfsetrectcap%
\pgfsetroundjoin%
\pgfsetlinewidth{2.007500pt}%
\definecolor{currentstroke}{rgb}{0.376471,0.407843,0.431373}%
\pgfsetstrokecolor{currentstroke}%
\pgfsetdash{}{0pt}%
\pgfpathmoveto{\pgfqpoint{0.729009in}{4.086707in}}%
\pgfpathlineto{\pgfqpoint{1.278028in}{4.155685in}}%
\pgfpathlineto{\pgfqpoint{1.969009in}{4.224664in}}%
\pgfpathlineto{\pgfqpoint{2.523466in}{4.197072in}}%
\pgfpathlineto{\pgfqpoint{3.540266in}{4.224664in}}%
\pgfpathlineto{\pgfqpoint{4.271866in}{4.086707in}}%
\pgfpathlineto{\pgfqpoint{4.971580in}{3.465901in}}%
\pgfpathlineto{\pgfqpoint{5.522831in}{2.720934in}}%
\pgfpathlineto{\pgfqpoint{5.670409in}{2.264225in}}%
\pgfpathlineto{\pgfqpoint{5.345688in}{1.672463in}}%
\pgfpathlineto{\pgfqpoint{4.991403in}{1.327571in}}%
\pgfpathlineto{\pgfqpoint{2.695024in}{0.637787in}}%
\pgfusepath{stroke}%
\end{pgfscope}%
\begin{pgfscope}%
\pgfpathrectangle{\pgfqpoint{0.729009in}{0.623991in}}{\pgfqpoint{6.200000in}{4.152500in}}%
\pgfusepath{clip}%
\pgfsetrectcap%
\pgfsetroundjoin%
\pgfsetlinewidth{2.007500pt}%
\definecolor{currentstroke}{rgb}{0.109804,0.909804,0.243137}%
\pgfsetstrokecolor{currentstroke}%
\pgfsetdash{}{0pt}%
\pgfpathmoveto{\pgfqpoint{4.945009in}{1.327571in}}%
\pgfusepath{stroke}%
\end{pgfscope}%
\begin{pgfscope}%
\pgfpathrectangle{\pgfqpoint{0.729009in}{0.623991in}}{\pgfqpoint{6.200000in}{4.152500in}}%
\pgfusepath{clip}%
\pgfsetbuttcap%
\pgfsetmiterjoin%
\definecolor{currentfill}{rgb}{0.109804,0.909804,0.243137}%
\pgfsetfillcolor{currentfill}%
\pgfsetlinewidth{1.003750pt}%
\definecolor{currentstroke}{rgb}{0.109804,0.909804,0.243137}%
\pgfsetstrokecolor{currentstroke}%
\pgfsetdash{}{0pt}%
\pgfsys@defobject{currentmarker}{\pgfqpoint{-0.062500in}{-0.062500in}}{\pgfqpoint{0.062500in}{0.062500in}}{%
\pgfpathmoveto{\pgfqpoint{0.000000in}{0.062500in}}%
\pgfpathlineto{\pgfqpoint{-0.062500in}{-0.062500in}}%
\pgfpathlineto{\pgfqpoint{0.062500in}{-0.062500in}}%
\pgfpathclose%
\pgfusepath{stroke,fill}%
}%
\begin{pgfscope}%
\pgfsys@transformshift{4.945009in}{1.327571in}%
\pgfsys@useobject{currentmarker}{}%
\end{pgfscope}%
\end{pgfscope}%
\begin{pgfscope}%
\pgfpathrectangle{\pgfqpoint{0.729009in}{0.623991in}}{\pgfqpoint{6.200000in}{4.152500in}}%
\pgfusepath{clip}%
\pgfsetrectcap%
\pgfsetroundjoin%
\pgfsetlinewidth{2.007500pt}%
\definecolor{currentstroke}{rgb}{0.109804,0.909804,0.243137}%
\pgfsetstrokecolor{currentstroke}%
\pgfsetdash{}{0pt}%
\pgfpathmoveto{\pgfqpoint{0.759123in}{1.327571in}}%
\pgfusepath{stroke}%
\end{pgfscope}%
\begin{pgfscope}%
\pgfpathrectangle{\pgfqpoint{0.729009in}{0.623991in}}{\pgfqpoint{6.200000in}{4.152500in}}%
\pgfusepath{clip}%
\pgfsetbuttcap%
\pgfsetroundjoin%
\definecolor{currentfill}{rgb}{0.109804,0.909804,0.243137}%
\pgfsetfillcolor{currentfill}%
\pgfsetlinewidth{1.003750pt}%
\definecolor{currentstroke}{rgb}{0.109804,0.909804,0.243137}%
\pgfsetstrokecolor{currentstroke}%
\pgfsetdash{}{0pt}%
\pgfsys@defobject{currentmarker}{\pgfqpoint{-0.062500in}{-0.062500in}}{\pgfqpoint{0.062500in}{0.062500in}}{%
\pgfpathmoveto{\pgfqpoint{0.000000in}{-0.062500in}}%
\pgfpathcurveto{\pgfqpoint{0.016575in}{-0.062500in}}{\pgfqpoint{0.032474in}{-0.055915in}}{\pgfqpoint{0.044194in}{-0.044194in}}%
\pgfpathcurveto{\pgfqpoint{0.055915in}{-0.032474in}}{\pgfqpoint{0.062500in}{-0.016575in}}{\pgfqpoint{0.062500in}{0.000000in}}%
\pgfpathcurveto{\pgfqpoint{0.062500in}{0.016575in}}{\pgfqpoint{0.055915in}{0.032474in}}{\pgfqpoint{0.044194in}{0.044194in}}%
\pgfpathcurveto{\pgfqpoint{0.032474in}{0.055915in}}{\pgfqpoint{0.016575in}{0.062500in}}{\pgfqpoint{0.000000in}{0.062500in}}%
\pgfpathcurveto{\pgfqpoint{-0.016575in}{0.062500in}}{\pgfqpoint{-0.032474in}{0.055915in}}{\pgfqpoint{-0.044194in}{0.044194in}}%
\pgfpathcurveto{\pgfqpoint{-0.055915in}{0.032474in}}{\pgfqpoint{-0.062500in}{0.016575in}}{\pgfqpoint{-0.062500in}{0.000000in}}%
\pgfpathcurveto{\pgfqpoint{-0.062500in}{-0.016575in}}{\pgfqpoint{-0.055915in}{-0.032474in}}{\pgfqpoint{-0.044194in}{-0.044194in}}%
\pgfpathcurveto{\pgfqpoint{-0.032474in}{-0.055915in}}{\pgfqpoint{-0.016575in}{-0.062500in}}{\pgfqpoint{0.000000in}{-0.062500in}}%
\pgfpathclose%
\pgfusepath{stroke,fill}%
}%
\begin{pgfscope}%
\pgfsys@transformshift{0.759123in}{1.327571in}%
\pgfsys@useobject{currentmarker}{}%
\end{pgfscope}%
\end{pgfscope}%
\begin{pgfscope}%
\pgfpathrectangle{\pgfqpoint{0.729009in}{0.623991in}}{\pgfqpoint{6.200000in}{4.152500in}}%
\pgfusepath{clip}%
\pgfsetrectcap%
\pgfsetroundjoin%
\pgfsetlinewidth{2.007500pt}%
\definecolor{currentstroke}{rgb}{0.109804,0.909804,0.243137}%
\pgfsetstrokecolor{currentstroke}%
\pgfsetdash{}{0pt}%
\pgfpathmoveto{\pgfqpoint{5.157580in}{1.327571in}}%
\pgfusepath{stroke}%
\end{pgfscope}%
\begin{pgfscope}%
\pgfpathrectangle{\pgfqpoint{0.729009in}{0.623991in}}{\pgfqpoint{6.200000in}{4.152500in}}%
\pgfusepath{clip}%
\pgfsetbuttcap%
\pgfsetmiterjoin%
\definecolor{currentfill}{rgb}{0.109804,0.909804,0.243137}%
\pgfsetfillcolor{currentfill}%
\pgfsetlinewidth{1.003750pt}%
\definecolor{currentstroke}{rgb}{0.109804,0.909804,0.243137}%
\pgfsetstrokecolor{currentstroke}%
\pgfsetdash{}{0pt}%
\pgfsys@defobject{currentmarker}{\pgfqpoint{-0.069444in}{-0.069444in}}{\pgfqpoint{0.069444in}{0.069444in}}{%
\pgfpathmoveto{\pgfqpoint{-0.000000in}{-0.069444in}}%
\pgfpathlineto{\pgfqpoint{0.069444in}{0.069444in}}%
\pgfpathlineto{\pgfqpoint{-0.069444in}{0.069444in}}%
\pgfpathclose%
\pgfusepath{stroke,fill}%
}%
\begin{pgfscope}%
\pgfsys@transformshift{5.157580in}{1.327571in}%
\pgfsys@useobject{currentmarker}{}%
\end{pgfscope}%
\end{pgfscope}%
\begin{pgfscope}%
\pgfpathrectangle{\pgfqpoint{0.729009in}{0.623991in}}{\pgfqpoint{6.200000in}{4.152500in}}%
\pgfusepath{clip}%
\pgfsetrectcap%
\pgfsetroundjoin%
\pgfsetlinewidth{2.007500pt}%
\definecolor{currentstroke}{rgb}{0.792157,0.298039,0.901961}%
\pgfsetstrokecolor{currentstroke}%
\pgfsetdash{}{0pt}%
\pgfpathmoveto{\pgfqpoint{4.271866in}{4.086707in}}%
\pgfusepath{stroke}%
\end{pgfscope}%
\begin{pgfscope}%
\pgfpathrectangle{\pgfqpoint{0.729009in}{0.623991in}}{\pgfqpoint{6.200000in}{4.152500in}}%
\pgfusepath{clip}%
\pgfsetbuttcap%
\pgfsetmiterjoin%
\definecolor{currentfill}{rgb}{0.792157,0.298039,0.901961}%
\pgfsetfillcolor{currentfill}%
\pgfsetlinewidth{1.003750pt}%
\definecolor{currentstroke}{rgb}{0.792157,0.298039,0.901961}%
\pgfsetstrokecolor{currentstroke}%
\pgfsetdash{}{0pt}%
\pgfsys@defobject{currentmarker}{\pgfqpoint{-0.062500in}{-0.062500in}}{\pgfqpoint{0.062500in}{0.062500in}}{%
\pgfpathmoveto{\pgfqpoint{0.000000in}{0.062500in}}%
\pgfpathlineto{\pgfqpoint{-0.062500in}{-0.062500in}}%
\pgfpathlineto{\pgfqpoint{0.062500in}{-0.062500in}}%
\pgfpathclose%
\pgfusepath{stroke,fill}%
}%
\begin{pgfscope}%
\pgfsys@transformshift{4.271866in}{4.086707in}%
\pgfsys@useobject{currentmarker}{}%
\end{pgfscope}%
\end{pgfscope}%
\begin{pgfscope}%
\pgfpathrectangle{\pgfqpoint{0.729009in}{0.623991in}}{\pgfqpoint{6.200000in}{4.152500in}}%
\pgfusepath{clip}%
\pgfsetrectcap%
\pgfsetroundjoin%
\pgfsetlinewidth{2.007500pt}%
\definecolor{currentstroke}{rgb}{0.792157,0.298039,0.901961}%
\pgfsetstrokecolor{currentstroke}%
\pgfsetdash{}{0pt}%
\pgfpathmoveto{\pgfqpoint{1.349009in}{4.086707in}}%
\pgfusepath{stroke}%
\end{pgfscope}%
\begin{pgfscope}%
\pgfpathrectangle{\pgfqpoint{0.729009in}{0.623991in}}{\pgfqpoint{6.200000in}{4.152500in}}%
\pgfusepath{clip}%
\pgfsetbuttcap%
\pgfsetroundjoin%
\definecolor{currentfill}{rgb}{0.792157,0.298039,0.901961}%
\pgfsetfillcolor{currentfill}%
\pgfsetlinewidth{1.003750pt}%
\definecolor{currentstroke}{rgb}{0.792157,0.298039,0.901961}%
\pgfsetstrokecolor{currentstroke}%
\pgfsetdash{}{0pt}%
\pgfsys@defobject{currentmarker}{\pgfqpoint{-0.062500in}{-0.062500in}}{\pgfqpoint{0.062500in}{0.062500in}}{%
\pgfpathmoveto{\pgfqpoint{0.000000in}{-0.062500in}}%
\pgfpathcurveto{\pgfqpoint{0.016575in}{-0.062500in}}{\pgfqpoint{0.032474in}{-0.055915in}}{\pgfqpoint{0.044194in}{-0.044194in}}%
\pgfpathcurveto{\pgfqpoint{0.055915in}{-0.032474in}}{\pgfqpoint{0.062500in}{-0.016575in}}{\pgfqpoint{0.062500in}{0.000000in}}%
\pgfpathcurveto{\pgfqpoint{0.062500in}{0.016575in}}{\pgfqpoint{0.055915in}{0.032474in}}{\pgfqpoint{0.044194in}{0.044194in}}%
\pgfpathcurveto{\pgfqpoint{0.032474in}{0.055915in}}{\pgfqpoint{0.016575in}{0.062500in}}{\pgfqpoint{0.000000in}{0.062500in}}%
\pgfpathcurveto{\pgfqpoint{-0.016575in}{0.062500in}}{\pgfqpoint{-0.032474in}{0.055915in}}{\pgfqpoint{-0.044194in}{0.044194in}}%
\pgfpathcurveto{\pgfqpoint{-0.055915in}{0.032474in}}{\pgfqpoint{-0.062500in}{0.016575in}}{\pgfqpoint{-0.062500in}{0.000000in}}%
\pgfpathcurveto{\pgfqpoint{-0.062500in}{-0.016575in}}{\pgfqpoint{-0.055915in}{-0.032474in}}{\pgfqpoint{-0.044194in}{-0.044194in}}%
\pgfpathcurveto{\pgfqpoint{-0.032474in}{-0.055915in}}{\pgfqpoint{-0.016575in}{-0.062500in}}{\pgfqpoint{0.000000in}{-0.062500in}}%
\pgfpathclose%
\pgfusepath{stroke,fill}%
}%
\begin{pgfscope}%
\pgfsys@transformshift{1.349009in}{4.086707in}%
\pgfsys@useobject{currentmarker}{}%
\end{pgfscope}%
\end{pgfscope}%
\begin{pgfscope}%
\pgfpathrectangle{\pgfqpoint{0.729009in}{0.623991in}}{\pgfqpoint{6.200000in}{4.152500in}}%
\pgfusepath{clip}%
\pgfsetrectcap%
\pgfsetroundjoin%
\pgfsetlinewidth{2.007500pt}%
\definecolor{currentstroke}{rgb}{0.792157,0.298039,0.901961}%
\pgfsetstrokecolor{currentstroke}%
\pgfsetdash{}{0pt}%
\pgfpathmoveto{\pgfqpoint{4.449009in}{4.086707in}}%
\pgfusepath{stroke}%
\end{pgfscope}%
\begin{pgfscope}%
\pgfpathrectangle{\pgfqpoint{0.729009in}{0.623991in}}{\pgfqpoint{6.200000in}{4.152500in}}%
\pgfusepath{clip}%
\pgfsetbuttcap%
\pgfsetmiterjoin%
\definecolor{currentfill}{rgb}{0.792157,0.298039,0.901961}%
\pgfsetfillcolor{currentfill}%
\pgfsetlinewidth{1.003750pt}%
\definecolor{currentstroke}{rgb}{0.792157,0.298039,0.901961}%
\pgfsetstrokecolor{currentstroke}%
\pgfsetdash{}{0pt}%
\pgfsys@defobject{currentmarker}{\pgfqpoint{-0.069444in}{-0.069444in}}{\pgfqpoint{0.069444in}{0.069444in}}{%
\pgfpathmoveto{\pgfqpoint{-0.000000in}{-0.069444in}}%
\pgfpathlineto{\pgfqpoint{0.069444in}{0.069444in}}%
\pgfpathlineto{\pgfqpoint{-0.069444in}{0.069444in}}%
\pgfpathclose%
\pgfusepath{stroke,fill}%
}%
\begin{pgfscope}%
\pgfsys@transformshift{4.449009in}{4.086707in}%
\pgfsys@useobject{currentmarker}{}%
\end{pgfscope}%
\end{pgfscope}%
\begin{pgfscope}%
\pgfpathrectangle{\pgfqpoint{0.729009in}{0.623991in}}{\pgfqpoint{6.200000in}{4.152500in}}%
\pgfusepath{clip}%
\pgfsetrectcap%
\pgfsetroundjoin%
\pgfsetlinewidth{2.007500pt}%
\definecolor{currentstroke}{rgb}{0.000000,0.000000,0.000000}%
\pgfsetstrokecolor{currentstroke}%
\pgfsetdash{}{0pt}%
\pgfpathmoveto{\pgfqpoint{0.746723in}{4.086707in}}%
\pgfusepath{stroke}%
\end{pgfscope}%
\begin{pgfscope}%
\pgfpathrectangle{\pgfqpoint{0.729009in}{0.623991in}}{\pgfqpoint{6.200000in}{4.152500in}}%
\pgfusepath{clip}%
\pgfsetbuttcap%
\pgfsetbeveljoin%
\definecolor{currentfill}{rgb}{0.000000,0.000000,0.000000}%
\pgfsetfillcolor{currentfill}%
\pgfsetlinewidth{1.003750pt}%
\definecolor{currentstroke}{rgb}{0.000000,0.000000,0.000000}%
\pgfsetstrokecolor{currentstroke}%
\pgfsetdash{}{0pt}%
\pgfsys@defobject{currentmarker}{\pgfqpoint{-0.066046in}{-0.056182in}}{\pgfqpoint{0.066046in}{0.069444in}}{%
\pgfpathmoveto{\pgfqpoint{0.000000in}{0.069444in}}%
\pgfpathlineto{\pgfqpoint{-0.015591in}{0.021460in}}%
\pgfpathlineto{\pgfqpoint{-0.066046in}{0.021460in}}%
\pgfpathlineto{\pgfqpoint{-0.025227in}{-0.008197in}}%
\pgfpathlineto{\pgfqpoint{-0.040818in}{-0.056182in}}%
\pgfpathlineto{\pgfqpoint{-0.000000in}{-0.026525in}}%
\pgfpathlineto{\pgfqpoint{0.040818in}{-0.056182in}}%
\pgfpathlineto{\pgfqpoint{0.025227in}{-0.008197in}}%
\pgfpathlineto{\pgfqpoint{0.066046in}{0.021460in}}%
\pgfpathlineto{\pgfqpoint{0.015591in}{0.021460in}}%
\pgfpathclose%
\pgfusepath{stroke,fill}%
}%
\begin{pgfscope}%
\pgfsys@transformshift{0.746723in}{4.086707in}%
\pgfsys@useobject{currentmarker}{}%
\end{pgfscope}%
\end{pgfscope}%
\begin{pgfscope}%
\pgfpathrectangle{\pgfqpoint{0.729009in}{0.623991in}}{\pgfqpoint{6.200000in}{4.152500in}}%
\pgfusepath{clip}%
\pgfsetrectcap%
\pgfsetroundjoin%
\pgfsetlinewidth{2.007500pt}%
\definecolor{currentstroke}{rgb}{0.000000,0.000000,0.000000}%
\pgfsetstrokecolor{currentstroke}%
\pgfsetdash{}{0pt}%
\pgfpathmoveto{\pgfqpoint{2.695024in}{0.651582in}}%
\pgfusepath{stroke}%
\end{pgfscope}%
\begin{pgfscope}%
\pgfpathrectangle{\pgfqpoint{0.729009in}{0.623991in}}{\pgfqpoint{6.200000in}{4.152500in}}%
\pgfusepath{clip}%
\pgfsetbuttcap%
\pgfsetbeveljoin%
\definecolor{currentfill}{rgb}{0.000000,0.000000,0.000000}%
\pgfsetfillcolor{currentfill}%
\pgfsetlinewidth{1.003750pt}%
\definecolor{currentstroke}{rgb}{0.000000,0.000000,0.000000}%
\pgfsetstrokecolor{currentstroke}%
\pgfsetdash{}{0pt}%
\pgfsys@defobject{currentmarker}{\pgfqpoint{-0.066046in}{-0.056182in}}{\pgfqpoint{0.066046in}{0.069444in}}{%
\pgfpathmoveto{\pgfqpoint{0.000000in}{0.069444in}}%
\pgfpathlineto{\pgfqpoint{-0.015591in}{0.021460in}}%
\pgfpathlineto{\pgfqpoint{-0.066046in}{0.021460in}}%
\pgfpathlineto{\pgfqpoint{-0.025227in}{-0.008197in}}%
\pgfpathlineto{\pgfqpoint{-0.040818in}{-0.056182in}}%
\pgfpathlineto{\pgfqpoint{-0.000000in}{-0.026525in}}%
\pgfpathlineto{\pgfqpoint{0.040818in}{-0.056182in}}%
\pgfpathlineto{\pgfqpoint{0.025227in}{-0.008197in}}%
\pgfpathlineto{\pgfqpoint{0.066046in}{0.021460in}}%
\pgfpathlineto{\pgfqpoint{0.015591in}{0.021460in}}%
\pgfpathclose%
\pgfusepath{stroke,fill}%
}%
\begin{pgfscope}%
\pgfsys@transformshift{2.695024in}{0.651582in}%
\pgfsys@useobject{currentmarker}{}%
\end{pgfscope}%
\end{pgfscope}%
\begin{pgfscope}%
\pgfsetrectcap%
\pgfsetmiterjoin%
\pgfsetlinewidth{1.003750pt}%
\definecolor{currentstroke}{rgb}{0.000000,0.000000,0.000000}%
\pgfsetstrokecolor{currentstroke}%
\pgfsetdash{}{0pt}%
\pgfpathmoveto{\pgfqpoint{0.729009in}{0.623991in}}%
\pgfpathlineto{\pgfqpoint{0.729009in}{4.776491in}}%
\pgfusepath{stroke}%
\end{pgfscope}%
\begin{pgfscope}%
\pgfsetrectcap%
\pgfsetmiterjoin%
\pgfsetlinewidth{1.003750pt}%
\definecolor{currentstroke}{rgb}{0.000000,0.000000,0.000000}%
\pgfsetstrokecolor{currentstroke}%
\pgfsetdash{}{0pt}%
\pgfpathmoveto{\pgfqpoint{6.929009in}{0.623991in}}%
\pgfpathlineto{\pgfqpoint{6.929009in}{4.776491in}}%
\pgfusepath{stroke}%
\end{pgfscope}%
\begin{pgfscope}%
\pgfsetrectcap%
\pgfsetmiterjoin%
\pgfsetlinewidth{1.003750pt}%
\definecolor{currentstroke}{rgb}{0.000000,0.000000,0.000000}%
\pgfsetstrokecolor{currentstroke}%
\pgfsetdash{}{0pt}%
\pgfpathmoveto{\pgfqpoint{0.729009in}{0.623991in}}%
\pgfpathlineto{\pgfqpoint{6.929009in}{0.623991in}}%
\pgfusepath{stroke}%
\end{pgfscope}%
\begin{pgfscope}%
\pgfsetrectcap%
\pgfsetmiterjoin%
\pgfsetlinewidth{1.003750pt}%
\definecolor{currentstroke}{rgb}{0.000000,0.000000,0.000000}%
\pgfsetstrokecolor{currentstroke}%
\pgfsetdash{}{0pt}%
\pgfpathmoveto{\pgfqpoint{0.729009in}{4.776491in}}%
\pgfpathlineto{\pgfqpoint{6.929009in}{4.776491in}}%
\pgfusepath{stroke}%
\end{pgfscope}%
\begin{pgfscope}%
\pgfsetfillopacity{0.600000}%
\pgfsetstrokeopacity{0.600000}%
\definecolor{textcolor}{rgb}{0.000000,0.000000,0.000000}%
\pgfsetstrokecolor{textcolor}%
\pgfsetfillcolor{textcolor}%
\pgftext[x=1.969009in,y=2.707139in,left,base]{\color{textcolor}\rmfamily\fontsize{20.000000}{24.000000}\selectfont Stable Static}%
\end{pgfscope}%
\begin{pgfscope}%
\pgfsetfillopacity{0.600000}%
\pgfsetstrokeopacity{0.600000}%
\definecolor{textcolor}{rgb}{0.000000,0.000000,0.000000}%
\pgfsetstrokecolor{textcolor}%
\pgfsetfillcolor{textcolor}%
\pgftext[x=1.969009in,y=2.431225in,left,base]{\color{textcolor}\rmfamily\fontsize{20.000000}{24.000000}\selectfont Solutions}%
\end{pgfscope}%
\begin{pgfscope}%
\pgfsetfillopacity{0.600000}%
\pgfsetstrokeopacity{0.600000}%
\definecolor{textcolor}{rgb}{0.000000,0.000000,0.000000}%
\pgfsetstrokecolor{textcolor}%
\pgfsetfillcolor{textcolor}%
\pgftext[x=4.714723in,y=4.362620in,left,base]{\color{textcolor}\rmfamily\fontsize{20.000000}{24.000000}\selectfont Unstable Static}%
\end{pgfscope}%
\begin{pgfscope}%
\pgfsetfillopacity{0.600000}%
\pgfsetstrokeopacity{0.600000}%
\definecolor{textcolor}{rgb}{0.000000,0.000000,0.000000}%
\pgfsetstrokecolor{textcolor}%
\pgfsetfillcolor{textcolor}%
\pgftext[x=4.714723in,y=4.086707in,left,base]{\color{textcolor}\rmfamily\fontsize{20.000000}{24.000000}\selectfont Solutions}%
\end{pgfscope}%
\begin{pgfscope}%
\definecolor{textcolor}{rgb}{0.000000,0.000000,0.000000}%
\pgfsetstrokecolor{textcolor}%
\pgfsetfillcolor{textcolor}%
\pgftext[x=4.643866in,y=0.982679in,left,base]{\color{textcolor}\rmfamily\fontsize{16.000000}{19.200000}\selectfont \(\displaystyle c_4=-1/2\)}%
\end{pgfscope}%
\begin{pgfscope}%
\definecolor{textcolor}{rgb}{0.000000,0.000000,0.000000}%
\pgfsetstrokecolor{textcolor}%
\pgfsetfillcolor{textcolor}%
\pgftext[x=4.626152in,y=0.775743in,left,base]{\color{textcolor}\rmfamily\fontsize{16.000000}{19.200000}\selectfont with \(\displaystyle \Lambda= 1/5 M^{1/3}_{\mathrm{Pl}}m^{2/3}\)}%
\end{pgfscope}%
\end{pgfpicture}%
\makeatother%
\endgroup%